\documentclass[3p]{elsarticle}

\usepackage{lineno}
\usepackage[hidelinks]{hyperref}
\usepackage{amssymb}
\usepackage{amsmath}
\usepackage{xcolor}
\usepackage{enumitem}
\usepackage{pifont}
\usepackage{caption}
\usepackage{subcaption}
\usepackage{float}
\usepackage[noabbrev,nameinlink]{cleveref}
\usepackage{algorithm}
\usepackage{algpseudocode}

\usepackage[nogroupskip,nonumberlist]{glossaries}
\usepackage[automake]{glossaries-extra}
\usepackage{nomencl}

\newlist{todolist}{itemize}{2}
\setlist[todolist]{label=$\square$}

\newcommand{\instructions}[1]{{\color{red}\tt #1}}
\renewcommand{\instructions}[1]{}

\newcommand{\algorithmicbreak}{\textbf{break}}

\journal{Elsevier}

\newglossary[ssg]{subscripts}{sbs}{ssc}{Subscripts and Sets}
\newglossary[prg]{parameters}{prm}{par}{Parameters}
\newglossary[vrg]{variables}{vrb}{var}{Variables}
\newglossary[abg]{acronyms}{abv}{abr}{Acronyms}
\makeglossaries

\setglossarypreamble[subscripts]{\vspace*{-\baselineskip}}
\setglossarypreamble[parameters]{\vspace*{-\baselineskip}}
\setglossarypreamble[variables]{\vspace*{-\baselineskip}}
\setglossarypreamble[acronyms]{\vspace*{-\baselineskip}}

\newglossarystyle{myglossary}%
{%
        {\begin{longtable}[l]{l lp{\glsdescwidth}}}%
        {\end{longtable}}%

}

\begin{document}
\begin{frontmatter}

\title{Modelling the Formation of Peer-to-Peer Trading Coalitions and Prosumer Participation Incentives in Transactive Energy Communities}

\author[1,2]{Ying Zhang}
\author[1,3,6]{Valentin Robu}
\author[1,4]{Sho Cremers}
\author[5]{Sonam Norbu}
\author[5]{Benoit Couraud}
\author[5]{Merlinda Andoni}
\author[5]{David Flynn}
\author[6]{H. Vincent Poor}
\address[1]{Intelligent and Autonomous Systems Group, CWI (National Research Centre for Mathematics and Computer Science), Amsterdam, The Netherlands}
\address[2]{Algorithmics Group, Delft University of Technology, Delft, The Netherlands}
\address[3]{Electrical Engineering Department, Eindhoven University of Technology, Eindhoven, The Netherlands}
\address[4]{Intelligent Electrical Power Grids Group, Delft University of Technology, Delft, The Netherlands}
\address[5]{James Watt School of Engineering, University of Glasgow, Glasgow, UK}
\address[6]{Electrical and Computer Engineering, Princeton University, Princeton, NJ, USA}

\begin{abstract}
  Peer-to-peer (P2P) energy trading and energy communities have garnered much attention over in recent years due to increasing investments in local energy generation and storage assets. Much research has been performed on the mechanisms and methodologies behind their implementation and realisation. However, the efficiency to be gained from P2P trading, and the structure of local energy markets raise many important challenges. To analyse the efficiency of P2P energy markets, in this work, we consider two different popular approaches to peer-to-peer trading: centralised (through a central market maker/clearing entity) vs. fully decentralised (P2P), and explore the comparative economic benefits of these models. We focus on the metric of Gains from Trade (GT), given optimal P2P trading schedule computed by a schedule optimiser. In both local market models, benefits from trading are realised mainly due to the diversity in consumption behaviour and renewable energy generation between prosumers in an energy community. 
Both market models will lead to the most promising P2P contracts (the ones with the highest Gains from Trade) to be established first. Yet, we find diversity decreases quickly as more peer-to-peer energy contracts are established and more prosumers join the market, leading to significantly diminishing returns. In this work, we aim to quantify this effect using real-world data from two large-scale smart energy trials in the UK, i.e. the Low Carbon London project and the Thames Valley Vision project. 
Our experimental study shows that, for both market models, only a small number of P2P contracts i.e. less than 10\% of the possible P2P contracts are required to achieve the majority of the maximal potential Gains from Trade. Similarly, only a fraction of prosumers are required to participate in energy trading to realise significant GT; namely we found that 60\% of the maximal GT can be realised with only 30\% of prosumers' participation, with the percentage of maximal GT reaching 80\% when participation increases to 50\% of prosumers. 
Finally, we study the effect that diversity in consumption profiles has on overall trading potential and dynamics in an energy community. We show that in a community with a DF(load diversity factor)=1, 80\% of potential maximal GT can be achieved by 10\% of prosumers engaging in P2P trading, while in a community with DF=1.5, it is beneficial for 40\% of the prosumers to trade. 

\end{abstract}


\end{frontmatter}

\newglossaryentry{time}{
type=variables,
name=\ensuremath{t},
description={time step}
}

\newglossaryentry{agent}{
type=subscripts,
name=\ensuremath{i}, 
description={for agents}
}

\newglossaryentry{agent_j}{
type=subscripts,
name=\ensuremath{j}, 
description={for agents}
}

\newglossaryentry{community}{
type=subscripts,
name=\ensuremath{N},
description={set of all agents in the community considered in the model}
}

\newglossaryentry{k}{
type=parameters,
name=\ensuremath{k},
description={number of peers in the negotiation framework}
}

\newglossaryentry{trading_community}{
type=subscripts,
name=\ensuremath{K},
description={[Sub]set of agents in the coalition $(K \subseteq N)$}
}

\newglossaryentry{time_horizon}{
type=parameters,
name=\ensuremath{T},
description={time horizon}
}

\newglossaryentry{export_tariff}{
type=parameters,
name=\ensuremath{\tau^s(\gls{time})}, 
description={selling price (i.e. export tariff) at \gls{time} [pence/kWh]}
}

\newglossaryentry{import_tariff}{
type=parameters,
name=\ensuremath{\tau^b(\gls{time})}, 
description={buying price (i.e. import tariff) at \gls{time} [pence/kWh]}
}

\newglossaryentry{PowerPVinstalled}{
type=parameters,
name=\ensuremath{P_{PV_{\gls{agent}}}}, 
description={Solar PV Power installed at agent \gls{agent} premises}
}

\newglossaryentry{BatteryCapacity}{
type=parameters,
name=\ensuremath{E_{bat_{\gls{agent}}}}, 
description={Battery capacity installed at agent \gls{agent} premises}
}

\newglossaryentry{net_demand}{
type=variables,
name=\ensuremath{e_{\gls{agent}}(\gls{time})}, 
description={net demand for agent \gls{agent}}
}

\newglossaryentry{energy_imported}{
type=variables,
name=\ensuremath{e^b_{\gls{agent}}(\gls{time})}, 
description={energy imported by agent \gls{agent}}
}

\newglossaryentry{energy_exported}{
type=variables,
name=\ensuremath{e^s_{\gls{agent}}(\gls{time})}, 
description={energy exported by agent \gls{agent}}
}

\newglossaryentry{energy_traded}{
type=variables,
name=\ensuremath{e^{p2p}_{\gls{agent}}(\gls{time})}, 
description={energy traded by agent \gls{agent}}
}

\newglossaryentry{net_energy_sold}{
type=variables,
name=\ensuremath{\widetilde{e^s_{\gls{agent}}}(\gls{time})},
description={energy sold by agent \gls{agent} after peer-to-peer trading}
}

\newglossaryentry{net_energy_bought}{
type=variables,
name=\ensuremath{\widetilde{e^b_{\gls{agent}}}(\gls{time})},
description={energy bought by agent \gls{agent} after peer-to-peer trading}
}
\newglossaryentry{net_energy_after_trade}{
type=variables,
name=\ensuremath{\widetilde{e_{\gls{agent}}}(\gls{time})},
description={net demand for agent \gls{agent} after peer-to-peer trading}
}

\newglossaryentry{energy_imported_c}{
type=variables,
name=\ensuremath{e^b_{\gls{trading_community}}(\gls{time})}, 
description={energy imported by \gls{trading_community}}
}

\newglossaryentry{energy_exported_c}{
type=variables,
name=\ensuremath{e^s_{\gls{trading_community}}(\gls{time})}, 
description={energy exported by \gls{trading_community}}
}

\newglossaryentry{demand}{
type=variables,
name=\ensuremath{d_{\gls{agent}}(\gls{time})}, 
description={demand for agent \gls{agent}}
}

\newglossaryentry{generation}{
type=variables,
name=\ensuremath{g_{\gls{agent}}(\gls{time})}, 
description={generation for agent \gls{agent}}
}

\newglossaryentry{demand_c}{
type=variables,
name=\ensuremath{d_{\gls{trading_community}}(\gls{time})}, 
description={aggregated demand for \gls{trading_community}}
}

\newglossaryentry{generation_c}{
type=variables,
name=\ensuremath{g_{\gls{trading_community}}(\gls{time})}, 
description={aggregated generation for \gls{trading_community}}
}

\newglossaryentry{battery_power}{
type=variables,
name=\ensuremath{p^{bat}(\gls{time})}, 
description={battery power at time \gls{time}}
}
\newglossaryentry{bill}{
type=variables,
name=\ensuremath{b_{\gls{agent}}(\gls{time_horizon})}, 
description={accumulated bill for agent \gls{agent} at time \gls{time_horizon}}
}

\newglossaryentry{bill_after_trade}{
type=variables,
name=\ensuremath{\widetilde{b_{\gls{agent}}}(\gls{time_horizon})},
description={accumulated bill for agent \gls{agent} at time \gls{time_horizon} after trading}
}

\newglossaryentry{bill_c}{
type=variables,
name=\ensuremath{b_{\gls{trading_community}}(T)}, 
description={accumulated bill for P2P coalition \gls{trading_community} at time \gls{time_horizon}}
}

\newglossaryentry{bill_n}{
type=variables,
name=\ensuremath{b_{\gls{community}}(T)}, 
description={accumulated bill for community \gls{community} at time \gls{time_horizon}}
}

\newglossaryentry{bill_k}{
type=variables,
name=\ensuremath{b_{\gls{trading_community}}(T)}, 
description={accumulated bill for coalition \gls{trading_community} at time \gls{time_horizon}}
}

\newglossaryentry{battery_degradation}{
type=variables,
name=\ensuremath{c^{bat}_{\gls{agent}}(\gls{time_horizon})},
description={battery degradation cost for agent \gls{agent} over \gls{time_horizon}}
}

\newglossaryentry{generator_cost}{
type=variables,
name=\ensuremath{c^g_{\gls{agent}}(\gls{time_horizon})},
description={generator costs for agent \gls{agent} over \gls{time_horizon}}
}

\newglossaryentry{battery_degradation_c}{
type=variables,
name=\ensuremath{c^{bat}_{\gls{trading_community}}(\gls{time_horizon})},
description={battery degradation cost for \gls{trading_community} over \gls{time_horizon}}
}

\newglossaryentry{generator_cost_c}{
type=variables,
name=\ensuremath{c^g_{\gls{trading_community}}(\gls{time_horizon})},
description={generator costs for \gls{trading_community} over \gls{time_horizon}}
}

\newglossaryentry{depreciation_factor}{
type=acronyms,
name=\ensuremath{DF},
description={Depreciation Factor}
}

\newglossaryentry{oracle}{
type=variables,
name=\ensuremath{\Theta},
description={peer-to-peer oracle}
}

\newglossaryentry{trades}{
type=variables,
name=\ensuremath{\theta},
description={Energy amounts traded decided by \gls{oracle}}
}
\newglossaryentry{trades_t}{
type=variables,
name=\ensuremath{\theta(\gls{time})},
description={Energy amount traded at time $t$ decided by \gls{oracle}}
}

\newglossaryentry{contract}{
type=variables,
name=\ensuremath{\omega},
description={energy contract tuple $(\gls{agent}, \gls{agent_j}, \gls{trades})$}}

\newglossaryentry{all_contracts}{
type=variables,
name=\ensuremath{\Omega},
description={Set of all possible energy contracts}}

\newglossaryentry{contracts}{
type=variables,
name=\ensuremath{\Omega_{\gls{agent}}},
description={Set of all possible energy contracts for agent $i$}}

\newglossaryentry{accepted_contracts}{
type=variables,
name=\ensuremath{\Omega^a_{\gls{agent}}},
description={energy contracts accepted by agent \gls{agent}}
}
\newglossaryentry{value_of_contract}{
type=variables,
name=\ensuremath{v_{\gls{agent}}(\gls{contract})},
description={Value of the contract \gls{contract} gained by agent \gls{agent}}
}

\newglossaryentry{state_of_charge}{
type=acronyms,
name=\ensuremath{SoC},
description={State of Charge of a battery in kWh}
}

\newglossaryentry{gains_from_trade}{
type=acronyms,
name=\ensuremath{GT},
description = {Gains from Trade}
}

\newglossaryentry{max_utility}{
type=variables,
name=\ensuremath{u_{\gls{agent}}^{max}},
description = {The maximum utility that can be gained by agent \gls{agent} from trading}
}

\newglossaryentry{round}{
type=variables,
name=\ensuremath{r},
description={The round number of negotiation}}

\newglossaryentry{offer}{
type=variables,
name=\ensuremath{o_{\gls{agent}}(\gls{round})},
description={The offer that agent \gls{agent} proposes in round \gls{round}}
}

\newglossaryentry{reservation}{
type=variables,
name=\ensuremath{rv_{\gls{agent}}},
description={The reservation value of agent \gls{agent}, i.e. the minimum utility the agent will accept.}
}

\newglossaryentry{deadline}{
type=parameters,
name=\ensuremath{dl},
description={The deadline of negotiation}
}

\newglossaryentry{res}{
type=acronyms,
name=\ensuremath{RES},
description={Renewable Energy Source}
}

\newglossaryentry{p2p}{
type=acronyms,
name=\ensuremath{P2P},
description={Peer-to-peer}
}

\newglossaryentry{lifetime}{
type=parameters,
name=\ensuremath{\lambda}, 
description={Lifetime of the generator}
}

\newglossaryentry{generatorCostPerKW}{
type=parameters,
name=\ensuremath{c^g_{kW}}, 
description={Cost of the generator [pence/kW]}
}

\newglossaryentry{batteryCostPerKWh}{
type=parameters,
name=\ensuremath{c^{bat}_{kWh}}, 
description={Cost of the battery [pence/kWh]}
}

\printglossary[type=subscripts,style=myglossary]
\printglossary[type=parameters,style=myglossary]
\printglossary[type=variables,style=myglossary]
\printglossary[type=acronyms,style=myglossary]
\section{Introduction}

Energy systems around the world are experiencing rapid changes. Until recently, energy systems were primarily centralised networks managed by large utility companies. However, currently a paradigm shift is taking place in energy systems across the world~\cite{tushar2019motivational,tushar_poor2023,Nizami_APEN,capper2021systematic,Schwidtal_review}. Recent years have seen many initiatives for creating local energy communities, typically low voltage networks consisting of 50-200 households \cite{lucas2016distribution}. Some notable examples of these initiatives include: the Brooklyn Microgrid project~\footnote{\url{https://www.brooklyn.energy/}}, the REflex smart energy demonstrator on the Orkney Islands~\footnote{\url{https://www.reflexorkney.co.uk/}}, the many local projects projects managed by Community Energy Scotland~\footnote{\url{https://communityenergyscotland.org.uk/}}, the SchoonSchip Community in the north of Amsterdam (location of the GridFriends Project)~\footnote{\url{https://amsterdamsmartcity.com/updates/project/grid-friends}}, the Thames Valley Vision project~\footnote{\url{https://eatechnology.com/resources/projects/new-thames-valley-vision-tvv/}}, the Low Carbon London demonstration project~\footnote{\url{https://innovation.ukpowernetworks.co.uk/projects/low-carbon-london/}} etc.  
However, there are many open challenges and knowledge gaps in setting up such local markets, starting with the best choices for market organisation and establishing contracts. There are currently two main approaches to this problem~\cite{capper2021systematic, sousa2019peer}.

The first approach involves centralised market clearing, in which a coordinator entity/market maker selects the most promising pairs of prosumers in the community to trade, and establishes contracts in order of decreasing gains from trade. The second approach involves purely decentralised peer-to-peer (P2P) systems where prosumers trade energy directly with one another - the selection of who to trade with is up to individual prosumers, and may be influenced by personal preferences, not just efficiency (e.g. trading energy with a neighbour may be preferable to trading with someone else in the community, who is not a close acquaintance). 

The centralised approach of establishing contracts benefits from performing coordinated optimisation and aggregation, leading to efficient use of resources and optimal social allocation for the community. However, centralised approaches may experience scalability issues especially when the number of participating prosumers increases. On the other hand, peer-to-peer systems benefit from their decentralised nature. They can often be privacy preserving and secure, and benefit from the development of new supporting technologies such as decentralised energy contracts and blockchain systems. Furthermore, they allow for a more individual choice and strategy, where prosumers are allowed complete freedom of choice who to trade with, in order to optimise their own benefits. 

While decentralization provides many tangible benefits, it raises many open questions. While with enough trades peer-to-peer systems may approach similar community benefits from a social welfare perspective as the centralized market approach, each trade has some associated cost, which may vary depending on various factors, e.g. the establishment and deployment of the decentralised platform, development, IT, security and administrative costs to name a few. This cost can make some trades with marginal benefits inefficient and not rational to set up. Hence, an important question is: How many P2P trading contracts are required to realise most of the potential benefits from trade? Another important question is that of prosumer participation. Recall that a community may be composed of up to several hundred prosumers. But are all of them required, and would they actually benefit from participation in P2P energy trading? 

This issue is often unaddressed in the literature, leaving the reader to think that every member in a community will benefit from being involved in peer-to-peer trading~\cite{capper2021systematic, sousa2019peer, lee2014direct}. 
However, the picture is more nuanced, after considering the different energy consumption profiles, physical constraints and the availability of home energy storage. In some cases, even if there are some marginal gains to be had, the potential costs of participating in a P2P trading scheme may outweigh the benefits. For example, consider 2  prosumers, each with their own micro-generation (e.g. solar panels on the roof) and battery. In theory, peer-to-peer trading should definitely provide benefits for these consumers, and this is certainly true at least some of the time. However, consider a setting in which  their energy demand profiles are very closely aligned, e.g. both households have similar work patterns, that involve not much consumption during the morning/day, and a consumption peak in the evening. In such a case, P2P energy trading might deliver rather insignificant benefits for the two prosumers, because they willl likely have excess energy to trade and residual demand at (mostly) the same time periods throughout each day.

Then, a question would be whether it would be possible to achieve a high percentage of the possible benefits (in our case, measured as fraction of maximal potential Gains from Trade) with only the most promising fraction of prosumers participating in peer-to-peer trading? Can we determine which prosumers would benefit the most, and which other prosumers should they trade with?

Finally, the value of bilateral energy contracts is dependent on the prosumers that establish them. Prosumers value energy quantities differently depending on their own consumption and generation profile with respect to others. The diversity of a community might affect the potential gains achieved from peer-to-peer trading. Thus, the effect of diversity factors in demand profiles across the community needs to be studied. While other authors have discussed these issues, to our knowledge, this paper is the first to examine these questions using a actual real data, from two large-scale UK trials, i.e. Thames Valley Vision and Low Carbon London.

To answer these questions, we consider two popular approaches to P2P markets. The first is a centralised approach to P2P that uses a central matching and clearing mechanism. The second is a decentralised approach using automated negotiation to establish P2P trading contracts. In both models, once a P2P trading contract is established, prosumers form a coalition, which produces a Gains from Trade for both parties (i.e. a financial benefit, in this case a reduction in bill from trading and operating assets jointly, compared to the bill when each agent uses only its own individual assets), and the agents split this benefit. The exact way energy is exchanged for each clearing period (in our case, we use half hourly clearing periods) throughout the day will be determined by a joint schedule optimiser, that takes into account the demand, generation and assets of both parties.

Our methodology allows modelling realistic peer-to-peer trading using large-scale data sets. Due to computational complexity, prior works often consider small communities with a few prosumers over short time windows, and without the use of large-scale, real world datasets. We argue a more detailed and data-driven approach is needed to provide insights in the effectiveness of peer-to-peer trading in real-world settings. Thus, in our framework, the benefit of energy exchanges are computed using a joint schedule optimiser. This optimiser computes the trading schedule with the maximal Gains from Trade for both parties. The prosumer agents themselves then decide whether to trade, and the redistribution of these benefits. In summary, our paper makes a number of the key new contributions:
\begin{itemize}
\item It performs a systematic comparison of two market models: centralized clearing (double auction-type format) vs. decentralized negotiations, in terms of dynamics and trading coalitions that form over time. In order to enable this comparison, we introduce a new negotiation framework, that incorporates a joint optimiser that computes the most profitable energy exchanges between any set (coalition) of prosumers in terms of the notion of Gains from Trade (i.e. joins that can be obtained by reaching a P2P contract and forming a coalition, compared to trading alone). This framework can be used in both types of markets, allowing us to model separately the market organisation layer (centralized clearing vs. negotiation) from the asset control and optimisation layer. 
\item In most energy community and micro-grid energy schemes, not all consumers are required to participate in peer-peer trading to reach market efficiency (in terms of potential Gains from Trade). In order to quantify this effect, the paper investigates whether the majority of the benefits can be achieved when only a subset of the most promising prosumers participate in peer-to-peer trading. Moreover, it includes a study of how many peer-to-peer contracts are needed to realise the benefits of energy trading in a real P2P market. Our analysis also includes an evaluation of diversity in demand profiles affects the potential benefits and the marginal value of the contracts. 
\item Finally, the developed simulation framework can use large-scale real data, and allows for a more realistic negotiation simulation with large communities over longer time periods, with high time granularity than possible in other prior works. Our results and conclusions are based on real demand and generation data sets from two large-scale energy trials in the UK. 
\end{itemize}

The remainder of the paper is organised as follows: In \cref{sec:related_work}, relevant literature is presented. \Cref{sec:modelling} outlines the models used for the prosumers, energy communities and \gls{p2p} trading. Two approaches to peer-to-peer trading are outlined in the following two sections, with \cref{sec:centralised} describing the central market clearing mechanism, and \cref{sec:negotiation} discussing the negotiation framework. In \cref{sec:experiments}, we perform an empirical analysis of the two methods compared to the community setting using realistic energy communities. Finally, \cref{sec:conclusion} concludes the paper and highlights some topics for future work.

\section{Related Work}
\label{sec:related_work}

Energy communities form an increasingly important area of research, with many earlier works studying different aspects and challenges. 
In recent years a number of projects and initiatives have published systematic reviews of the field. 
Tushar et al.~\cite{tushar2019motivational} review the state of the art in connected energy communities, provide extensive background on different aspects of P2P sharing, and describe a number of relevant pilot projects across the globe. Capper et al.~\cite{capper2021systematic} provide a systematic review and classification of market design and energy trading models based on 139 recent peer-reviewed journal articles, in research sponsored by the global observatory on peer-to-peer, community self-consumption and transactive energy models\footnote{\url{https://userstcp.org/task/peer-to-peer-energy-trading/}}. Sousa et al.~\cite{sousa2019peer} provide a classification of P2P energy trading markets based on the degree of decentralization and P2P topology into: full P2P markets, community-based markets and hybrid markets (combination of the two designs, each operating at different layers). Schwidtal et al.~\cite{Schwidtal_review} discuss the feasibility of different business models for profitable energy community and transactive energy projects. 

In this paper, following prior literature, we consider two prominent market models. First, we model markets that use centralised clearing, through a market maker or other centralised entity, that selects the contracts with the optimal Gains from Trade. 
Double auction is a common mechanism used within this approach, widely used in many applications ranging from wholesale energy markets to the stock market. Being widely understood, it is also the most studied mechanism for local energy markets and has seen many practical applications.
Sioshansi~\cite{sioshansi2013evolution} discusses double auction mechanisms and their use in electricity markets, while Wang et al.~\cite{wang2020distributed} and Capper et al.~\cite{capper2021systematic} provide examples of double auction use in P2P energy markets and energy communities. 
In double auctions, buyers and sellers relay their preferences and utility to a central entity or broker. Then, this central entity will decide the matching of parties, price and will clear the market, which guarantees market efficiency. Wang et al.~\cite{wang2020distributed} show that such an auction mechanism can provide these benefits while preserving privacy using a two-level transactional model.
 Our centralised mechanism matches, essentially, a double auction, where contracts are sorted and cleared in order of decreasing Gains from Trade (GT).

The second model of market organisation we consider is a decentralised peer-to-peer energy market, that allows direct trading and establishing of contract between prosumers, forgoing the need for an intermediary. As opposed to markets relying on a clearing entity/market maker, peer-to-peer markets allow prosumers to select who they trade with - for example they could make or request offers with several other prosumers of their own choice (e.g. neighbours, friends in the community etc.), and select the most favourable one. 

The most widely used mechanism to automate decentralised P2P markets is automated negotiation. 
Pinto et al.~\cite{pinto2019decision} proposed a multi-agent simulation platform to support negotiations of small players in transactive electricity markets, including local energy markets, bilateral contracts and participation in wholesale energy markets. Another example of automated negotiations is provided by Saxena and Abhyankar~\cite{saxena2019agent} focusing in electricity markets for distribution grid management. 
In automated negotiations, prosumers are represented by autonomous agents that negotiate over several topics, e.g. price per unit, energy quantities, etc. Often, the main topic negotiation is concerned about is the price per unit, such as in the works by Guo et al.~\cite{guo2021asynchronous} and Imran et al.~\cite{imran2020bilateral}. Etukudor et al.~\cite{etukudor2020automated} discuss a framework where prosumers negotiate the prices for a single day ahead together with the energy quantities traded for multiple periods within the day. Another approach is discussed by Chakraborty et al.~\cite{chakraborty2020automated}, who explore an energy lending scheme, where prosumers negotiate over energy quantities that they lend and the time the borrowed energy is returned. 

Since peer-to-peer negotiation only considers bilateral agreements, as opposed to an auction mechanism where a market-wide price is set, an important process is also peer selection, i.e. how a prosumer decides with whom to negotiate. Concurrent negotiation approaches have been explored, where sellers negotiate with multiple sellers concurrently and vice versa \cite{dang2011wholesale}. However, due to computational complexity, systems favour a selection of peers before the negotiation process. A popular approach is the facilitation of peer selection by a central institution (often the platform provider) \cite{kalbantner2021p2pedge}. 
Khorasany et al.~\cite{khorasany2020new} propose a single peer selection approach, where prosumers select peers greedily based on maximum expected profit. Other recent works focus on optimising economic objectives when taking physical constraints and storage into account \cite{su2020optimization}, and the fair redistribution of benefits in energy coalitions. Long et al.~\cite{long2019game} provide a P2P trading scheme for a community microgrid that focused on the optimality and fairness between participating prosumers. Norbu et al.~\cite{norbu2021modelling} provide a model for the fair redistribution of benefits achieved by community-owned assets to individual prosumers. Robu et al.~\cite{VPP_energy} look at the fair allocation of gains achieved by virtual power plant formation, while while Kota et al.~\cite{Kota_ECAI} study this issue for demand-side management cooperatives. The concept of Shapley value has been used to ensure fairness, such as in the works of Cremers et al.~\cite{Cremers_Shapley_APEN} and Robu et al.~\cite{groupbuying_TSG}. 
Finally, other works take a social science perspective, and examine the perception of consumers to P2P energy trading schemes~\cite{pumphrey_perception}.

However, although great progress has been made by related works, there are still key remaining questions. Firstly, many peer-to-peer approaches, especially negotiation, suffer from high complexity, and generalizing them to large communities has been difficult~\cite{moret2018negotiation}. Thus, the main contribution of earlier works often lies in their methodologies, since their experiments consider only a small simulation-based setup, and often do not rely on real prosumer demand data. Further research into how these models generalise to larger settings is required. 
Secondly, while peer-to-peer markets will approach the efficiency of a central market when enough contracts have been established, each contract has some associated costs. Hence, how many P2P energy trading contracts need to be established to achieve the majority of the maximal potential Gains from Trade is still an open question.
Thirdly, these works focus on the total economic or technical benefits for either the community or an individual prosumer. However, within communities, especially those based on locality, some prosumers benefit to different degrees (i.e. asymmetrically) from peer-to-peer trading. Efforts have been made to ensure Pareto optimality, i.e. that no prosumer is worse off in favour of another prosumer. But the benefits earned by each prosumer are not equal, since they depend on their sice, and crucially on the energy consumption and generation behaviour of the prosumer compared to those of the community, e.g. a prosumer that consumes with a different pattern is likely will benefit more. Thus, there is a need to investigate the effects of prosumer participation and diversity on the benefits of trading. 
This paper aims to model these settings - and thus address these challenges - using a large-scale real-world data set. 
\section{Modelling prosumers in an energy community}
\label{sec:modelling}

In this section, we first introduce a formal model of a prosumer, by modelling their respective demand and generation profile. Prosumers are also assumed to have a source of energy storage (i.e. a battery).

\subsection{Single prosumer model}
A prosumer $\gls{agent} \in \gls{community}$, where \gls{community} are the prosumers connected to the local energy grid, has access to the battery and some renewable energy resource (\gls{res}) (e.g. wind turbine). 
The generation profile of this prosumer \gls{agent} is represented as $\gls{generation}, \forall \gls{time} \in [1,\gls{time_horizon}]$ for the power generated at time step \gls{time}, where \gls{time_horizon} corresponds to the time horizon for the operation window. 
Similarly, the demand profile is represented as $\gls{demand}, \forall \gls{time} \in [0,\gls{time_horizon}]$. 
Furthermore, for each prosumer,  we compute the optimal battery usage $\gls{battery_power}$, where a positive value indicates charging and a negative value discharging.
Finally, a prosumer also has access to the power grid and is able to export and import energy, represented as \gls{net_demand}, where it is positive when importing from the grid and it is negative when exporting to the grid.

The energy demand of the prosumer should always be satisfied, either due to self-consumption or by buying power from the grid. Furthermore, excess energy should be either sold or stored. So, the following constraint should always hold:
\begin{equation}
\label{eq:modelling:net-demand}
    \gls{net_demand} = \gls{demand} - \gls{generation} + \gls{battery_power}
\end{equation}

\subsection{Battery Control Algorithm}
\label{sec:modelling:battery-control-algorithm}
Battery usage at each time step was controlled using a heuristic-based algorithm adapted from Norbu et al.~\cite{norbu2021modelling}. 
Battery charging and discharging follow directly from the energy generation and consumption. If at some time, the generation exceeds the consumption the remaining energy is used for charging the battery. However, if the battery has reached its maximum capacity, excess energy is instead sold to the grid.

Instead, if consumption exceeds the generation, we instead discharge the battery.
If at any time the battery cannot provide sufficient energy, energy is bought from the central power grid.

This battery control algorithm provides a tractable method to approximate an optimal control strategy. Our heuristic-based strategy performs well (essentially close to optimal) under the assumption that flat import and export tariffs are used and that using energy storage is more cost-effective than importing and exporting (see~\cite{couraud_ISGT} for a discussion). In the current energy market, these assumptions are realistic, since import tariffs are high, while export tariffs are being phased out \cite{ofgem2020fit}. Given this aspect is technically complex, and not the main focus of this paper, we provide the full details of the battery control algorithm we use in our experiments in in~\ref{app_subsec:battery_control}.

\subsection{Cost Computation}
The costs for a single prosumer can be calculated as the sum of the costs of importing energy from the grid, the revenue from selling excess energy to the grid, and the depreciation costs of the private energy assets, i.e. the battery and the generator. It can be computed as follows:

\begin{equation}
    \gls{bill} = \sum_{\gls{time}=1}^{\gls{time_horizon}}\gls{energy_imported}\gls{import_tariff} - \sum_{\gls{time}=1}^{\gls{time_horizon}}\gls{energy_exported}\gls{export_tariff} + \gls{battery_degradation} + \gls{generator_cost}
\end{equation}

where the \gls{energy_imported} and \gls{energy_exported} are the energy bought from and sold to the grid at time \gls{time} respectively. \gls{import_tariff} and \gls{export_tariff} are the import and export tariffs, and \gls{battery_degradation} and \gls{generator_cost} are the depreciation costs for the battery and \gls{res} respectively. Note that, in our model, both individual prosumers and communities are modeled as price-takers w.r.t. the central grid or power distribution company, i.e. they are too small to influence the wholesale market prices. Hence, the import/export tariffs from the grid are inputs for our pricing model.

The depreciation cost of the \gls{res} is expressed as follows:
\begin{equation}
    \gls{generator_cost} = \frac{\gls{PowerPVinstalled}\cdot 
    \gls{generatorCostPerKW}\cdot \gls{time_horizon}}{\gls{lifetime}}
\end{equation}

where \gls{generatorCostPerKW} is the cost per rated kW and \gls{lifetime} is the lifetime of the generator.

The depreciation cost of the battery can be computed as follows:
\begin{equation}
    \gls{battery_degradation} = \frac{\gls{BatteryCapacity}\cdot \gls{batteryCostPerKWh}\cdot \gls{time_horizon}}{\max\left(\frac{1}{\text{\gls{depreciation_factor}}},\,\gls{lifetime}\right)}
\label{eq:batteryCost}
\end{equation}

where $\gls{state_of_charge}_{\gls{agent}}^{max}$ is the battery capacity, \gls{batteryCostPerKWh} is the price per kWh of capacity, and \gls{depreciation_factor} is the depreciation factor of the battery based on the usage, calculated using ``rain-flow'' cycle counting \cite{ke2015control, norbu2021modelling}. We include the depreciation factor in this calculation since the battery's lifetime can be influenced by the number of cycles and depth of discharge. See~\ref{app_subsec:battery_degradation} for full details on the computation of \gls{depreciation_factor}.

The cost computation in this section refers primarily to the use of energy assets and the resulting energy bill. In practical implementations, prosumers engaging in trading with peers in an energy community also face other costs, such as the development and maintenance of the energy trading platform, costs of control systems of energy assets, IT and administrative costs for contract deployment, etc. Furthermore, these costs depend on several factors, such as the technological solution selected for the trading platform, the complexity of contracts deployed, the security level and general market conditions. Due to this complexity, in this work we focus on the costs that are directly related to the energy bill and use of energy assets deployed.

\subsection{Coalition Model}

The above energy community model corresponds to a coalitional model where $N$ prosumers are players that can share their energy assets, resulting in a joint cost reduction. The joint cost function of this coalition is defined by the joint bill that prosumers that form the coalition will have to pay. Formally, we consider a coalition $\gls{trading_community} \subseteq \gls{community}$ where each prosumer \gls{agent} has their own energy assets. However, once two agents agree to trade (enter a P2P trading contract), they form a coalition, where each prosumer can use the other members' assets if excess generation/storage capacity exists, and they are not being used by the party owning them. Thus, in this model, we can aggregate the generation, demand and storage to form the community's generation, demand and storage, i.e.
\begin{equation}
    \gls{demand_c} = \sum_{\gls{agent} \in \gls{trading_community}} \gls{demand}\qquad\forall \gls{time} \in [0,\gls{time_horizon}]
\end{equation}
\begin{equation}
    \gls{generation_c} = \sum_{\gls{agent} \in \gls{trading_community}} \gls{generation}\qquad\forall \gls{time} \in [0,\gls{time_horizon}]
\end{equation}
\begin{equation}
    \gls{state_of_charge}_{\gls{trading_community}}^{max} = \sum_{\gls{agent} \in \gls{trading_community}} \gls{state_of_charge}_{\gls{agent}}^{max}
\end{equation}

Similarly, we can compute the bill for the energy community given these aggregate profiles as follows:
\begin{equation}
    \gls{bill_c} = \sum_{\gls{time}=1}^{\gls{time_horizon}}\gls{energy_imported_c}\gls{import_tariff} - \sum_{\gls{time}=1}^{\gls{time_horizon}}\gls{energy_exported_c}\gls{export_tariff} + \gls{battery_degradation_c} + \gls{generator_cost_c}
\end{equation}

Hence the term ``coalition" in this paper denotes a set of agents that agree to exchange energy according to a schedule that optimises the use of their joint energy assets (battery, generation) and joint demand. The term is borrowed by cooperative or coalitional game theory, where coalitions are cooperative groups formed by agents, in order to increase the joint benefits achieved (or in this case, reduce the joint costs or bill). The combined battery and generation assets are then controlled jointly in an optimal way to satisfy their joint demand. This minimises their imports from the grid (or utility company) resulting in a mutual saving, we call the Gains from Trade (c.f. Section~\ref{sect:gains_trade}).

\subsection{The Peer-to-Peer Trading Model}
We consider a peer-to-peer model where a prosumer \gls{agent} trades energy with another prosumer \gls{agent_j} using a joint profile optimiser \gls{oracle}, which computes the optimal energy exchange at each time step. The contents of the computation of these trades are dependent on the net demand of the prosumers. So we record all trades in \emph{energy contracts} $\gls{contract} = (\gls{agent}, \gls{agent_j}, \gls{trades})$, where \gls{agent} is the energy-receiving prosumer and \gls{agent_j} is the energy-providing prosumer participating in the trade, and \gls{trades} are the energy trades found by the profile optimiser $\Theta$ for each time step $0..\gls{time_horizon}$.
We denote the set of trades by prosumer \gls{agent} as \gls{accepted_contracts}. Note that a coalition can be formed through multiple contracts, as agents iteratively join and trade with existing prosumers inside the coalition.

The total energy exchanged at time $t$ by a prosumer $i$ is given by
\begin{equation}
    \gls{energy_traded} =  \sum_{(\gls{agent}, \gls{agent_j}, \gls{trades}) \in \gls{accepted_contracts}} \gls{trades_t} - \sum_{(\gls{agent_j}, \gls{agent}, \gls{trades}) \in \gls{accepted_contracts}} \gls{trades_t}, \qquad \forall \gls{agent_j} \in \gls{trading_community}
\end{equation}

The exchanged energy \gls{energy_traded} may be positive or negative, in which case a positive value indicates prosumer $i$ received more energy than sent to other prosumers at time $t$, while a negative value depicts $i$ giving more energy than receiving.

The net demand of the prosumer after trading is given by
\begin{equation}
    \gls{net_energy_after_trade} = \gls{net_demand} - \gls{energy_traded}
\end{equation}

Similarly to \autoref{eq:modelling:net-demand}, we can identify two regimes where $\gls{net_energy_bought} = \gls{net_energy_after_trade} > 0$, i.e. residual demand and where $\gls{net_energy_sold} = \gls{net_energy_after_trade} < 0$, i.e. excess generation.

The bill after trading can be computed as follows:
\begin{equation}
    \gls{bill_after_trade} = \sum_{\gls{time}=1}^{\gls{time_horizon}}\gls{net_energy_bought}\gls{import_tariff} - \sum_{\gls{time}=1}^{\gls{time_horizon}}\gls{net_energy_sold}\gls{export_tariff} + \gls{battery_degradation} + \gls{generator_cost} - \sum_{\gls{contract} \in \gls{accepted_contracts}} \gls{value_of_contract}
\end{equation}
where $v_i$ is the function that gives the financial value gained by prosumer \gls{agent} from contract \gls{contract}. This value is decided based on the contract mechanism. In this paper, we investigate two mechanisms, i.e. a centralised market, and bilateral, concurrent negotiations.

\subsection{Gains from Trade in energy trading}
\label{sect:gains_trade}

Due to the diversity of demand and energy profiles, aggregation of profiles and assets leads to each coalition requiring less energy import and exports from the central power grid. Furthermore, considering all the distributed residential batteries as a combined battery capacity that could be centrally controlled to optimise the gains of the community leads to more flexibility, further decreasing reliance on the grid, and reducing the joint bill of the community for the residual demand over some time period $\gls{bill_n}$. Considering all distributed storage assets of the community as a single combined battery capacity controlled centrally in both the centralised community model and the peer-to-peer model leads to a cost reduction for the prosumers (as compared to the case where each prosumer uses and controls its own individual assets such as individual battery independently, and thus pays all the imported electricity to his/her main electricity supplier). Using terminology from market design/economics, we call this reduction in costs the `Gains from Trade' (gains from trading energy between prosumers in the community), as opposed to using only one's own individual assets). Formally, the Gains from Trade (also abbreviated GT) for a community of $N$ prosumers over a time horizon $T$ is defined as:
\begin{equation}
    \gls{gains_from_trade}_{\gls{community}}(\gls{time_horizon}) = \sum_{\gls{agent} \in \gls{community}} \gls{bill} -  \gls{bill_n}
    \label{eq:GT_coalition}
\end{equation}
where, for all agents $i\in \gls{community}$, $b_i(T)$ denotes the bill of each agent, when controlling its assets individually, and $\gls{bill_after_trade}$ denotes its bill after trading based on the schedule recommended by the joint profile optimiser.
Intuitively, $\gls{gains_from_trade}$ is the difference between the sum of the prosumers' bills to the utility company (for residual demand not covered by their own renewable generation) that agents would have to pay jointly, minus the bill for the whole community if they share and control their energy assets together (i.e. appearing a single prosumer to the utility company).  Note that here $\gls{bill_n}=  \sum_{\gls{agent} \in \gls{community}} \gls{bill_after_trade}$, i.e. the sum of the bills of the $N$ prosumers trading inside the community is necessarily the total bill of the community.
It is always the case that $\gls{bill_n} \leq \sum_{\gls{agent} \in \gls{community}} \gls{bill}$, hence the $\gls{gains_from_trade} \geq 0$ value is greater than zero, since in the worst case, the best control is for everyone to exclusively use their individual assets. However, in most realistic cases, there will be times (clearing periods) when pooling of assets is better.

The above definition is provided for a community of $N$ agents. But actually, in peer-to-peer trading, we can apply this concept the same way for any subset (or coalition) of prosumers $K \subseteq N$ in this community trading over period $T$ (i.e. any subset greater than $|K| \geq 2$ agents). The $\gls{gains_from_trade}$ for any coalition $K$ is
\begin{equation}
    \gls{gains_from_trade}_{\gls{trading_community}}(\gls{time_horizon}) = \sum_{\gls{agent} \in \gls{trading_community}} \gls{bill} - \gls{bill_k}
\end{equation}

Note that unlike conventional game-theoretic approaches, the Gains from Trades that underpin the coalition formation are derived by the results of the optimisation process and energy assets utilisation, rather than a clearly mathematically defined characteristic function. This allows for a more realistic representation and pragmatic approach of the model developed. Furthermore, note that $\gls{gains_from_trade}_{\gls{trading_community}}(\gls{time_horizon}) \geq 0$, because no P2P contract would be established, and no trading would occur if the agents are worse off by pooling their assets together in a coalition. Similarly, here $\gls{bill_k}=  \sum_{\gls{agent} \in \gls{trading_community}} \gls{bill_after_trade}$, i.e. the sum of the bills of the $K$ prosumers inside the coalition is the total bill of the coalition. Also important to note that the Gains from Trade for a subset (sub-coalition) of agents in the community will always be less or equal to that of the whole community, i.e. $\gls{gains_from_trade}_{\gls{trading_community}}(\gls{time_horizon}) \leq \gls{gains_from_trade}_{\gls{community}}(\gls{time_horizon}), \forall K \subseteq N$).
However, the growth of $\gls{gains_from_trade}$ is definitely not linear or proportional to the number of prosumers joining the coalition $K$, in fact quite the opposite: for some prosumers/contracts, their effect on the  gains from trade growth is high, while for some it is very marginal (for example, small consumers or with a non-diverse diverse demand profile, who have little scope to trade),
In most of the experiments in this paper, it is a more general methodology to report the relative ratio:
$$
\frac{\gls{gains_from_trade}_{\gls{trading_community}}(\gls{time_horizon})}{\gls{gains_from_trade}_{\gls{community}}(\gls{time_horizon})} \text{  for possible coalitions } K \subseteq N
$$
in percentage terms, rather than the raw $\gls{gains_from_trade}$ numbers, that is often specific to the trial, location in the UK, or energy tariff levels in time period when the trial data was collected.

\subsubsection{Joint Profile Optimiser}

The goal of peer-to-peer trading is to maximise the Gain from Trade. However, since every prosumer has different demand and generation profiles, it is not trivial to find the best exchange that minimises the joint total cost. Furthermore, even if we find the optimal (maximal) Gains from Trade, we need to distribute these gains over the prosumers in the contract. 

To address the optimisation of the energy exchange, we propose a method using a joint profile optimiser that can compute the energy that should be exchanged between two prosumers.  Recall that in the setting considered in this paper, once consumers decide to trade, either through centralised or automated negotiation approach, they agree to share their assets, i.e. their flexible assets will be controlled by the Joint Profile Optimiser. Therefore, in both scenarios, consumers allow the Joint Profile Optimiser to use their excess energy generation/storage by other prosumers, at times they do not need it themselves. Hence, we can find the optimal energy contract, minimizing the combined bill of prosumers in a coalition, using the aggregated demand, generation and battery of the prosumers in that coalition.

However, we still need to calculate the demand, generation and battery usage of the individual prosumers in the coalition, that would guarantee them the maximum Gains from Trade. This calculation requires satisfying a number of constraints. 

First, the net demand at each time step given by
\begin{equation}
    e(t) = d(t) - g(t) + p^{bat}(t)
\end{equation}
for the coalition of two agents is equal to the sum of net demands of the individual agents after trading, i.e.
\begin{equation}
    e_{ij}(t) = e_i(t) + e_j(t)
\end{equation}
Since this would mean that the total energy between the two consumers individually is the same as the expected energy in the coalition.

Second, the state of charge (\gls{state_of_charge}) of the battery should be equal for the coalition and the sum of individuals, i.e.
\begin{equation}
    \gls{state_of_charge}_{\gls{agent}\gls{agent_j}}(\gls{time}) = \gls{state_of_charge}_{\gls{agent}}(\gls{time}) + \gls{state_of_charge}_{\gls{agent_j}}(\gls{time})
\end{equation}

These constraints combined ensure that the system of two individuals behave as a coalition, while preserving their individual profiles. To ensure that these constraints are satisfied we can use our battery control algorithm that balances the demand and generation of each prosumer at each time step.

\begin{figure}[t]
     \centering
     \begin{subfigure}[b]{0.49\textwidth}
         \centering
         \includegraphics[width=\textwidth]{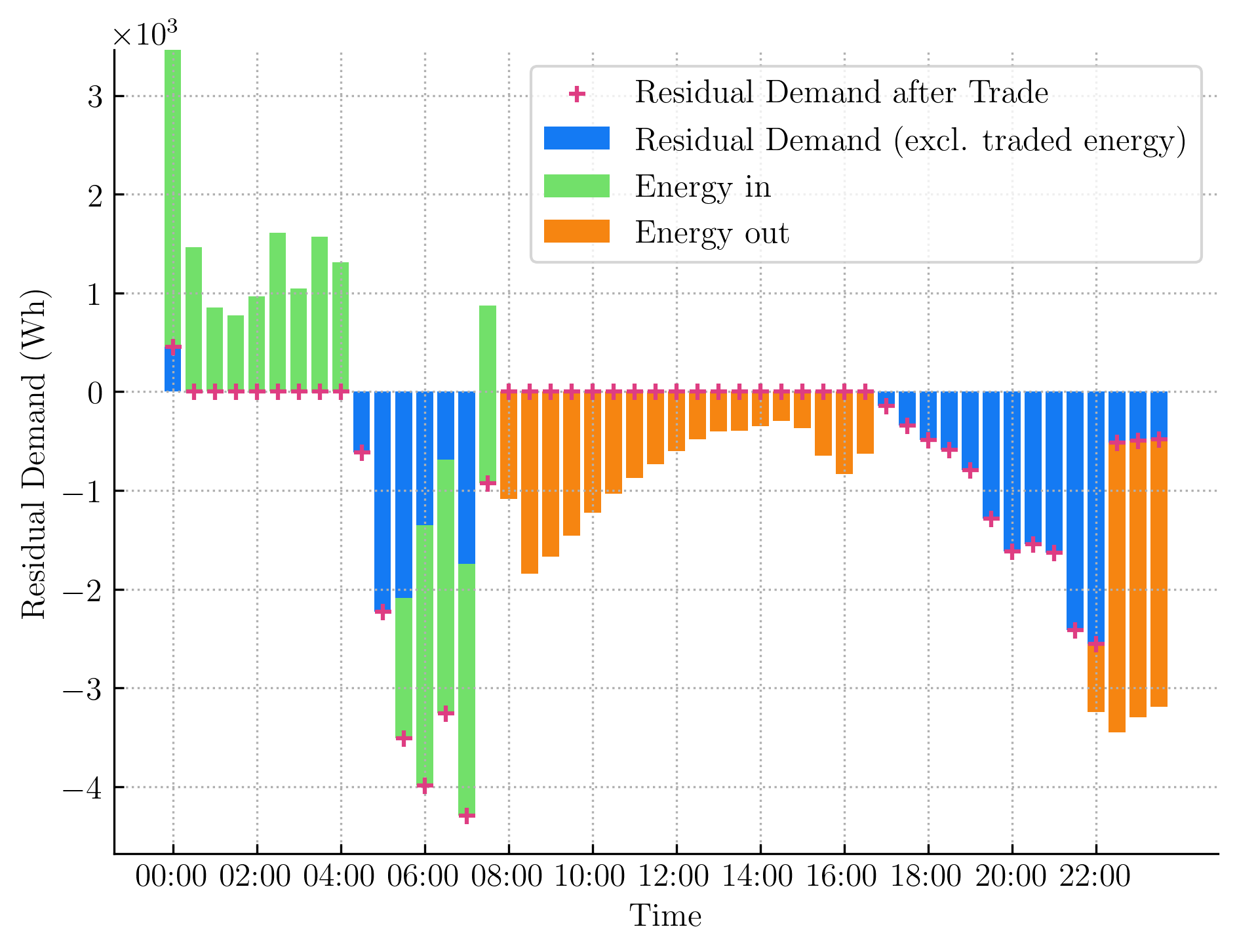}
         \caption{Prosumer A profile after P2P trading}
         \label{fig:p2p_prosumerA}
     \end{subfigure}
     \hfill
     \begin{subfigure}[b]{0.49\textwidth}
         \centering
         \includegraphics[width=\textwidth]{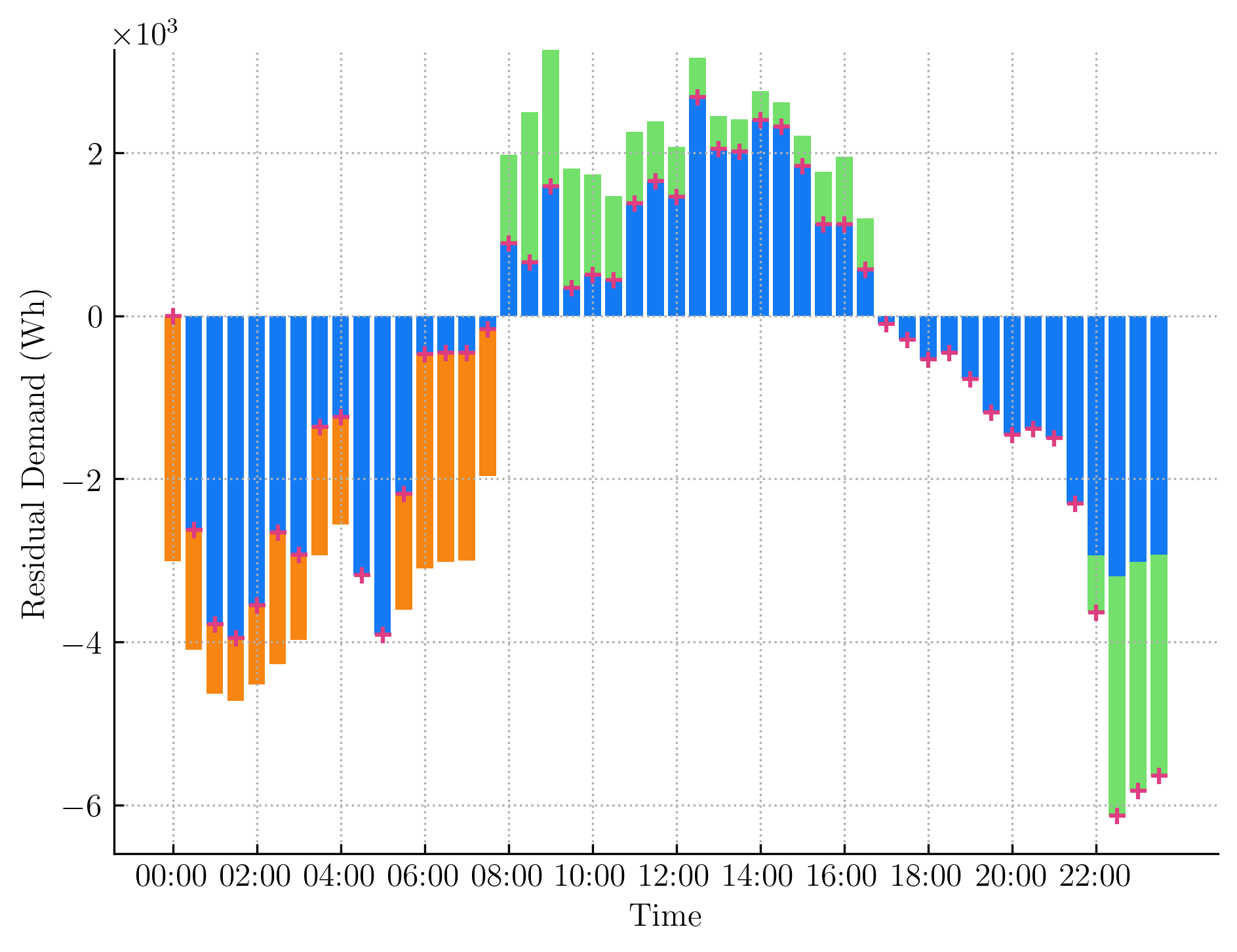}
         \caption{Prosumer B profile after P2P trading}
         \label{fig:p2p_prosumerB}
     \end{subfigure}
        \caption{Example of trading profiles between 2 prosumers: Prosumer A and Prosumer B, who decide to trade and form a coalition, over one example day (48 half-hourly time slots). Note that the imports (green sections) of one prosumer correspond to the orange sections (exports) of the other prosumer. In this case, it appears Prosumer A is a ``night owl" type prosumer (i.e. a prosumer who is working and using energy during the night)  - we note this is the real [anonymised] profile of a consumer from the Low Carbon London trial data. Important to note that, in the case of these two specific profiles substantial trading occurs, but for most profile pairs in the dataset there can be much less (or even no) P2P trading occurring, depending on the fit between the specific demand profiles.}
        \label{fig:p2p_trading_example}
\end{figure}

The basic intuition of the profile optimiser is that, when two prosumer agents decide to trade, their optimal joint generation and demand (load) profile is the same as if they were a larger agent, controlling the assets of both parties. Then, this joint agent would perform optimal control w.r.t. the joint assets, and the aggregated load/generation of both prosumers. In figure~\ref{fig:p2p_trading_example} we provide an example illustration of trade that optimises the joint Gains from Trade between two prosumer agents, labelled Prosumer A and Prosumer B (note these are two real, anonymised prosumers, picked from the Low Carbon London dataset used in this paper). While the joint profile optimiser can operate on two individual prosumer agents (as in this example), it can also operate on the coalition of the two prosumers, which is the case considered in this work when 2 agents agree to trade, either through centralised matching or through negotiation. In this case, the generation/demand profile used for the coalition is the aggregate of the generations/demands of all prosumers in that coalition. Thus, the profile optimiser can be applied to iteratively build larger coalitions, scalable to any number of agents.

\section{Centralised Matching and Clearing}
\label{sec:centralised}

\begin{figure}
    \centering
    \includegraphics[width=.8\textwidth]{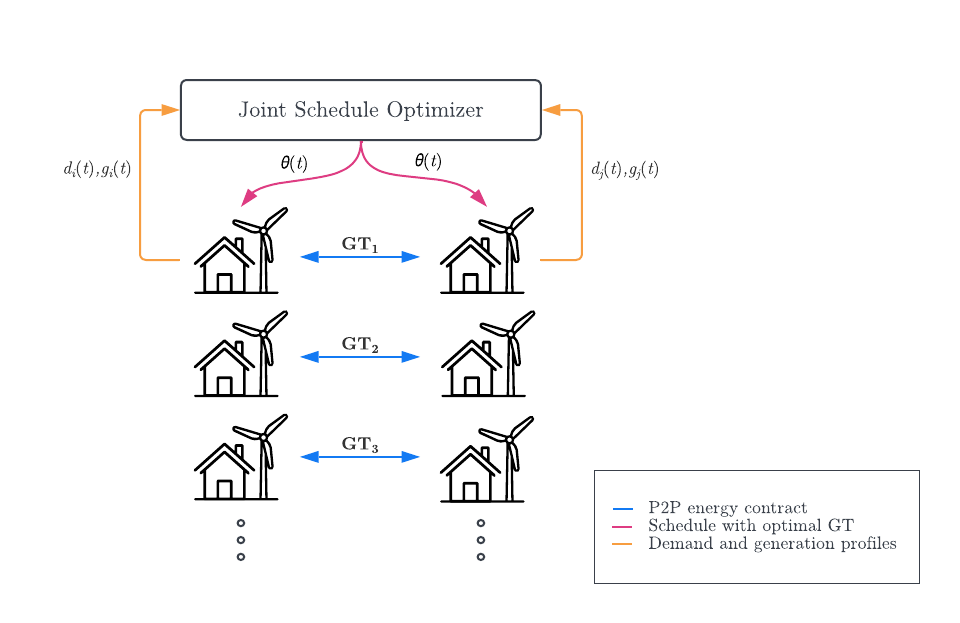}
    \caption{Centralised matching and clearing; note that contracts are matched in order of decreasing Gains from Trade, i.e. $GT_1 \geq GT_2 \geq GT_3 \geq ...$}
    \label{fig:centralised_market_schematics}
\end{figure}

The centralised market mechanism tries to clear the market by accepting the contracts in order of most Gains from Trade that can be achieved by the agents trading (as computed by the joint schedule optimiser). For a visual overview of the process, see \cref{fig:centralised_market_schematics}.

To find out which contracts are available, we compute all contracts in the contract space by computing all pairs of prosumers and the resulting trades \gls{trades} produced by the joint profile optimiser \gls{oracle}. The contract with the highest gains from trade is accepted and the involved prosumers trade their energy if the power flow resulting from the energy quantities in the contract do not violate any local grid constraint. This verification is done either directly by the DSO or by the Joint Schedule Optimizer. The gains of the accepted contract are split equally between the two prosumers. However, this results in the net demand profiles of these prosumers changing, i.e. the contract space changes. This requires a re-computation of the contract space before a new contract is accepted.

Hence, we can define a single round of accepting a contract as follows:
\begin{enumerate}
    \item Compute the contract space given the prosumers net demands profiles \gls{net_energy_after_trade} ( \gls{net_demand} for the initial round)
    \item Sort the contracts space in descending order by potential Gains from Trade (GT)
     \item Run a powerflow  on the local network with the quantities from the trade with the highest potential GT
    \item If the local grid constraints are met, accept the contract with the highest GT, otherwise, go the next contract
    \item Split the gains equally between the two prosumers
\end{enumerate}

This process continues until no more trades can be made, or the gains of trade drop below a specified threshold. Section \ref{sec:experiments} provides an overview of the gains of trades evolution for different numbers of rounds. 
Agents who establish trades with their peers form a coalition.

The proposed method attempts to minimise the number of trades (contracts) between individual prosumers. However, it can be extended to instead minimise the number of prosumers participating in trading. We can model the trading coalitions that form around the agents that trade with each other. Note that, within these trading coalitions, not every member trades with each other, but only those who have established a trade contract.
We can look at the gains those coalitions could realistically produce on their own and model them as a single prosumer for future contracts. This process will minimise the number of agents participating in trade since gains are maximised for groups first.

\section{Peer-to-Peer Negotiation Framework}
\label{sec:negotiation}

\begin{figure}
    \centering
    \includegraphics[width=.8\textwidth]{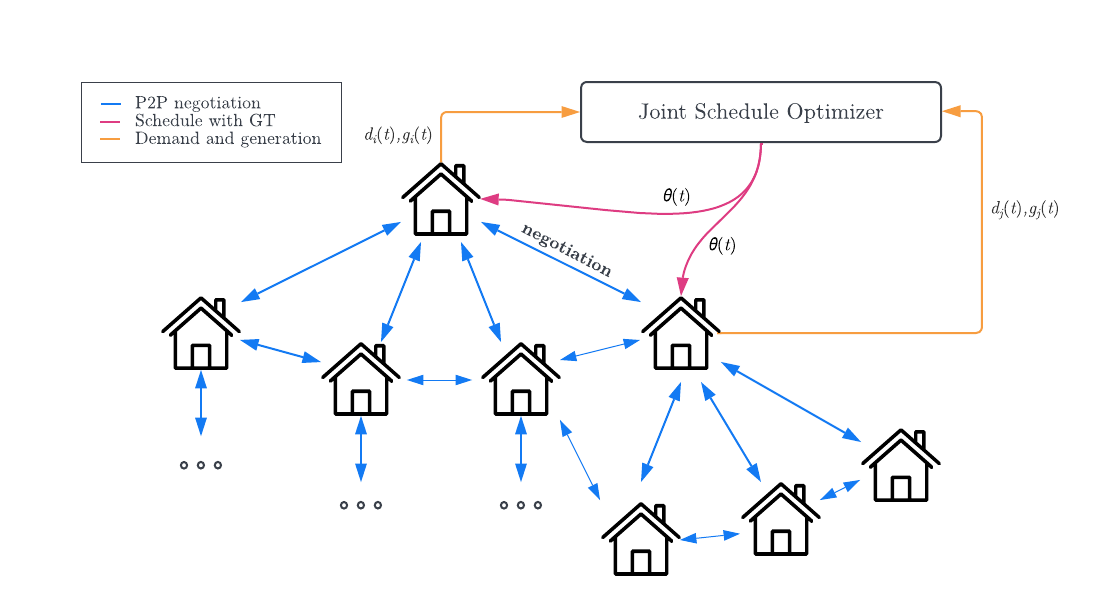}
    \caption{Decentralised P2P negotiation between several prosumers, note that interaction between the joint profile optimiser and prosumers happen for each bilateral negotiation thread. In the figure, one example is shown.}
    \label{fig:negotiation_schematics}
\end{figure}

Modeling automated P2P energy negotiations among multiple agents has proven complex \cite{etukudor2020automated, chakraborty2020automated, khorasany2020new}, with earlier works often limiting the negotiation domain to a very small number of time steps and quantities. Our framework consists of two main steps, i.e. peer selection and negotiation.  During the peer selection step, we select a number of prosumers based on the highest possible savings (Gains from Trade). Finally, during negotiation, the distribution of the gains is decided. These steps are then iterated over until no more new contracts can be established or a specified deadline has been reached. While earlier works have considered multi-issue negotiation over the amount of energy to be exchanged at every time point - resulting in an extremely large negotiation space, we instead consider a negotiation mechanism where agents can use the joint profile optimiser, that computes their joint optimal energy exchange schedule. Prosumers, therefore, do not need to negotiate over how much energy needs to be exchanged at each time step, as in the proposed setting, the community embraces the best strategy proposed by the profile optimiser considering all batteries aggregated. A graphical illustration of this process is provided in Figure~\ref{fig:negotiation_schematics}.

The profile optimiser returns an exchange schedule and its expected joint Gains from Trade (compared to the sum of individual optimisations), but will not specify \emph{how the financial gains will be divided} for each prosumer. Thus, prosumers will need to negotiate only over the \emph{redistribution} of the expected gains from trade inside the coalition being formed. 
The introduction of a joint profile optimiser enables a reduction of complexity, and modeling larger domains. It allows us to consider trades over very long periods, with many time steps - and hence be able to build large data-driven simulations. While agents have to share their generation/consumption with the optimiser, they do not need to share their full generation/demand information with other agents directly, which preserves privacy. 

\subsection{Negotiation}
The negotiation protocol is based on the concurrent alternating offers protocol. It consists of two phases: An offer phase and a commit phase.
In the offer phase, agents keep exchanging offers until at least one is accepted. While in the commit phase, a single most promising offer is accepted in the market.

\subsubsection{Offer phase}
The bidding phase is described in \cref{alg:bidding}. This phase consists of multiple rounds. During each round, every agent provides a price offer to multiple different agents along with the energy quantities defined by the joint profile optimiser. Note that the price offered by an agent to its partners is the same for all in a single negotiation round. 
The receiving agent then can accept or reject the offer. In case the offer is accepted, this is made public to other market participants, and a new coalition is formed by the agents reaching the deal. Otherwise, each bidding agent will create another offer during the next round. Note that, if a deal is accepted, the newly formed coalition can continue to trade in the market as a single player, to reach P2P deals with other prosumer agents and/or other coalitions, and the schedule optimiser will compute the Gains from Trade of the new coalition compared to the one already formed.

\begin{algorithm}[H]
\caption{Offer exchange phase}
\label{alg:bidding}
\begin{algorithmic}
\Require Contract space \gls{all_contracts}, sets of potential trading partners $P_{\gls{agent}}$ for each agent $\gls{agent} \in \gls{community}$, agent strategy $s_{\gls{agent}}$ for each agent \gls{agent}
\Ensure Set of accepted offers and contracts (can be more than one)
\State $\gls{round} \gets 0$
\State $\gls{all_contracts}^a \gets \emptyset$
\State $O^a \gets \emptyset$
\While{$r < deadline$ AND $\Omega^a = \emptyset$}
\ForAll{$\gls{agent} \in \gls{community}$}
\ForAll{$\gls{agent_j} \in P_i$}
\State select $\gls{contract} = (\gls{agent}, \gls{agent_j}, \gls{trades}) \in \gls{all_contracts}$ \Comment{Precomputed by the profile optimiser}
\State make-offer$_{\gls{agent} \rightarrow \gls{agent_j}}$(\gls{round}, \gls{value_of_contract}, $s_{\gls{agent}}$)
\EndFor
\EndFor
\ForAll{$o \in $ receive-offers$_{\gls{agent}}(\gls{round})$}
\If{accept-offer$_{\gls{agent}}$(\gls{round}, $o$, $s_{\gls{agent}})$}
\State select $\gls{contract} = (\gls{agent}, \gls{agent_j}, \gls{trades}) \in \gls{all_contracts}$  \Comment{Find the contract belonging to the offer}
\State $\gls{all_contracts}^a \gets\gls{all_contracts}^a \cup \{\gls{contract}\}$
\State $O^a \gets O^a \cup \{o\}$
\EndIf
\EndFor
\State $\gls{round} \gets \gls{round} + 1$
\EndWhile
\State \Return $\gls{all_contracts}^a$, $O^a$
\end{algorithmic}
\end{algorithm}

\subsubsection{Commit phase}
In the commit phase, shown in \cref{alg:commit}, the market looks at all accepted offers and approves the one with the highest Gains of Trade after validation through a power flow analysis that the local grid constraints are not exceeded (voltage and lines thermal limits). If the offer with the highest gains results in the violation of grid constraints, the next highest gains' offer is selected and assessed through power flow study, until an offer is compatible with the grid state or until no offer remains. All other offers in that specific round are then rejected, to prevent large regret, as some players in the market have changed. Specifically, as a new coalition has been formed (by the agents with the highest GT exchange having reached a contract, and now participating as a coalition), this new coalition may offer a better deal to some other agent. 
Hence the acceptance process is iterative, with exactly one coalition being formed in each round. After acceptance, the payments are settled and energy is exchanged. After the energy exchange the contract space is updated.

\begin{algorithm}[H]
\caption{Commit phase}
\label{alg:commit}
\begin{algorithmic}
\Require Contract space \gls{all_contracts}, all prosumers $\gls{agent} \in \gls{community}$, set of all accepted contracts $\gls{all_contracts}^a$ and offers $O^a$ in the offer phase
\Ensure Accepted contract $\gls{contract}^a$ and new contract space \gls{all_contracts}$'$
\State Sort $(\gls{contract}, o) \in \gls{all_contracts}^a \times O^a$ on $\gls{value_of_contract}$ in descending order
\For{$\gls{agent} \in \gls{community}$}
\State $\gls{contract}^{a,\gls{agent}}, o^{a,\gls{agent}} \gets \gls{all_contracts}^a \times O^a$.pop() \Comment{The $i^{th}$ contract with the most value}
\If{$\gls{contract}^{a,\gls{agent}}$meets local grid constraints}
\State $ (\gls{agent}, \gls{agent_j}, \gls{trades}) = \gls{contract}^{a,\gls{agent}}$
\State $\hat{\gls{agent}} \gets$ update-profiles(\gls{agent}, \gls{trades})
\State $\hat{\gls{agent_j}} \gets$ update-profiles(\gls{agent_j}, \gls{trades})
\State exchange-payment$_{\gls{agent} \rightarrow \gls{agent_j}}$($o^{a,\gls{agent}}$)
\State \gls{all_contracts}$' \gets \gls{all_contracts}$
\algorithmicbreak
\EndIf
\EndFor
\ForAll{$\gls{contract} = (\gls{agent}, k, \gls{trades}) \in \gls{all_contracts}$} \Comment{Contracts including agents in $\gls{contract}^a$ should be updated}
\State Compute $\gls{contract}'$ using the profile optimiser
\State  \gls{all_contracts}$' \gets \gls{all_contracts}' \setminus \{\gls{contract}\} \cup \{\gls{contract}'\}$
\EndFor
\State \Return $\gls{contract}^a$ , \gls{all_contracts}$'$
\end{algorithmic}
\end{algorithm}

\subsection{Peer selection}
During the peer selection process, the mechanism selects for each prosumer a subset of prosumers to trade with. This procedure is shown in \cref{alg:peer}.

In this model, we assume that each prosumer sends offers to a limited number of $k$ promising prosumers it negotiates with (in most simulations in this paper $k=5$, which basically means that each prosumer negotiates simultaneously with 5 prosumers that are most promising in terms of potential Gains from Trade).
The negotiation partners of a prosumer are sorted based on the potential gains as computed by Eq. \ref{eq:GT_coalition}, and the \gls{k} peers with the highest potential gain are selected as partners to send offers to. The prosumer is able to negotiate with each partner concurrently. Note the negotiating power of different prosumers may be asymmetric: considering some prosumer, it is not guaranteed that for each partner in its top \gls{k} preferred partners, the considered prosumer is also in those partners' top \gls{k}. This is natural, as some prosumers have more power to trade, more flexibility to offer, or a more attractive profile than others. On the other hand, if there are a lot of prosumers with similar generation/demand profiles, their negotiating power is limited, as there are a lot of alternative choices.

\begin{algorithm}[H]
\caption{Peer selection algorithm for agent $i$}
\label{alg:peer}
\begin{algorithmic}
\Require Set of all possible energy contracts \gls{contracts}, number of desired trading partners $k$
\Ensure List of the top-$k$ trading partners
\State Sort $\gls{contract} \in \gls{contracts}$ on $\gls{value_of_contract}$ in descending order
\State \Return $\gls{contracts} [1..k]$
\end{algorithmic}
\end{algorithm}

\subsection{Agent strategies}
The negotiation strategy dictates the offer and acceptance behaviour of an agent. 
Different negotiation strategies affect how the Gains from Trade inside a trading coalition is redistributed. But, it would not significantly affect the overall benefits of peer-to-peer trading. For this work, we model agents as having a linear concession strategy for dividing the expected gains during negotiations, but other strategies could also be employed.

In more detail, in the first round the agent proposes an offer that is equal to the maximum utility \gls{max_utility} that can be gained from trading with all partners combined. This represents the market power of the agent. The agent also defines a reservation value \gls{reservation}, which is the minimum utility the agent will accept. The bid in consequent rounds is determined as a linear function of the round number \gls{round} that decays to the minimum utility when the deadline $d$ is reached.
Formally, this function is defined as
\begin{equation}
    \gls{offer} = \gls{reservation} + (\gls{max_utility} - \gls{reservation}) \cdot (1 - \frac{\gls{round}}{\gls{deadline}})
\end{equation}

The agent will accept a bid, if a received bid has a value higher or equal to its last bid.

It is important to observe this decentralised P2P trading process is much more computationally intensive than the approach managed by a central market maker. Essentially, at each time step, each agent sends and accepts/rejects offers from $k$ peers or negotiation partners (for example, in our simulations $k=5$), after which the accepted offer with the highest GT in the market is cleared, a coalition forms and the process repeats. In the centralised clearing process, also one coalition is formed per time step, but the market maker selects and clears the deal with the highest GT, leading to much less offers being exchanged. Of course, we mostly look at market efficiency, but decentralised negotiation is computationally more expensive, especially if implemented in a decentralised system, such as a blockchain.

\section{Experimental Comparisons using Real World Datasets}
\label{sec:experiments}

To compare the effectiveness of peer-to-peer trading in a realistic community, we consider two scenarios, using data from real prosumers of two energy trials in the United Kingdom. The first scenario uses anonymised data from the Thames Valley Vision trial, where we compare the methods in an average-size community of ~200 households~\cite{ukerc2017new}.
The second scenario makes use of a much larger set of prosumers, using anonymised data from the Low Carbon London trial, providing data from several thousand households. Thus we can check whether the results hold up in larger settings as well as the effect that different compositions of a smaller community may have on the effectiveness of trading.
In both scenarios, we model the energy assets of a prosumer as follows:
\begin{itemize}
    \item Each prosumer has access to a lithium-ion battery. The price of the battery was fixed to \pounds150/kWh and its lifetime to 20 years. The battery life cycle data was obtained from \cite{xu2018modeling}. For each prosumer, the battery capacity was computed as the battery capacity that minimizes the prosumer's bill when used for self-consumption only.
    \item Each prosumer has access to a share of a wind turbine, where the share was computed by minimizing the yearly bill given no access to any storage. The price of the wind turbine was set to \pounds1072 /kW and its lifetime to 20 years. The wind speed to power curve has been interpolated using data for a typical community wind turbine (i.e. Enercon E-33).
\end{itemize}
Furthermore, both scenarios consider a flat import tariff of 16p/kWh (i.e. 16 pence or 0.16 British pounds per kWh) and an export tariff of 0p/kWh.
We consider the simulation of these energy communities for a 1 year period with 30-minute granularity, resulting in $365 * 48 = 17520$ time steps. 
Finally, we note that during the simulations conducted with real load profiles (from the two communities described above) and the European low voltage test feeder network provided by IEEE~\cite{IEEE_bus_system}, we found no grid constraints were violated. As a result, the trades or contracts with the highest gains were always accepted by the grid operator.

\subsection{Scenario 1: Thames Valley Vision}
The first scenario considers a simulation of 200 households. The demand data has been retrieved from the Thames Valley Vision trial \cite{ukerc2017new}. 

\begin{figure}
     \centering
     \begin{subfigure}[b]{0.48\textwidth}
         \centering
         \includegraphics[width=\textwidth]{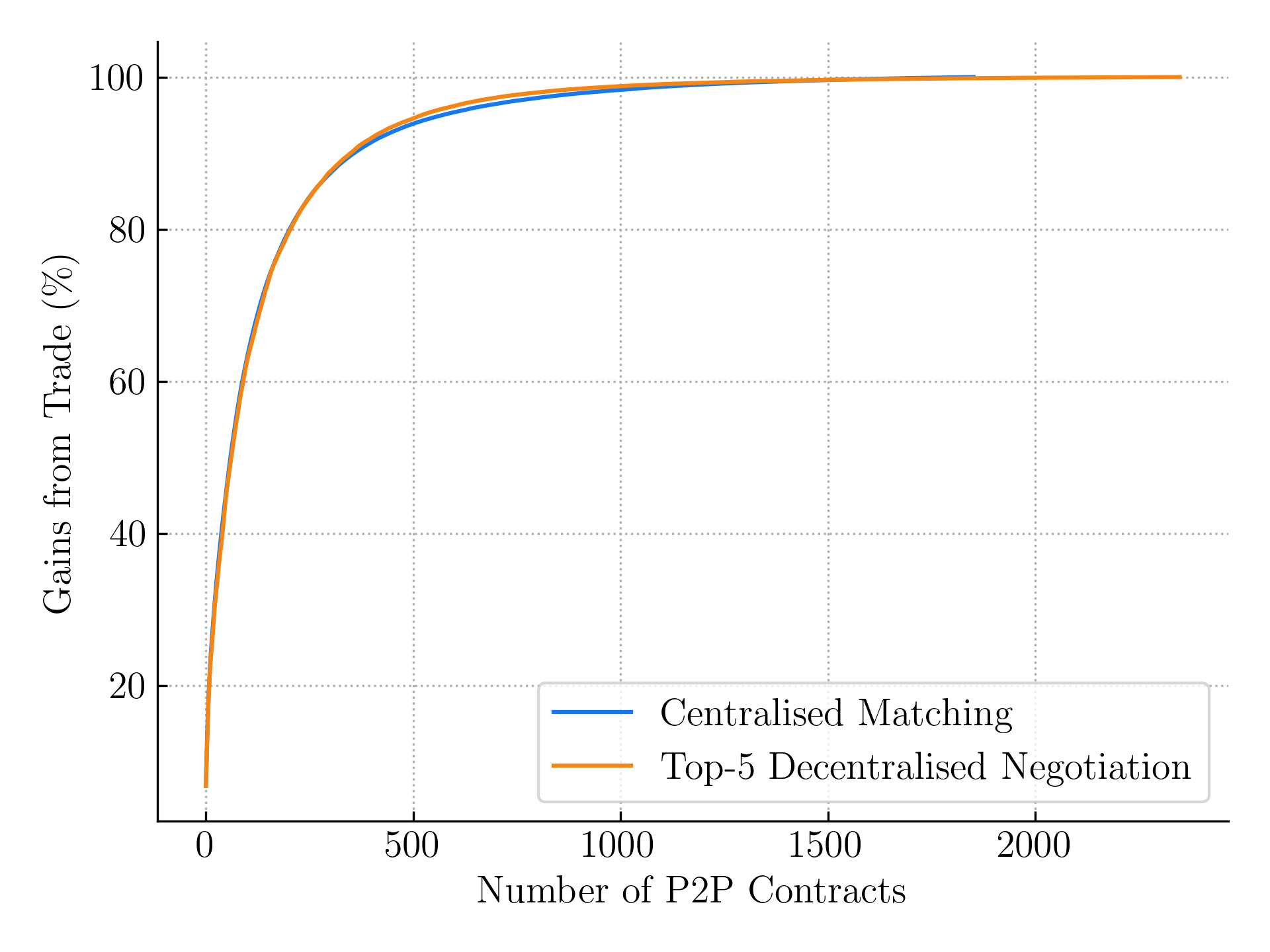}
         \caption{P2P Energy Contracts}
         \label{fig:experiments:thames:trade_for_gains}
     \end{subfigure}
     \hfill
     \begin{subfigure}[b]{0.48\textwidth}
         \centering
         \includegraphics[width=\textwidth]{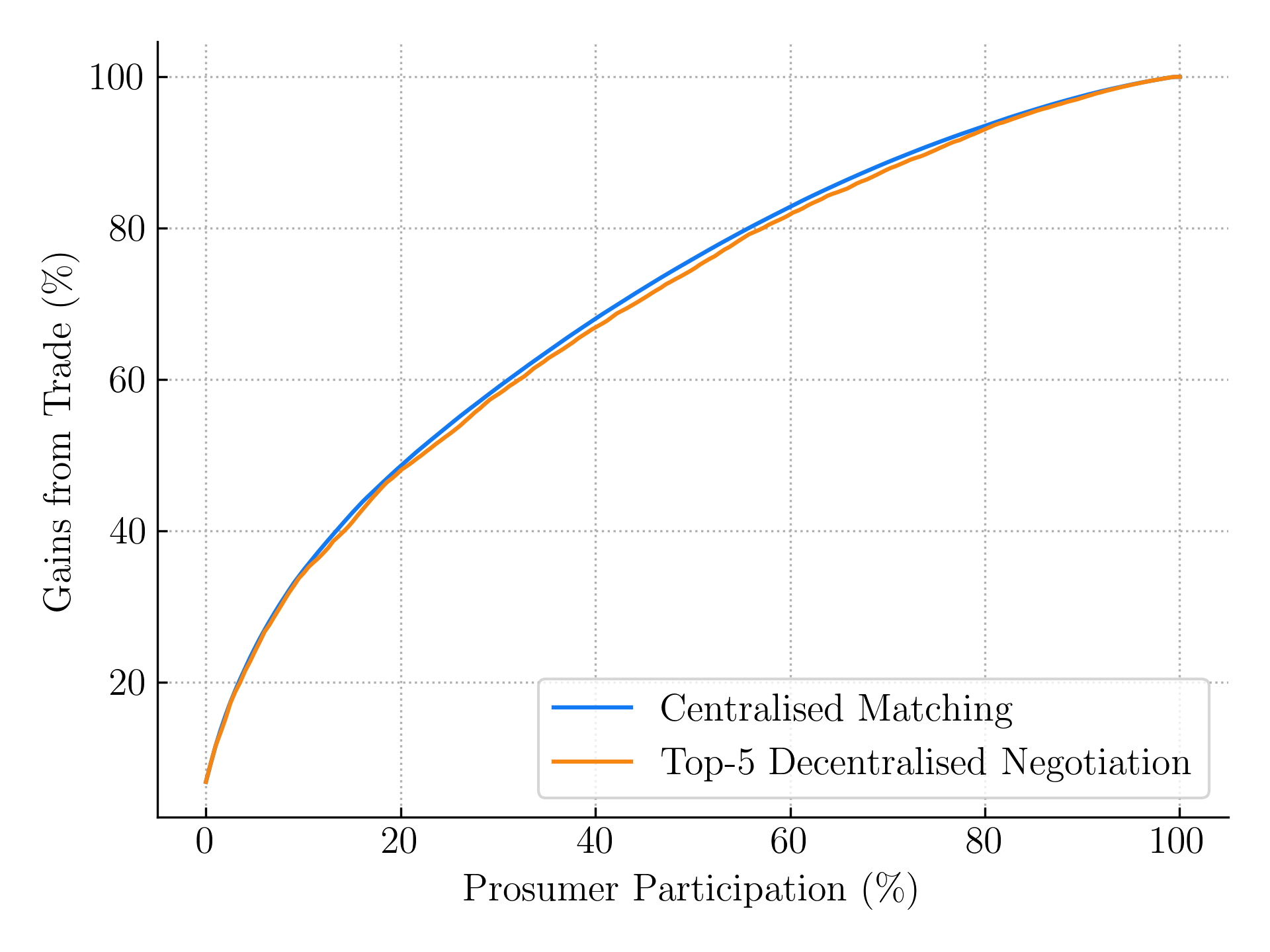}
         \caption{Prosumer participation}
         \label{fig:experiments:thames:community_for_gains}
     \end{subfigure}
        \caption{Convergence of the Gains from Trade (as a percentage of the maximum possible Gains from Trade)}
        \label{fig:experiments:thames:gains_convergence}
\end{figure}

\Cref{fig:experiments:thames:trade_for_gains} shows the Gains from Trade achieved by the community as more and more contracts (and thus P2P trading coalitions) are established. This figure displays the convergence of the GT towards the theoretical limit of maximal GT possible in that community (this limit is reached when every single prosumer in the community participates in trading). As expected, this graph shows that it eventually does converge to this theoretical limit as the number of P2P energy contracts increases.

However, we find that only a small number of P2P energy contracts are actually required, for both the centralised matching and clearing models and the decentralised negotiation model. Both models achieve close to the maximum possible Gains from Trade after about 1000 established contracts, while with a community of 200 prosumers there would be a theoretical total of $\frac{200^{2}}{2}-200=19900$ different pairings. This might suggest that peer-to-peer models achieve maximal efficiency relatively quickly and that only a small number of bilateral contracts need to be established. Between centralised matching and clearing, the difference between convergence is insignificant. This supports that the converging behaviour is generally applicable to different peer-to-peer trading models.

\Cref{fig:experiments:thames:community_for_gains} shows the convergence of the Gains from Trade in terms of prosumer participation. Notably, also here we do need all prosumers to participate in peer-to-peer trading to achieve maximum efficiency. A noticeable portion of the gains can be achieved with less participation, as 60\% of the gains are reached with 30\% of prosumers' participation. Furthermore, the marginal contribution provided by increased prosumer participation diminishes quickly for both models, which can mostly be explained by the fact that the set up proposed in this work first selects the trades with the highest gains for the community, leaving trades with lower gains for later rounds. Both models have similar behaviour as they both use optimal schedules computed by the joint profile optimiser.

To investigate further why the marginal contribution of a contract diminishes so quickly and why both models show similar behaviours, we take a closer look at the dynamics of the peer-to-peer trading mechanism, by looking at the trading coalitions that form between prosumers in each negotiation/matching round.
A trading coalition can be formed when prosumers have agreed on a contract with each other either directly, but it is also formed when a prosumer is indirectly connected by another prosumer, e.g. prosumer $i$ has a contract with prosumer $j$ and prosumer $i$ has a contract with prosumer $k$, then the trading coalition would be $\{i, j, k\}$, and the profile optimiser will jointly optimise the assets of the 3 prosumers. 

\begin{figure}[H]
     \centering
     \begin{subfigure}[b]{0.48\textwidth}
         \centering
         \includegraphics[width=\textwidth]{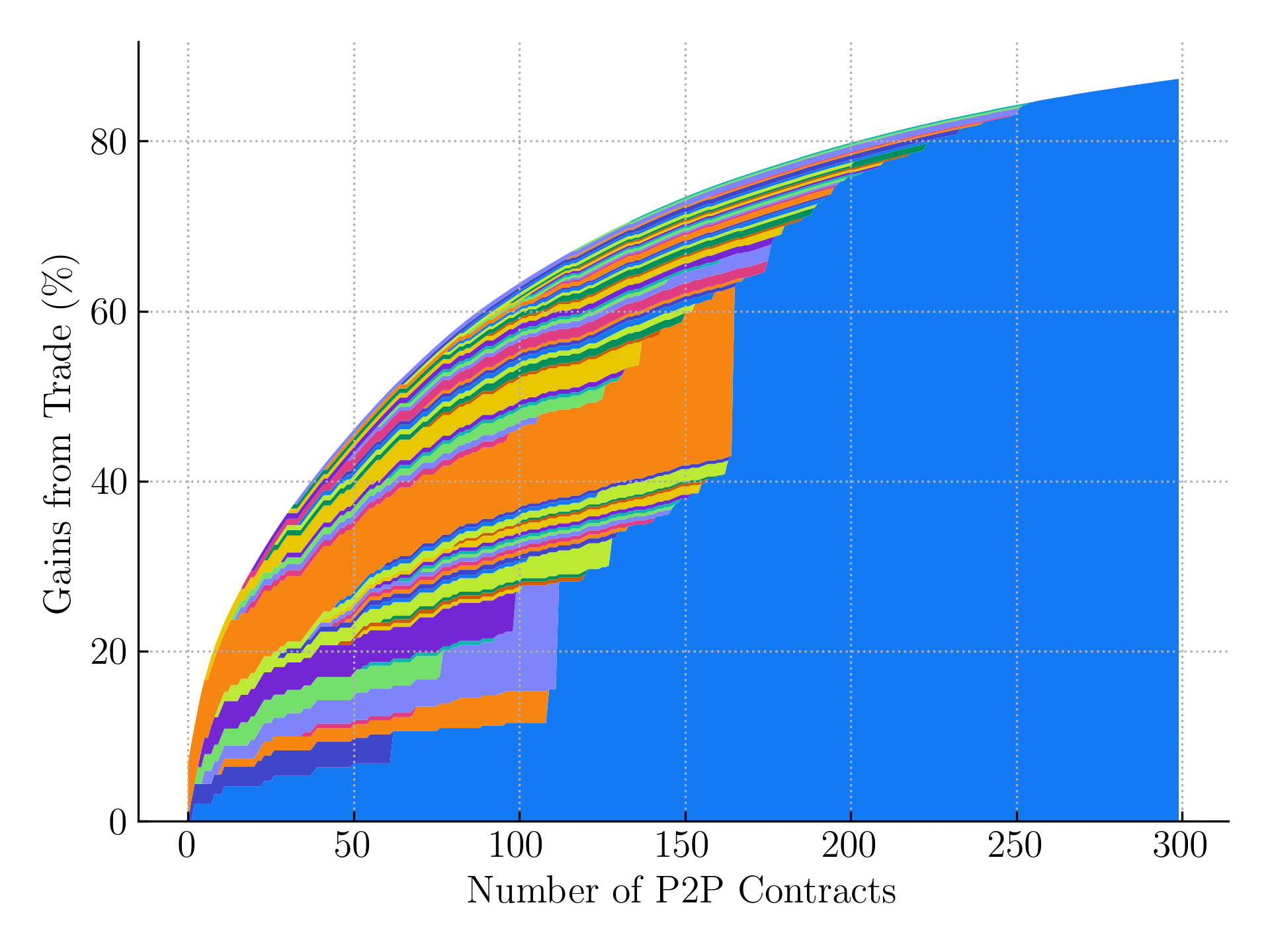}
         \caption{Centralised matching and clearing}
         \label{fig:experiments:thames:dynamics:tgc:cmc}
     \end{subfigure}
     \hfill
     \begin{subfigure}[b]{0.48\textwidth}
         \centering
         \includegraphics[width=\textwidth]{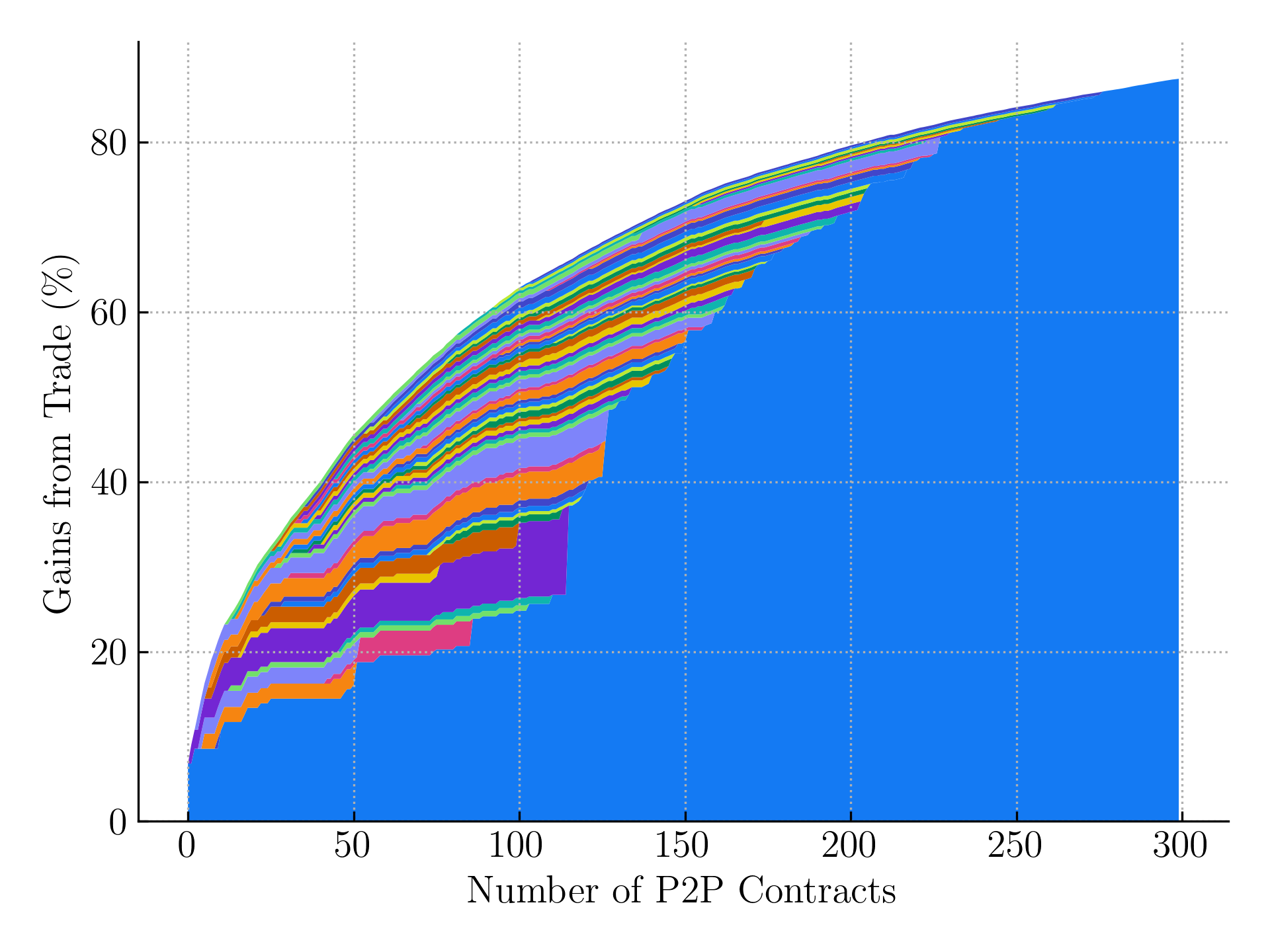}
         \caption{Top-5 decentralised negotiation}
         \label{fig:experiments:thames:dynamics:tgc:nego}
     \end{subfigure}
        \caption{Gains per energy trading coalition as a function of the number of P2P contracts between prosumers. Different colours represent coalitions formed by P2P contracts.}
        \label{fig:experiments:thames:dynamics:tgains_convergence}
    \begin{subfigure}[b]{0.48\textwidth}
         \centering
         \includegraphics[width=\textwidth]{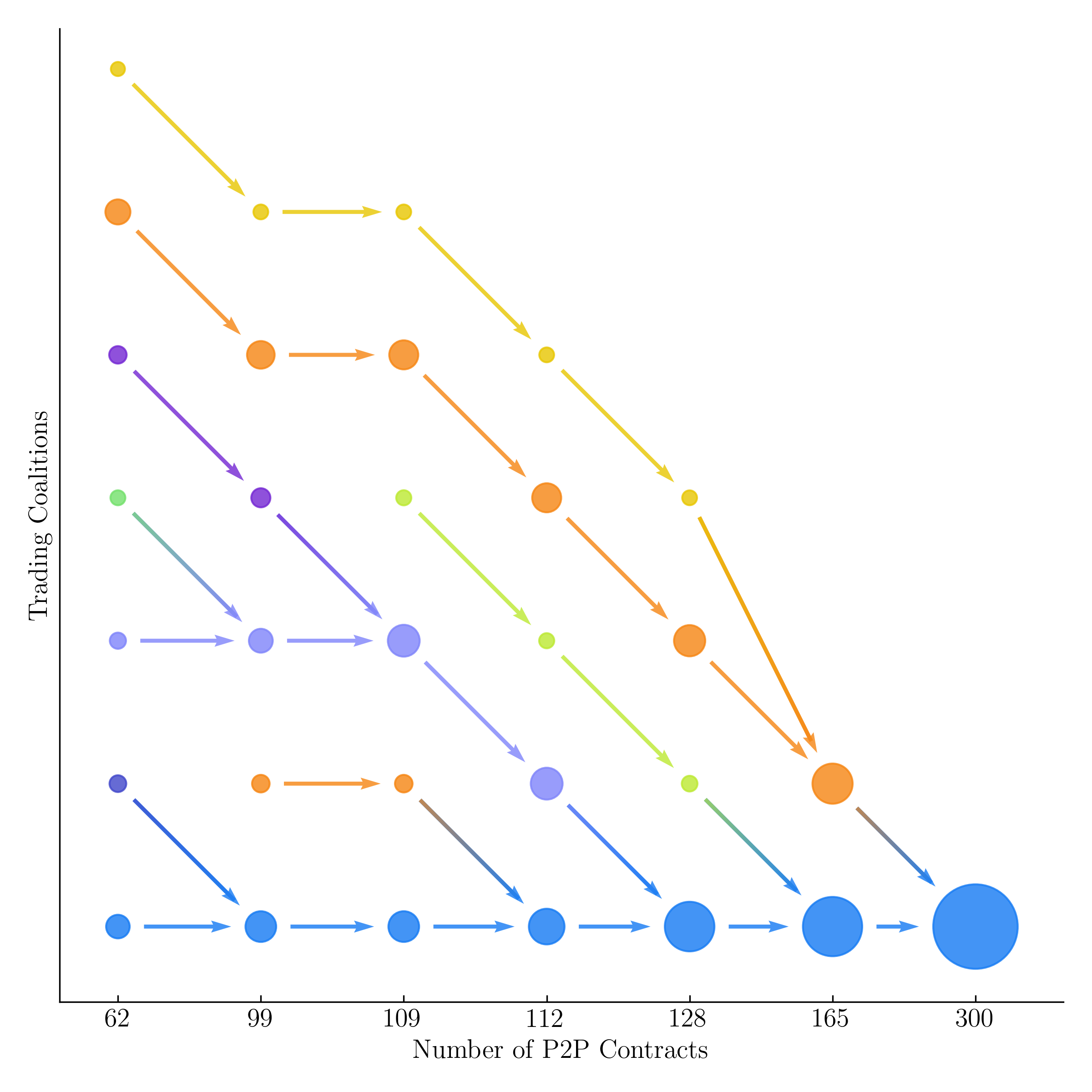}
         \caption{Centralised matching and clearing}
         \label{fig:experiments:thames:flow:trades:cmc}
     \end{subfigure}
     \hfill
     \begin{subfigure}[b]{0.48\textwidth}
         \centering
         \includegraphics[width=\textwidth]{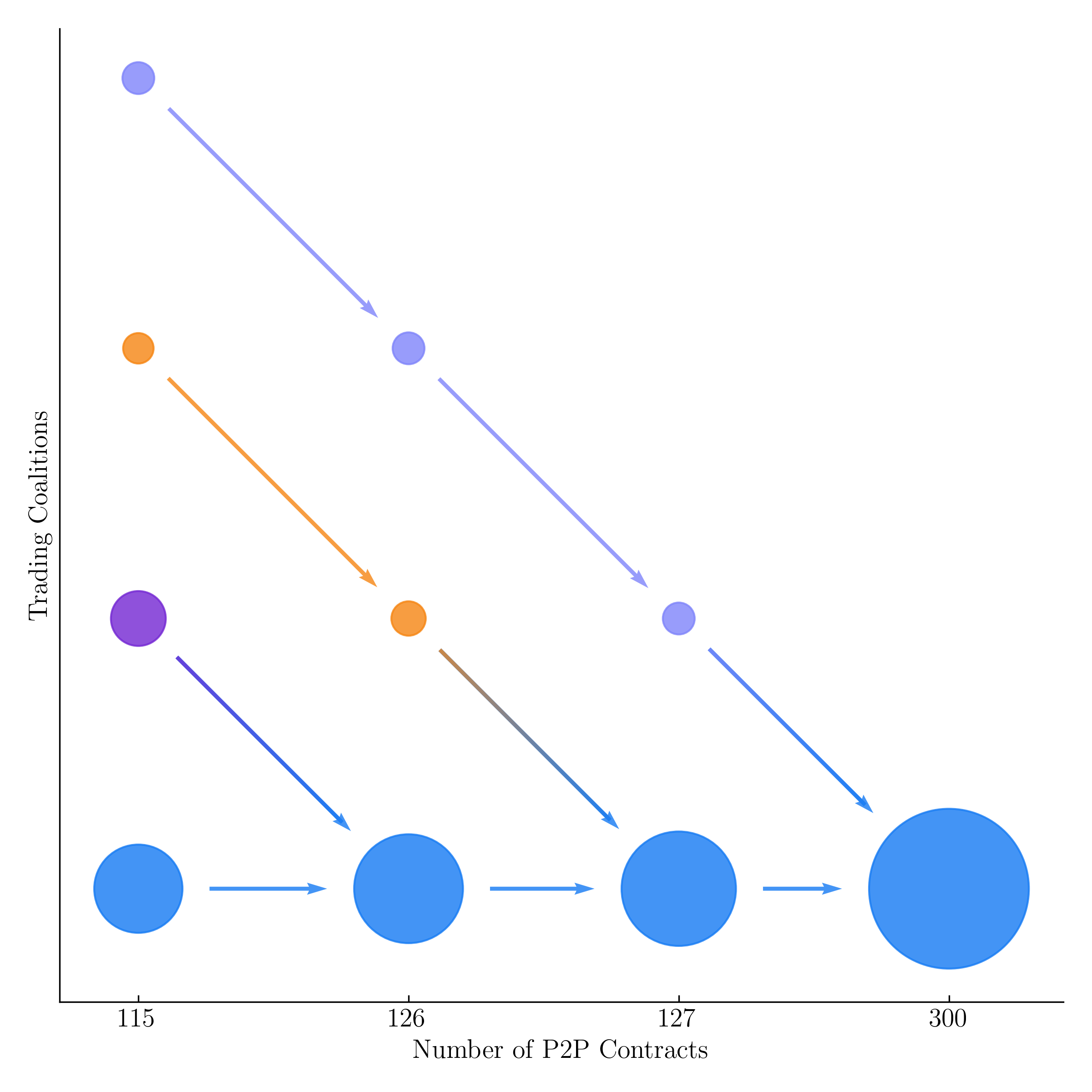}
         \caption{Top-5 decentralised negotiation}
         \label{fig:experiments:thames:flow:trades:nego}
     \end{subfigure}
        \caption{Merging of different energy trading coalitions based on the number of P2P contracts established. The area of the disks represents the total gains from trade the trading coalition achieves. While a new coalition merge happens in each step, only time steps where coalitions that contribute at least 3\% of the Gains from Trade are merged are shown in the illustration, and only the coalitions that contribute at least 2\% to the maximal GT are shown.}
        \label{fig:experiments:thames:flow:trades}
\end{figure}

\Cref{fig:experiments:thames:dynamics:tgains_convergence} shows the Gains of Trade for each trading coalition, where the x-axis represents the number of contracts accepted and the y-axis represents the total GT achieved. Different colours represent a different disjoint trading coalition. Merging of coalitions can be seen in the figure when one coalition of a certain colour that has accepted the contract is taken over by the colour of the coalition that has offered the contract. Initially, we see that there are many small coalitions, but eventually, every prosumer is included within the same trading coalition, i.e. forming the grand coalition. 
Here we can see clear differences between the dynamics experienced, by the centralised matching and clearing approach as opposed to the decentralised negotiation approach.
The centralised matching approach creates a larger trading coalition before merging them together into the grand coalition, while for decentralised negotiation trading coalitions stay relatively small. 

In \cref{fig:experiments:thames:flow:trades}, we show the merging process of the trading coalitions for both centralised matching and decentralised negotiation. The x-axis represents the time steps in which merging happened in terms of how many P2P contracts were established. The area of the disk represents the size of the coalition, and the colours correspond to the different coalitions in \Cref{fig:experiments:thames:dynamics:tgains_convergence}. Only coalitions with significant size (contributing to at least 2\% of Gains from Trade) and their merges are presented in the figure.
Here we can see that the centralised matching approach creates larger and more trading coalitions.
This seems to suggest that centralised matching prefers contracts between prosumers that are already part of a trading coalition, before merging them into the larger trading coalition. Furthermore, when the trading coalitions are merging into the largest coalition, the majority of the potential Gains from Trade has already been achieved. This might suggest that the gains primarily come from matching the prosumers that have very different demand profiles since those contracts are established first.
After those gains have been realised, the trading coalitions have compounded enough minor differences in demand profiles for them to lead to a significant gain. 
For decentralised negotiation, while this effect can also be observed, since lots of small trading coalitions form, they tend to not grow larger. This is likely due to these small trading coalitions being quite diverse as opposed to the larger coalitions, giving them more market power and therefore laying claim to a larger fraction of the benefits. Larger coalitions have less market power and are therefore more likely to accept a contract proposed by a small coalition, leading to small trading coalitions merging into the largest.

\begin{figure}[H]
     \centering
     \begin{subfigure}[b]{0.48\textwidth}
         \centering
         \includegraphics[width=\textwidth]{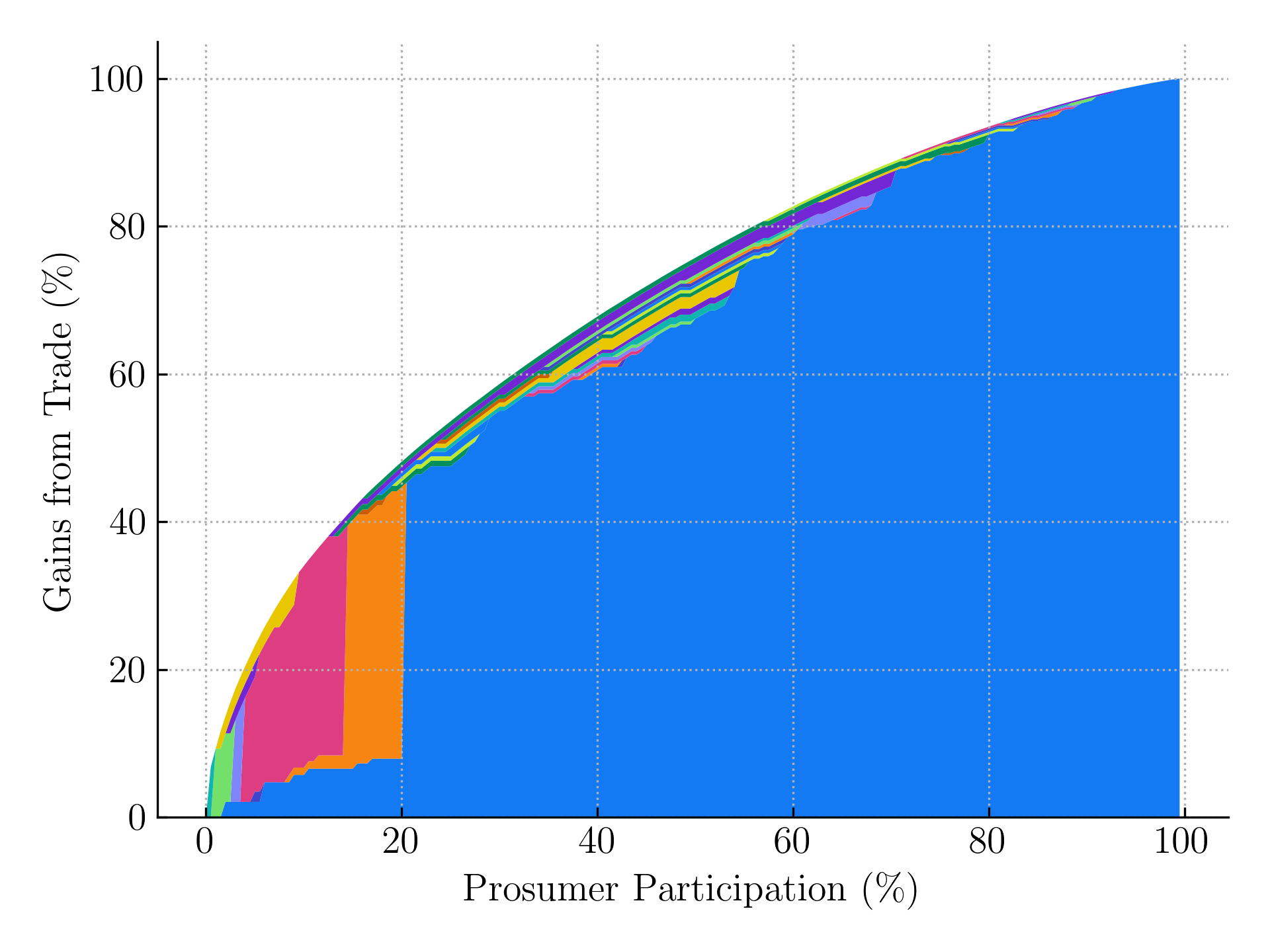}
         \caption{Centralised matching and clearing}
         \label{fig:experiments:thames:dynamics:gc:cmc}
     \end{subfigure}
     \hfill
     \begin{subfigure}[b]{0.48\textwidth}
         \centering
         \includegraphics[width=\textwidth]{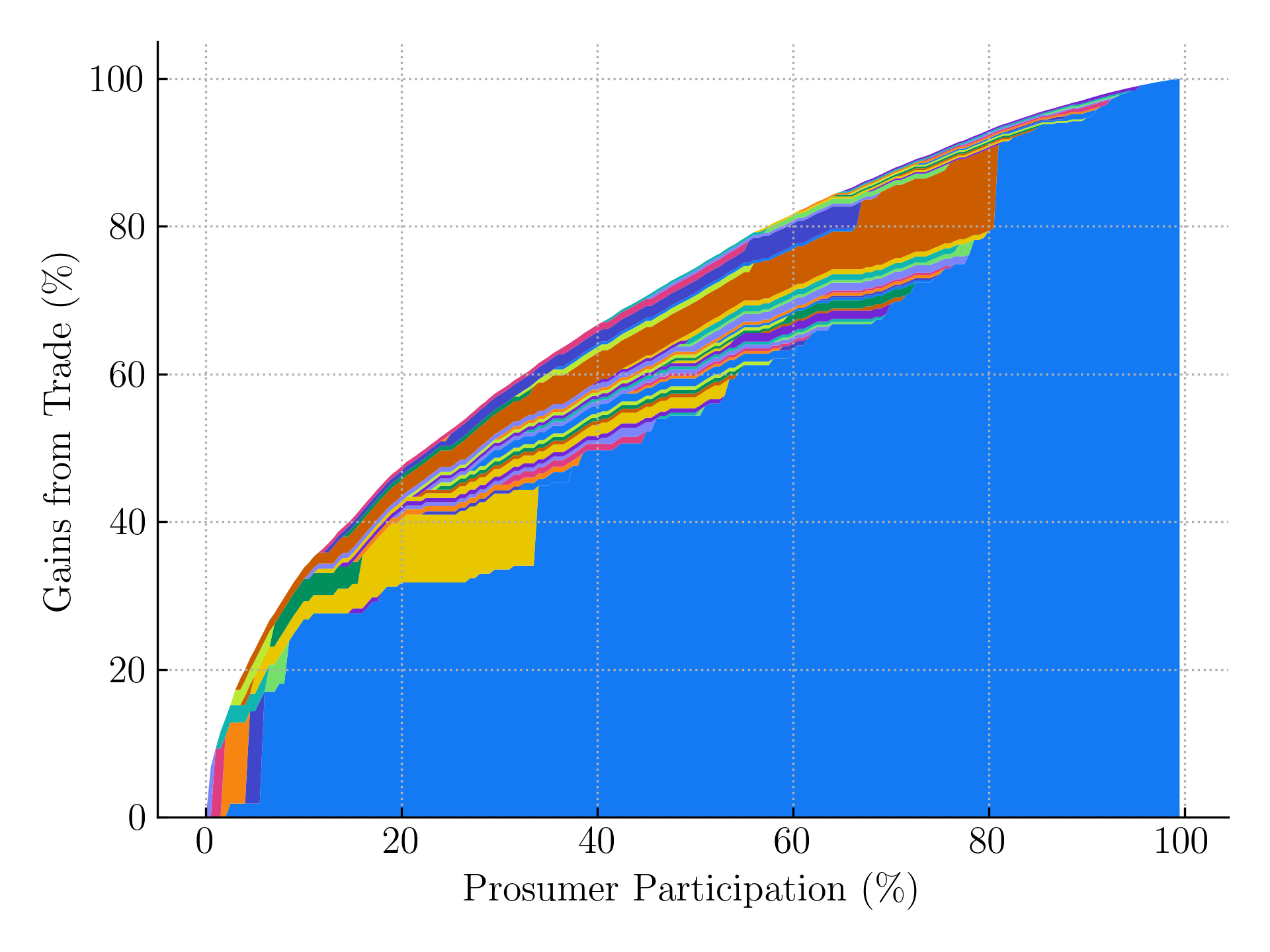}
         \caption{Top-5 decentralised negotiation}
         \label{fig:experiments:thames:dynamics:gc:nego}
     \end{subfigure}
        \caption{Gains per trading coalition as a function of prosumers that participate in trading. Different colours represent coalitions formed by P2P contracts.}
        \label{fig:experiments:thames:dynamics:gains_convergence}
             \begin{subfigure}[b]{0.48\textwidth}
         \centering
         \includegraphics[width=\textwidth]{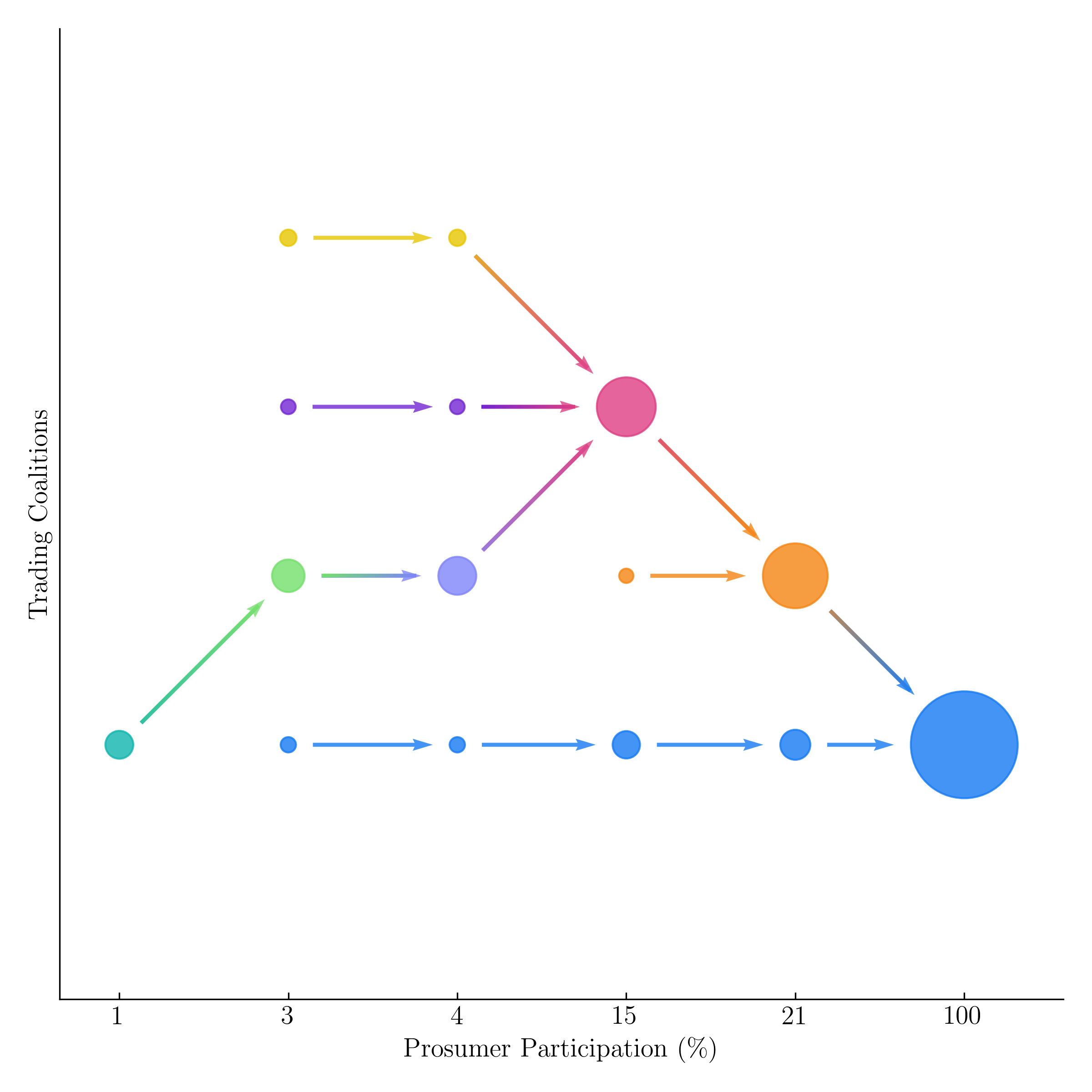}
         \caption{Centralised matching and clearing}
         \label{fig:experiments:thames:flow:cmc}
     \end{subfigure}
     \hfill
     \begin{subfigure}[b]{0.48\textwidth}
         \centering
         \includegraphics[width=\textwidth]{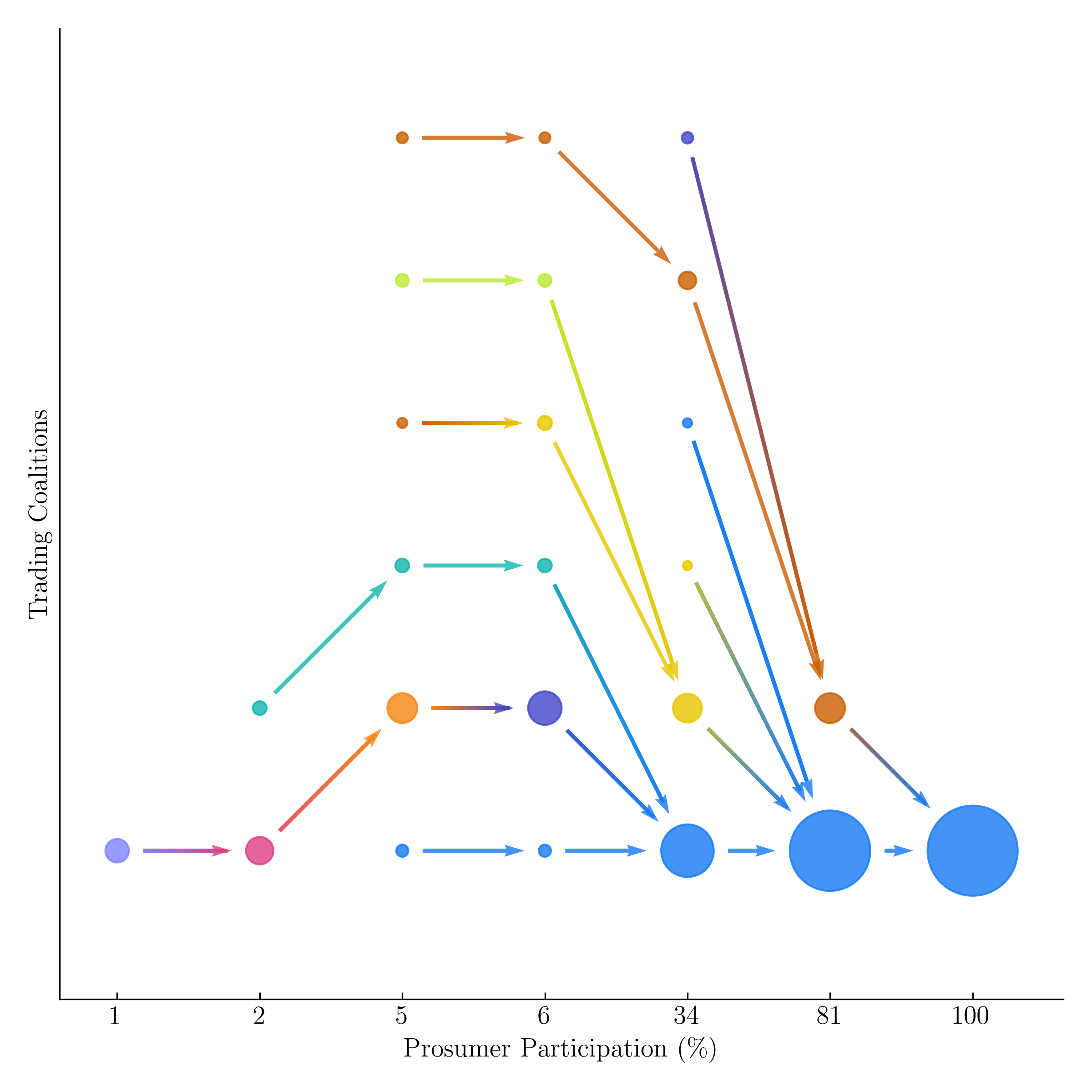}
         \caption{Top-5 decentralised negotiation}
         \label{fig:experiments:thames:flow:nego}
     \end{subfigure}
        \caption{Merging of different energy trading coalitions based on the percentage of prosumers that participate in trade. The size (area) of the disks represents the total Gains from Trade that specific trading coalition achieves. Only timesteps where coalitions that contribute at least 5\% of the GT are included in the visualisation, and only the coalitions that contribute at least 1\% to the gains from trades are shown.}
        \label{fig:experiments:thames:flow}
\end{figure}

\Cref{fig:experiments:thames:dynamics:gains_convergence} shows the Gains from Trade of trading coalitions related to prosumer participation. The x-axis represents the percentage of prosumer participation in the P2P model which continues to increase as more contracts are established, and the y-axis represents the total GT achieved, as percentage of the maximal. The colours represent trading coalitions, similarly to \Cref{fig:experiments:thames:dynamics:tgains_convergence}. Here we observe that for the centralised matching and clearing there are fewer trading coalitions, while for decentralised negotiation we observe a lot more trading coalitions. 

In \Cref{fig:experiments:thames:flow}, we show the trading coalitions of significant size (contributing to at least 5\% of the maximal GT) and their merging processes, where the size of the disk represents coalition size and the colour corresponds to the coalitions from \Cref{fig:experiments:thames:dynamics:gains_convergence}. Here we can also see for centralised matching that there are a small number of significantly sized trading coalitions and that they tend to merge into the largest coalition. Yet, the merging process for decentralised negotiation involves a lot more groups.
These effects seem quite different from those observed when looking at the dynamics when looking at the accepted contracts, However, they can be explained by the fact that the most diverse consumers first form a trading coalition, leading to prosumers that join at a later stage either joining the diverse consumers in the larger coalition, or having to form themselves a smaller trading coalition. 
We see that some of these smaller coalitions do form, however, the majority do join this larger coalition since no other single consumer is diverse enough. For decentralised negotiation, this plays a smaller role, since these diverse consumers are more likely to not join together initially due to them having a higher market power and being more likely to form smaller trading coalitions.

\subsection{Scenario 2: Low Carbon London}
In the second scenario, we consider a larger dataset with over 5000 households. The demand data has been retrieved from the Low Carbon London trial \cite{ukpower2014smart}. This dataset provides a larger and more representative set of households for the London region. However, since realistically sized communities behind an LV substation often are smaller sized, e.g. around 50-200~\cite{lucas2016distribution}, instead of considering the whole dataset at once, we create experimental scenarios with realistic diversity and sizes closely fitting the distributions in the large data set.

\begin{figure}[t]
     \centering
     \begin{subfigure}[b]{0.32\textwidth}
         \centering
         \includegraphics[width=\textwidth]{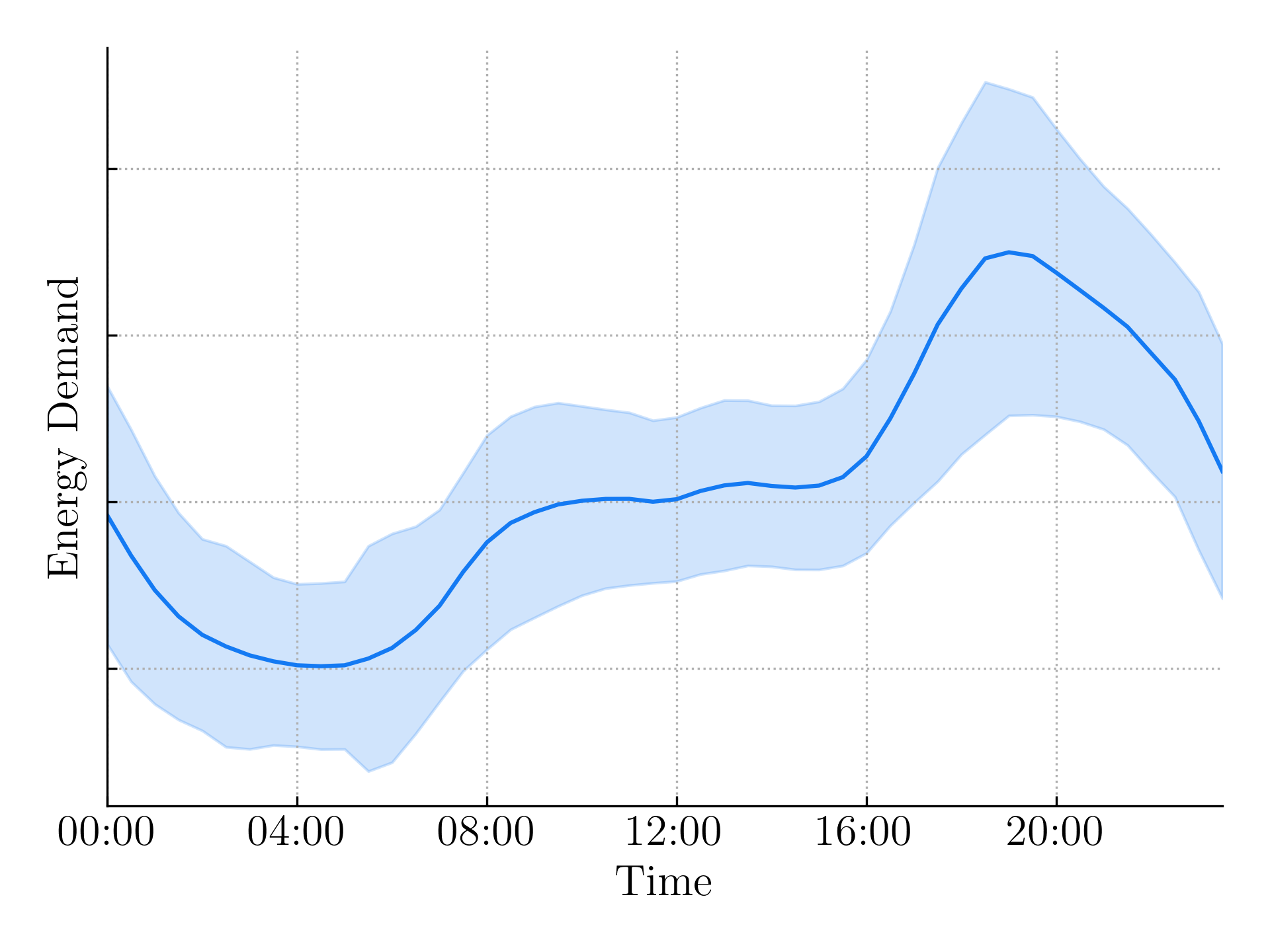}
         \caption{Evening peak (3910 profiles)}
         \label{fig:experiments:london:clusters:ep}
     \end{subfigure}
     \begin{subfigure}[b]{0.32\textwidth}
         \centering
         \includegraphics[width=\textwidth]{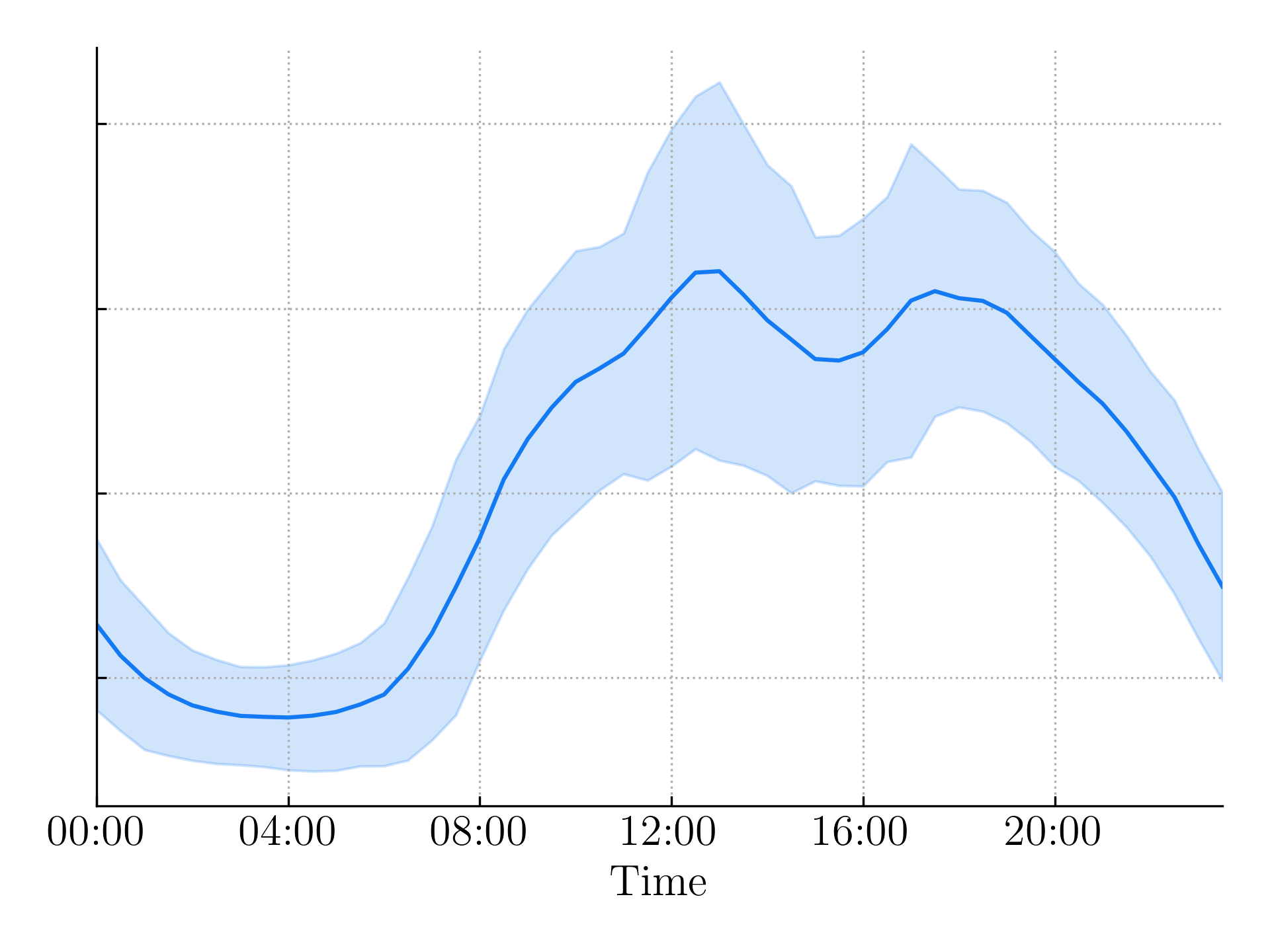}
         \caption{Work from home (599 profiles)}
         \label{fig:experiments:london:clusters:daily}
     \end{subfigure}
     \begin{subfigure}[b]{0.32\textwidth}
         \centering
         \includegraphics[width=\textwidth]{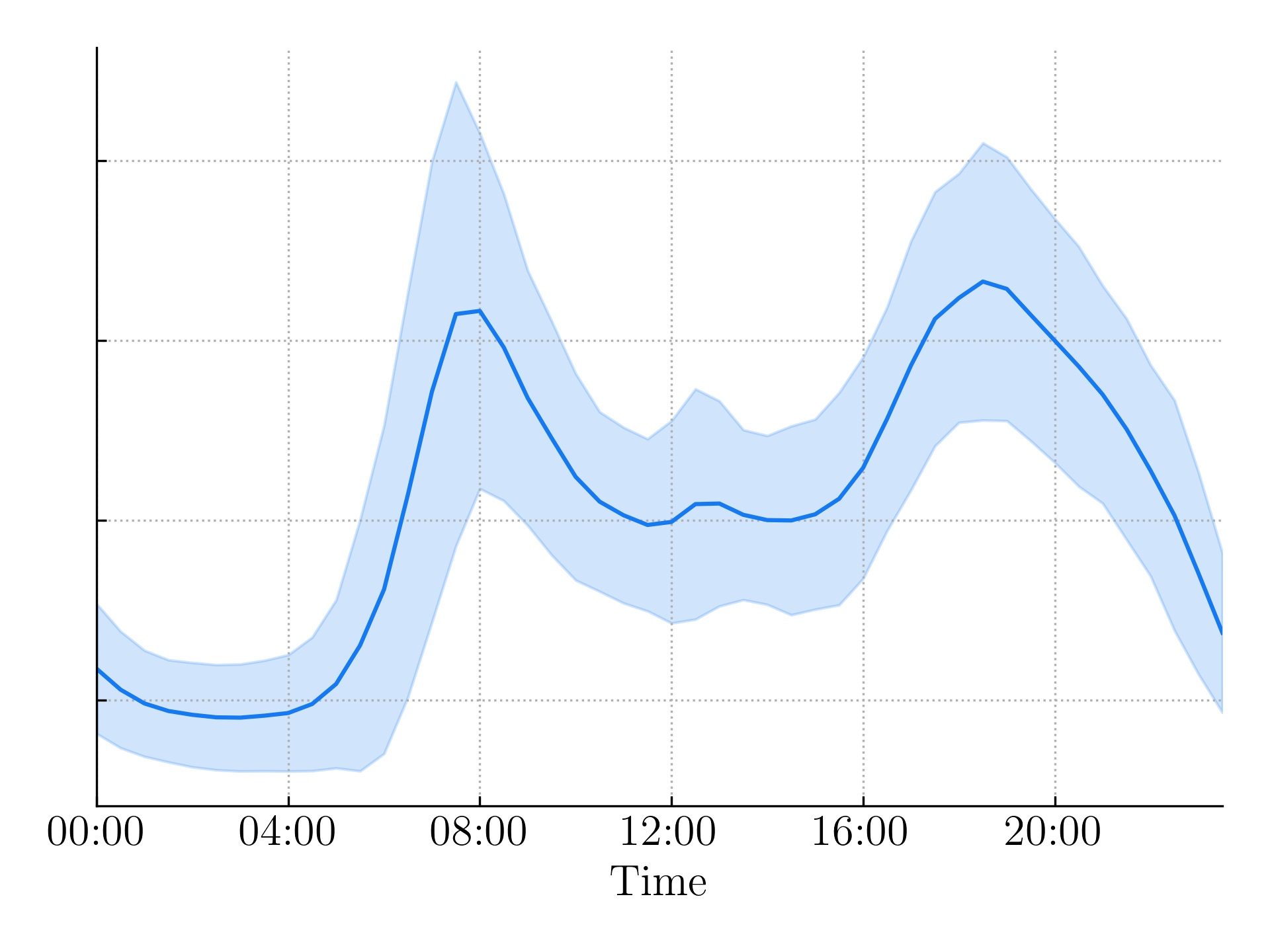}
         \caption{Morning + Evening peak (526 profiles)}
         \label{fig:experiments:london:clusters:m}
     \end{subfigure}
     \begin{subfigure}[b]{0.32\textwidth}
         \centering
         \includegraphics[width=\textwidth]{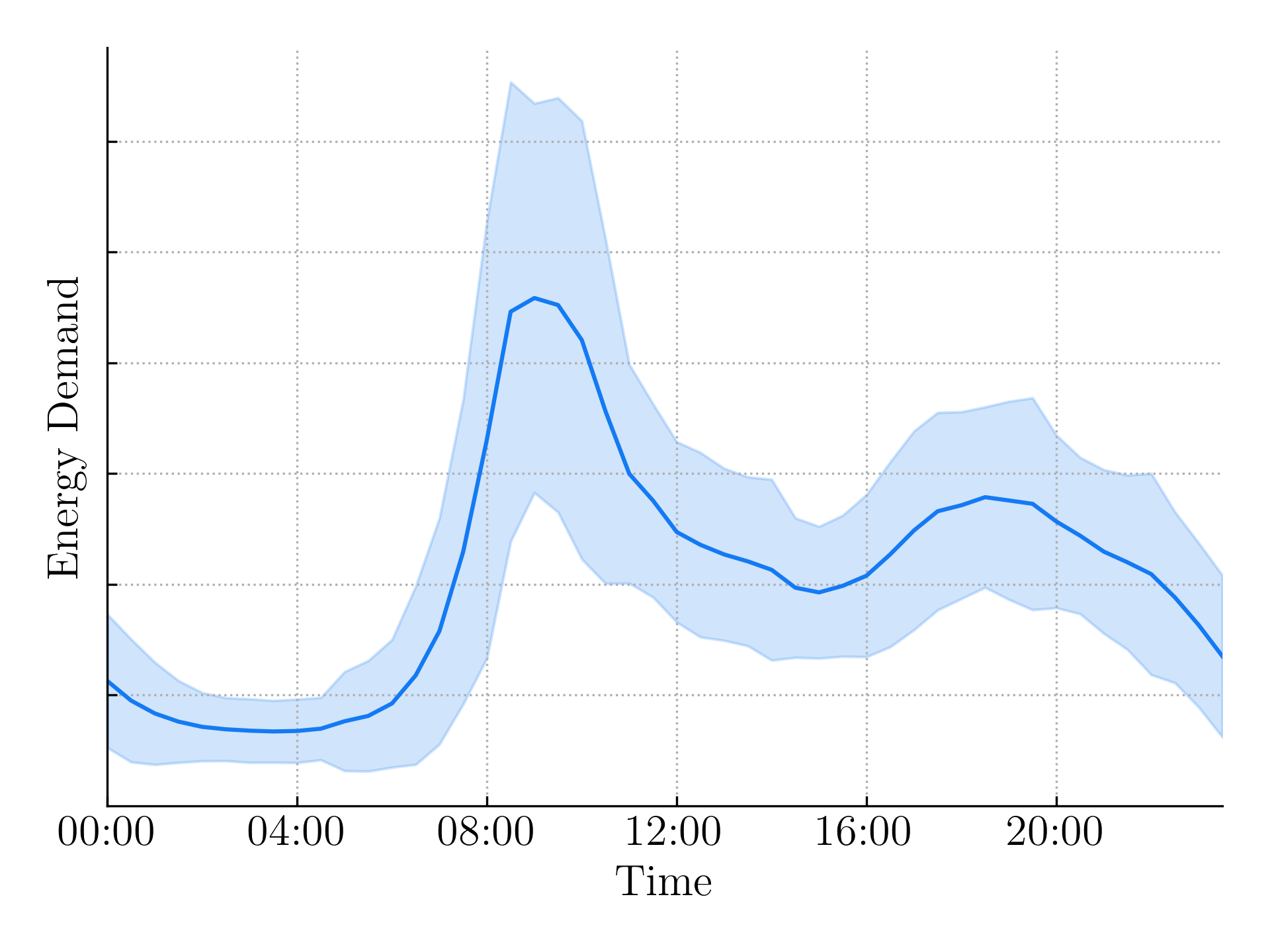}
         \caption{Morning peak (169 profiles)}
         \label{fig:experiments:london:clusters:mp}
     \end{subfigure}
     \begin{subfigure}[b]{0.32\textwidth}
         \centering
         \includegraphics[width=\textwidth]{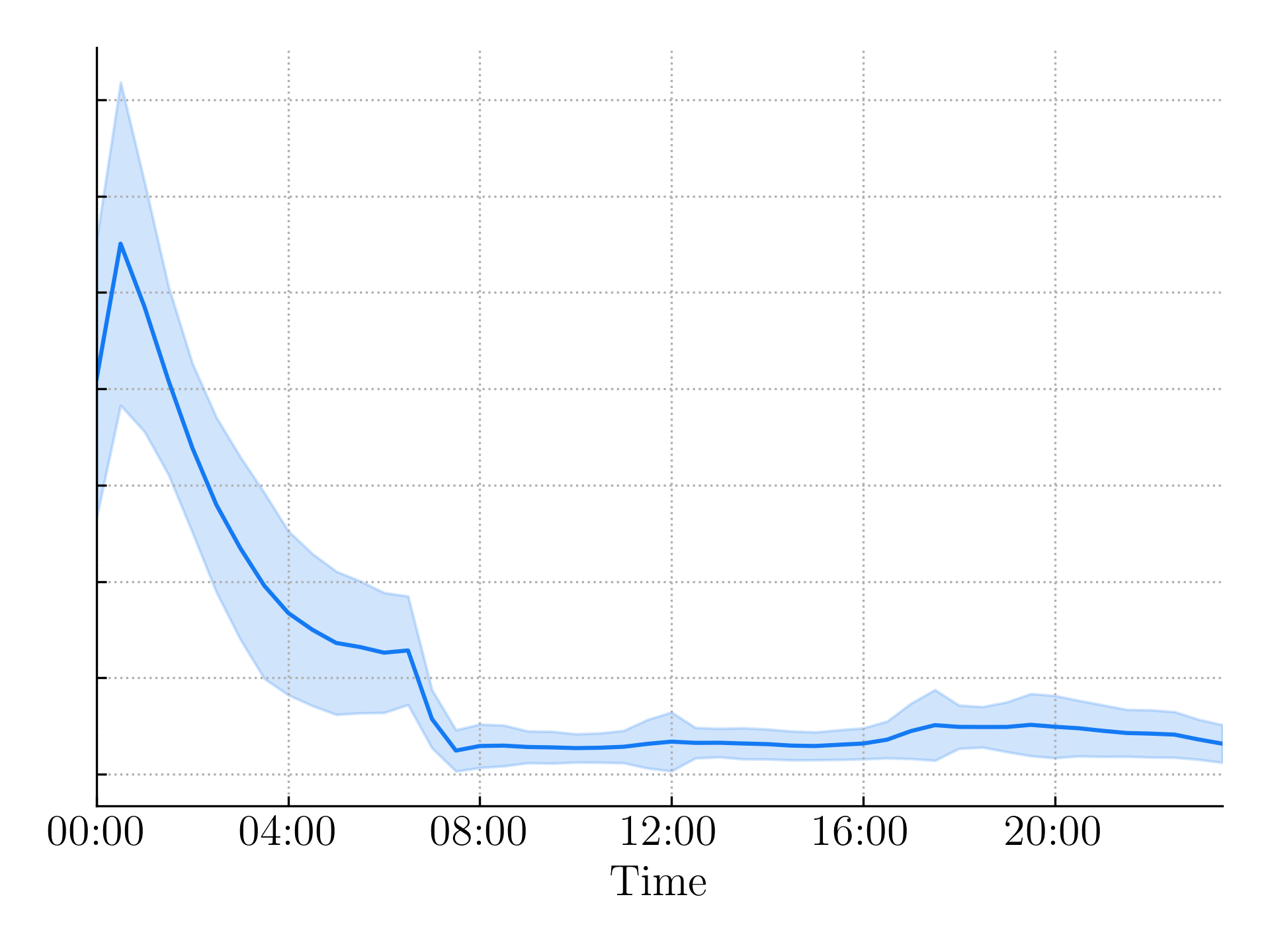}
         \caption{Night Owl (47 profiles)}
         \label{fig:experiments:london:clusters:no}
     \end{subfigure}
        \caption{Average daily demands for different consumer profile clusters. Error bars represent the standard deviation of the demand profiles within the cluster.}
        \label{fig:experiments:london:clusters}
\end{figure}

To ensure that these scenarios reflect the whole community correctly, we apply stratified sampling of consumer profiles from the larger pool. A widely used method for identifying these strata in energy demand modelling is clustering on energy demand profiles \cite{electronics10030290, kwac2014household}. In this study, we use K-means clustering on daily averaged consumption in the winter months of the prosumers. The full methodology for performing the clustering is described in \ref{appendix:clustering}. 
From our clustering analysis, we identify and select 5 distinct demand profiles from the resulting clusters. These are shown in \cref{fig:experiments:london:clusters}.

Analogous to earlier works, we find that the majority of profiles follow an evening peak pattern, where a noticeable peak in consumption can be observed during the early evening, often as a result of prosumers coming home from work. \Cref{fig:experiments:london:clusters:daily} shows another much occurring pattern, where energy demands stay consistent throughout the day while still being low during the night. Since this is likely an effect of prosumers working from home and therefore using energy throughout the day, we call this group `work from home'. The final large group consists of prosumers that consume large quantities of energy during the morning and evening, likely pertaining to users that have a similar lifestyle to the `Evening peak' group, but also use a lot of energy in the mornings.
The remaining two clusters do not have this peak consumption during the evening but at different moments during the day. The `morning peak' group consumes the majority of energy during the early mornings, while the `night owl' group consumes the majority of their energy overnight. 

\begin{figure}[H]
    \centering
     \includegraphics[width=0.7\textwidth]{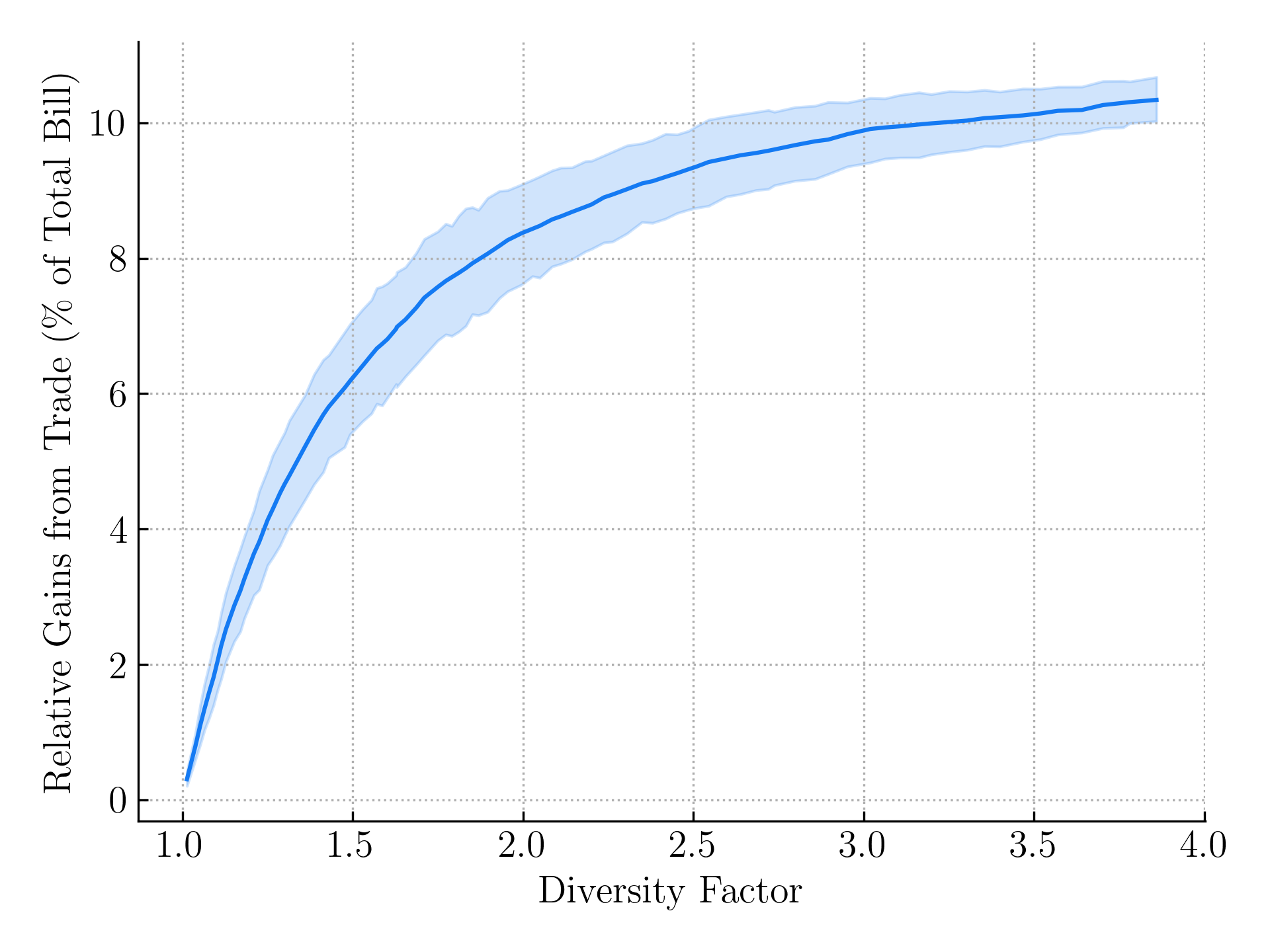}
    \caption{Influence of diversity factor of the energy community on total Gains from Trade}
    \label{fig:experiments:london:diversity_factor:gt}
\end{figure}

Experiments for convergence of the Gains from Trade and the trading dynamics show results very similar to those found for the Thames Valley Vision data set, which is not surprising, given that both trials study consumers from roughly the same area of the UK (southern England/London area).

However, we also explore scenarios where change the composition of the community (in terms of demand profiles or patterns), and here we find slightly different results. To quantify how the composition changes, we evaluate the efficiency with regard to the diversity of the community. The diversity of a community can be expressed using the diversity factor.
 
\Cref{fig:experiments:london:diversity_factor:gt} shows the Gains from Trade as a percentage of the total bill compared to the diversity factor. Multiple (i.e. 15) different communities have been generated for several (i.e. 100) diversity values. 
We find that the gains from trade increase with diversity. Diverse communities contain prosumers that have many different demand profiles, making it more likely that another prosumer can cover residual demands. However, this effect is less noticeable the more diverse a community is.

\begin{figure}[H]
     \centering
     \begin{subfigure}[b]{0.48\textwidth}
         \centering
         \includegraphics[width=\textwidth]{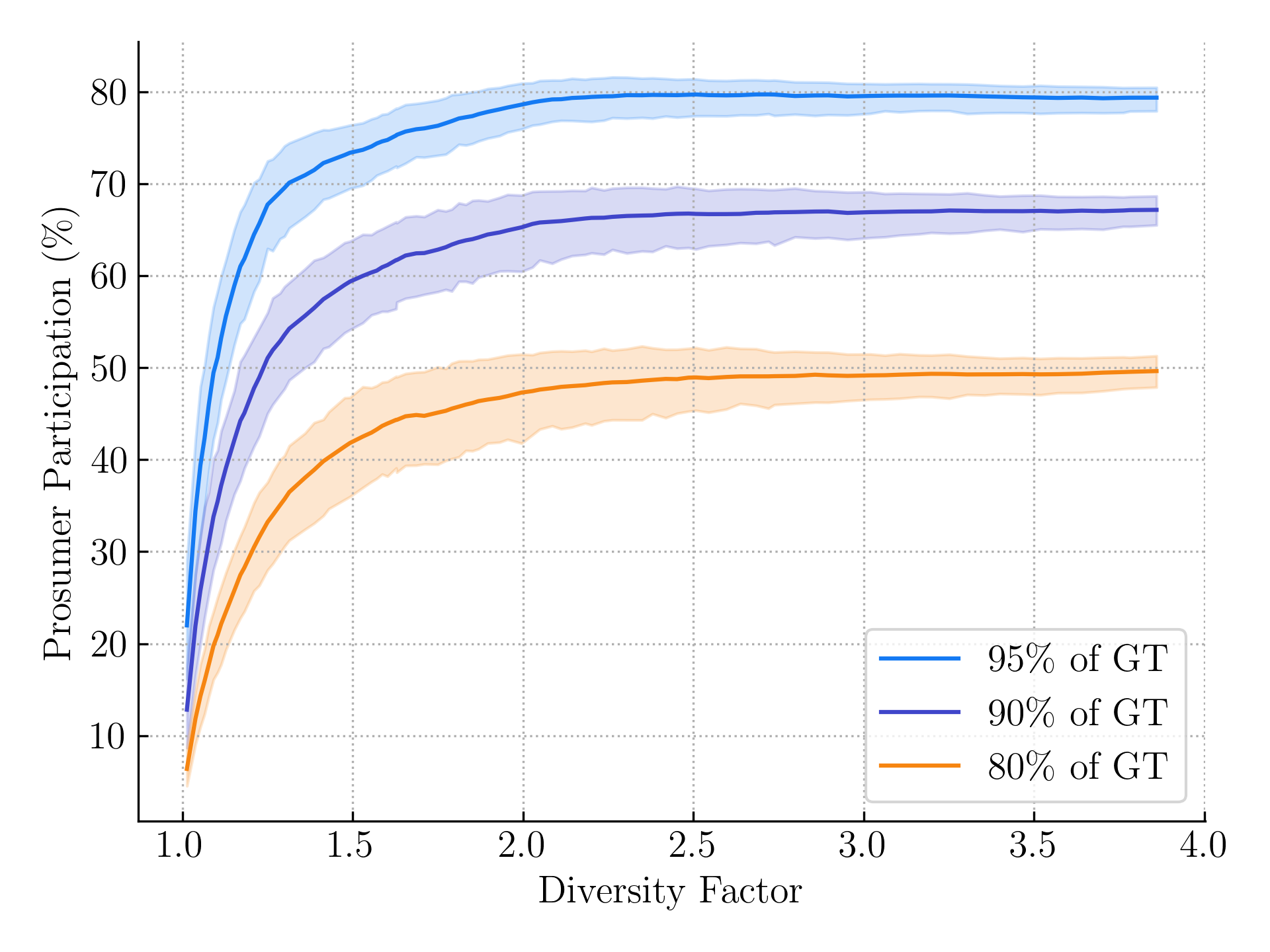}
         \caption{Centralised matching and clearing}
         \label{fig:experiments:london:diversity_factor:agents:cmc}
     \end{subfigure}
     \hfill
     \begin{subfigure}[b]{0.48\textwidth}
         \centering
         \includegraphics[width=\textwidth]{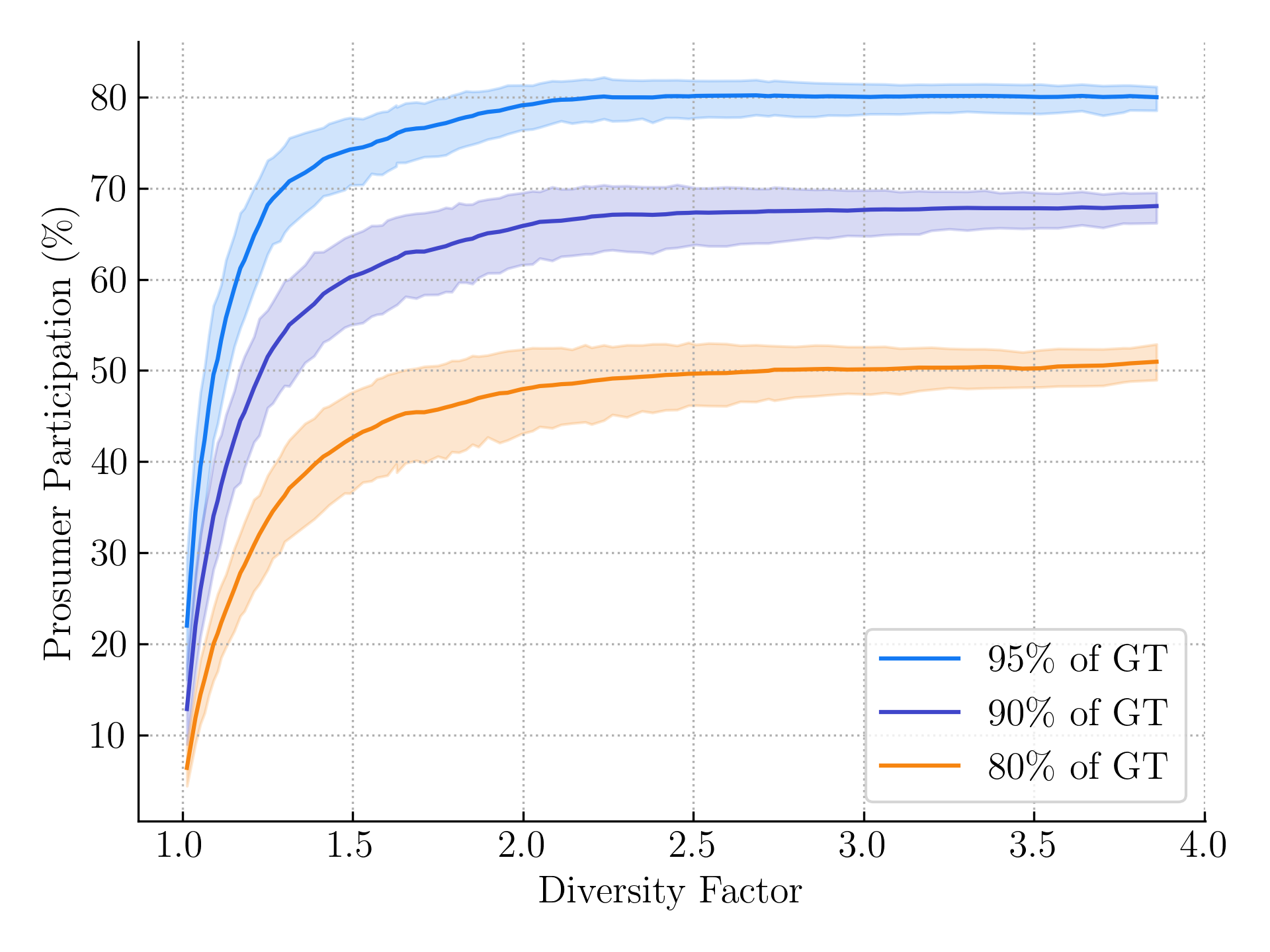}
         \caption{Top-5 decentralised negotiation}
         \label{fig:experiments:london:diversity_factor:agents:nego}
     \end{subfigure}
        \caption{Influence of diversity factor of the energy community on the prosumer participation required for achieving the majority of the gains from trade}
        \label{fig:experiments:london:diversity:gains_convergence}
\end{figure}

\Cref{fig:experiments:london:diversity:gains_convergence} shows the prosumer participation required for different fractions of the Gains from Trade. Interestingly, these thresholds grow quickly for non-diverse communities. The difference effect of increasing diversity for an already decently diverse community seems to have negligible effects.

\subsection{Discussion}
\instructions{
\begin{todolist}
\item Convergence to coalitional setting
\item Low number of trades
\item Diversity, (different profiles gain a lot from trading)
\end{todolist}
}

Results from both the Thames Valley Vision and Low Carbon London datasets suggest that only a small number of contracts need to be established to realise the bulk of the potential Gains from Trade from establishing a P2P or community energy trading scheme. Since generation profiles between prosumers are very similar due to the locality of the community, only requiring a small number of contracts can be attributed due to the majority of the benefits being achieved by prosumers that have significantly different demand profiles from the other agents. After these prosumers establish contracts, they likely still have some excess generation and residual demand, however, these values are closer to the mean and, therefore, less of a factor for future contracts. Similarly, we can see that for the majority of the Gains from Trade, not all prosumers need to participate, with those prosumers that exhibit common consumption behaviour having less marginal contribution. This points to the benefits of considering the exact demand profiles in promoting participation of prosumers in P2P trading schemes.

The convergence of the maximal Gains from Trade happens at a fast initial rate and then levels off, for both forms of market organisation (centralised vs. peer-to-peer models). However, trade dynamics do reveal that there is a difference in how both models achieve similar gains. Both of these models prioritise different contracts in the beginning, when most contracts are likely to be formed, with the peer-to-peer model resulting initially in a larger number of small coalitions.

These results are further confirmed by the scenarios generated using the large-scale Low Carbon London data set. Furthermore, the Low Carbon London data set provides some insights on how this effect generalises to other compositions of communities. Experiments show that while the Gains from Trade increase based on the diversity of the community, prosumer participation required for realising the majority of the benefit stays fairly consistent, with around 50\% of the prosumers being required for 80\% of the maximal potential GT. However, a lack of diversity within a community suggests that the majority of the gains can be achieved by a very small number of contracts. This makes sense, as only those that differ significantly from the majority benefit significantly from participating in peer-to-peer trading.

In addition to these quantitative results, it is important to discuss the replicability of these energy communities in real implementations. In \cite{socialstudiesenergycommunities}, the acceptability of such schemes of energy communities was assessed through a survey filled by 617 citizens from Finland, and showed that only 20\% of the respondents were not interested in participating in an energy community, whereas the incentive of citizens interested where mostly the potential economic gains through buying and selling electricity, but also the reduction of their environmental impact, and the reduction of their dependence to big energy companies. However, it was shown that energy communities should not require any change in end-users' consumption nor any technical challenge. This feedback is similar to the one that was collected within the ReFLEX (Responsive FLEXibility) project in the Orkney Islands, one of the largest Smart Grid demonstrator in the UK \cite{ReflexRSER}, that highlighted the fact that social acceptability is a critical driver for energy communities adoption, and that energy trades should be transparent for the end-users. This can be achieved by enabling a post-delivery local energy market clearing (centralised or decentralised) that would result in validated trades that will be deducted from the energy supplier bills of the prosumers and consumers. This enables a seamless integration of energy communities within the current energy system and only requires an energy community manager to operate the local energy market and send the resulting trades to the DSO or the energy suppliers, as is the case in several countries in Europe \cite{energycommunitieseurope}. However, further work on social acceptability and end-users' preferences for a centralised or fully decentralised approach should be conducted through surveys and focus groups. 

\section{Conclusions \& Future work}
\label{sec:conclusion}

Peer-to-peer trading has a lot of potential in local energy markets. However, there is no consensus on the best organization of a local energy market. Centralised market clearing models (such as double auction-type models) seem to be more efficient for extracting the total social benefit, while decentralised peer-to-peer models provide more flexibility and individual choice.
In this paper, we examine the efficiency of peer-to-peer markets vs. centralised markets, in terms of contracts established and market participation using large-scale data from several trials in the UK. 
We study this by using 2 models grounded in earlier research, each respectively representing centralised and decentralised markets. Furthermore, to ensure that the results are representative, we develop a new framework that uses a joint profile optimiser to allow for simulating large-scale peer-to-peer processes.

Results suggest that irrespective of the model used peer-to-peer markets can achieve maximum benefits using a very small number of contracts, i.e. less than 10\% of the possible P2P contracts are required. Furthermore, a majority of the benefits can be extracted even when only a fraction of the community participates in trade, e.g. only about 50\% needs to trade for about 80\% of the benefits. This effect is more noticeable given a less diverse community, requiring only a small fraction of the community to trade. However, less diverse communities also have less potential for trading, as the overall benefits increase when diversity of prosumer demand profiles increases.

Results in this work are based on two large-scale public datasets from the UK. However, future work may consider replicating these results for other regions around the world, as consumption habits and generation profiles are likely to differ by location. Furthermore, we include only prosumers that have access to a wind turbine, but the effect of solar panels, heat pumps and other forms of renewable generation could be considered in future work. Further investigation could be made for a more complex community, consisting of individuals that exclusively consume energy and prosumers that use different kinds of renewable energy sources such as photovoltaic.

As a result of our experiments, we find that the marginal Gains from Trade from established contracts is high in the beginning (for the most promising prosumers in the community), but then tapers of drastically as more prosumers are added. 
While costs of establishing contracts and P2P coalition are not explicitly modelled here (e.g. administrative, IT or convenience costs), it may be that the costs associated with establishing such a contract outweighing the benefits it can provide, for many prosumers. In future work, these costs could be explicitly modelled and used to derive an equilibrium point, where establishing more contracts would be counterproductive. Furthermore, this point may be used to determine a practical limit for a fraction of benefits that can be extracted from peer-to-peer energy trading.

Another promising extension - especially important for practical application - is extending the community energy model to consider physical network constraints, such as voltage and power limits (an aspect considered only in a limited number of prior works, such as~\cite{chapman_constraints1,chapman_constraints2,norbu_constraints}. Physical constraints may make specific energy contracts more or less valuable depending on the logistics and network used costs of transporting the energy, and thus influence the participation in the peer-to-peer trading scheme for prosumers at specific times.

Finally, another technology that is increasingly used for peer-peer and transactive energy systems are blockchain-enabled smart contracts~\cite{Kirli_etal_blockchain,Andoni_etal_blockchain,couraud_blockchain}. Implementing our energy exchange and coalition formation protocols on a blockchain (e.g. through a smart contract) could also be a relevant avenue for further work.

\vspace{-0.1cm}
\section*{Acknowledgements}
\vspace{-0.25cm}
Valentin Robu acknowledges the support of the project ``TESTBED2: Testing and Evaluating Sophisticated information and communication Technologies for enaBling scalablE smart griD Deployment'', funded by the European Union under the Horizon2020 Marie Skłodowska-Curie Actions (MSCA) [Grant agreement number: 872172]. Benoit Couraud, Merlinda Andoni and David Flynn acknowledge the support of the InnovateUK Responsive Flexibility (ReFLEX) project [ref: 104780]. Merlinda Andoni and David Flynn also acknowledge the support of the UK Engineering and Physical Science Research Council in the project DecarbonISation PAThways for Cooling and Heating (DISPATCH) [grant: EP/V042955/1]. Sonam Norbu acknowledges the support of the InnovateUK Knowledge Transfer Partnerships (KTP) Project based at The Crichton Trust in Dumfries [KTP-13052]. H. Vincent Poor acknowledges the support of the U.S. National Science Foundation under Grant ECCS-2039716 and a grant from the C3.ai Digital Transformation Institute.

\bibliographystyle{elsarticle-num}

\begin{thebibliography}{10}
\expandafter\ifx\csname url\endcsname\relax
  \def\url#1{\texttt{#1}}\fi
\expandafter\ifx\csname urlprefix\endcsname\relax\def\urlprefix{URL }\fi
\expandafter\ifx\csname href\endcsname\relax
  \def\href#1#2{#2} \def\path#1{#1}\fi

\bibitem{tushar2019motivational}
W.~Tushar, T.~K. Saha, C.~Yuen, T.~Morstyn, M.~D. McCulloch, H.~V. Poor, K.~L.
  Wood, A motivational game-theoretic approach for peer-to-peer energy trading
  in the smart grid, Applied Energy 243 (2019) 10--20.
\newblock \href {http://dx.doi.org/10.1016/j.apenergy.2019.03.111}
  {\path{doi:10.1016/j.apenergy.2019.03.111}}.

\bibitem{tushar_poor2023}
W.~Tushar, S.~Nizami, M.~I. Azim, C.~Yuen, D.~B. Smith, T.~Saha, H.~V. Poor,
  Peer-to-peer energy sharing: A comprehensive review, Foundations and Trends®
  in Electric Energy Systems 6~(1) (2023) 1--82.
\newblock \href {http://dx.doi.org/10.1561/3100000031}
  {\path{doi:10.1561/3100000031}}.

\bibitem{Nizami_APEN}
S.~Nizami, W.~Tushar, M.~Hossain, C.~Yuen, T.~Saha, H.~V. Poor, Transactive
  energy for low voltage residential networks: A review, Applied Energy 323
  (2022) 119556.
\newblock \href {http://dx.doi.org/10.1016/j.apenergy.2022.119556}
  {\path{doi:10.1016/j.apenergy.2022.119556}}.

\bibitem{capper2021systematic}
T.~Capper, A.~Gorbatcheva, M.~A. Mustafa, M.~Bahloul, J.~M. Schwidtal,
  R.~Chitchyan, M.~Andoni, V.~Robu, M.~Montakhabi, I.~J. Scott, C.~Francis,
  T.~Mbavarira, J.~M. Espana, L.~Kiesling, Peer-to-peer, community
  self-consumption, and transactive energy: A systematic literature review of
  local energy market models, Renewable and Sustainable Energy Reviews 162
  (2022) 112403.
\newblock \href {http://dx.doi.org/10.1016/j.rser.2022.112403}
  {\path{doi:10.1016/j.rser.2022.112403}}.

\bibitem{Schwidtal_review}
J.~Schwidtal, P.~Piccini, M.~Troncia, R.~Chitchyan, M.~Montakhabi, C.~Francis,
  A.~Gorbatcheva, T.~Capper, M.~Mustafa, M.~Andoni, V.~Robu, M.~Bahloul,
  I.~Scott, T.~Mbavarira, J.~España, L.~Kiesling, Emerging business models in
  local energy markets: A systematic review of peer-to-peer, community
  self-consumption, and transactive energy models, Renewable and Sustainable
  Energy Reviews 179 (2023) 113273.
\newblock \href {http://dx.doi.org/10.1016/j.rser.2023.113273}
  {\path{doi:10.1016/j.rser.2023.113273}}.

\bibitem{lucas2016distribution}
A.~Lucas, G.~Prettico, A.~Mengolini, G.~Fulli, F.~Gangale, Distribution system
  operators observatory : from European electricity distribution systems to
  reference network, Publications Office, 2016.
\newblock \href {http://dx.doi.org/10.2790/471701} {\path{doi:10.2790/471701}}.

\bibitem{sousa2019peer}
T.~Sousa, T.~Soares, P.~Pinson, F.~Moret, T.~Baroche, E.~Sorin, Peer-to-peer
  and community-based markets: A comprehensive review, Renewable and
  Sustainable Energy Reviews 104 (2019) 367--378.
\newblock \href {http://dx.doi.org/10.1016/j.rser.2019.01.036}
  {\path{doi:10.1016/j.rser.2019.01.036}}.

\bibitem{lee2014direct}
W.~Lee, L.~Xiang, R.~Schober, V.~W.~S. Wong, Direct electricity trading in
  smart grid: A coalitional game analysis, IEEE Journal on Selected Areas in
  Communications 32~(7) (2014) 1398--1411.
\newblock \href {http://dx.doi.org/10.1109/JSAC.2014.2332112}
  {\path{doi:10.1109/JSAC.2014.2332112}}.

\bibitem{sioshansi2013evolution}
F.~P. Sioshansi, Evolution of Global Electricity Markets: New paradigms, new
  challenges, new approaches, Academic Press, 2013.
\newblock \href {http://dx.doi.org/10.1016/C2012-0-00444-9}
  {\path{doi:10.1016/C2012-0-00444-9}}.

\bibitem{wang2020distributed}
Z.~Wang, X.~Yu, Y.~Mu, H.~Jia, A distributed peer-to-peer energy transaction
  method for diversified prosumers in urban community microgrid system, Applied
  Energy 260 (2020) 114327.
\newblock \href {http://dx.doi.org/10.1016/j.apenergy.2019.114327}
  {\path{doi:10.1016/j.apenergy.2019.114327}}.

\bibitem{pinto2019decision}
T.~Pinto, R.~Faia, M.~A.~F. Ghazvini, J.~Soares, J.~M. Corchado, Z.~Vale,
  Decision support for small players negotiations under a transactive energy
  framework, IEEE Transactions on Power Systems 34~(5) (2019) 4015--4023.
\newblock \href {http://dx.doi.org/10.1109/TPWRS.2018.2861325}
  {\path{doi:10.1109/TPWRS.2018.2861325}}.

\bibitem{saxena2019agent}
K.~Saxena, A.~R. Abhyankar, Agent based bilateral transactive market for
  emerging distribution system considering imbalances, Sustainable Energy,
  Grids and Networks 18 (2019) 100203.
\newblock \href {http://dx.doi.org/10.1016/j.segan.2019.100203}
  {\path{doi:10.1016/j.segan.2019.100203}}.

\bibitem{guo2021asynchronous}
Z.~Guo, P.~Pinson, Q.~Wu, S.~Chen, Q.~Yang, Z.~Yang, An asynchronous online
  negotiation mechanism for real-time peer-to-peer electricity markets, IEEE
  Transactions on Power Systems 37~(3) (2022) 1868--1880.
\newblock \href {http://dx.doi.org/10.1109/TPWRS.2021.3111869}
  {\path{doi:10.1109/TPWRS.2021.3111869}}.

\bibitem{imran2020bilateral}
K.~Imran, J.~Zhang, A.~Pal, A.~Khattak, K.~Ullah, S.~M. Baig, Bilateral
  negotiations for electricity market by adaptive agent-tracking strategy,
  Electric Power Systems Research 186 (2020) 106390.
\newblock \href {http://dx.doi.org/10.1016/j.epsr.2020.106390}
  {\path{doi:10.1016/j.epsr.2020.106390}}.

\bibitem{etukudor2020automated}
C.~Etukudor, B.~Couraud, V.~Robu, W.-G. Fr{\"u}h, D.~Flynn, C.~Okereke,
  Automated negotiation for peer-to-peer electricity trading in local energy
  markets, Energies 13~(4) (2020) 920.
\newblock \href {http://dx.doi.org/10.3390/en13040920}
  {\path{doi:10.3390/en13040920}}.

\bibitem{chakraborty2020automated}
S.~Chakraborty, T.~Baarslag, M.~Kaisers, Automated peer-to-peer negotiation for
  energy contract settlements in residential cooperatives, Applied Energy 259
  (2020) 114173.
\newblock \href {http://dx.doi.org/10.1016/j.apenergy.2019.114173}
  {\path{doi:10.1016/j.apenergy.2019.114173}}.

\bibitem{dang2011wholesale}
J.~Dang, Y.~Lu, P.~Zhao, M.~Jafari, Wholesale power trading through concurrent
  multiple-issue negotiation, Transactions of the Institute of Measurement and
  Control 33~(3-4) (2011) 386--405.
\newblock \href {http://dx.doi.org/10.1177/0142331208100101}
  {\path{doi:10.1177/0142331208100101}}.

\bibitem{kalbantner2021p2pedge}
J.~Kalbantner, K.~Markantonakis, D.~Hurley-Smith, R.~N. Akram, B.~Semal,
  {P2PEdge}: A decentralised, scalable {P2P} architecture for energy trading in
  real-time, Energies 14~(3) (2021) 606.
\newblock \href {http://dx.doi.org/10.3390/en14030606}
  {\path{doi:10.3390/en14030606}}.

\bibitem{khorasany2020new}
M.~Khorasany, A.~Paudel, R.~Razzaghi, P.~Siano, A new method for peer matching
  and negotiation of prosumers in peer-to-peer energy markets, IEEE
  Transactions on Smart Grid 12~(3) (2021) 2472--2483.
\newblock \href {http://dx.doi.org/10.1109/TSG.2020.3048397}
  {\path{doi:10.1109/TSG.2020.3048397}}.

\bibitem{su2020optimization}
H.~Huang, S.~Nie, J.~Lin, Y.~Wang, J.~Dong, Optimization of peer-to-peer power
  trading in a microgrid with distributed pv and battery energy storage
  systems, Sustainability 12~(3) (2020) 923.
\newblock \href {http://dx.doi.org/10.3390/su12030923}
  {\path{doi:10.3390/su12030923}}.

\bibitem{long2019game}
C.~Long, Y.~Zhou, J.~Wu, A game theoretic approach for peer to peer energy
  trading, Energy Procedia 159 (2019) 454--459, renewable Energy Integration
  with Mini/Microgrid.
\newblock \href {http://dx.doi.org/10.1016/j.egypro.2018.12.075}
  {\path{doi:10.1016/j.egypro.2018.12.075}}.

\bibitem{norbu2021modelling}
S.~Norbu, B.~Couraud, V.~Robu, M.~Andoni, D.~Flynn, Modelling the
  redistribution of benefits from joint investments in community energy
  projects, Applied Energy 287 (2021) 116575.
\newblock \href {http://dx.doi.org/10.1016/j.apenergy.2021.116575}
  {\path{doi:10.1016/j.apenergy.2021.116575}}.

\bibitem{VPP_energy}
V.~Robu, G.~Chalkiadakis, R.~Kota, A.~Rogers, N.~R. Jennings, Rewarding
  cooperative virtual power plant formation using scoring rules, Energy
  (Elsevier) 117 (2016) 19--28.
\newblock \href {http://dx.doi.org/10.1016/j.energy.2016.10.077}
  {\path{doi:10.1016/j.energy.2016.10.077}}.

\bibitem{Kota_ECAI}
R.~Kota, G.~Chalkiadakis, V.~Robu, A.~Rogers, N.~R. Jennings, {Cooperatives for
  demand side management}, Proceedings of the 20th European Conference on
  Artificial Intelligence (2012) 969--974\href
  {http://dx.doi.org/10.3233/978-1-61499-098-7-969}
  {\path{doi:10.3233/978-1-61499-098-7-969}}.

\bibitem{Cremers_Shapley_APEN}
S.~Cremers, V.~Robu, P.~Zhang, M.~Andoni, S.~Norbu, D.~Flynn, Efficient methods
  for approximating the shapley value for asset sharing in energy communities,
  Applied Energy 331 (2023) 120328.
\newblock \href {http://dx.doi.org/10.1016/j.apenergy.2022.120328}
  {\path{doi:10.1016/j.apenergy.2022.120328}}.

\bibitem{groupbuying_TSG}
V.~Robu, M.~Vinyals, A.~Rogers, N.~R. Jennings, Efficient buyer groups with
  prediction-of-use electricity tariffs, IEEE Transactions on Smart Grid 9~(5)
  (2018) 4468--4479.
\newblock \href {http://dx.doi.org/10.1109/TSG.2017.2660580}
  {\path{doi:10.1109/TSG.2017.2660580}}.

\bibitem{pumphrey_perception}
K.~Pumphrey, S.~L. Walker, M.~Andoni, V.~Robu, Green hope or red herring?
  examining consumer perceptions of peer-to-peer energy trading in the united
  kingdom, Energy Research and Social Science 68 (2020) 101603.
\newblock \href {http://dx.doi.org/10.1016/j.erss.2020.101603}
  {\path{doi:10.1016/j.erss.2020.101603}}.

\bibitem{moret2018negotiation}
F.~Moret, T.~Baroche, E.~Sorin, P.~Pinson, Negotiation algorithms for
  peer-to-peer electricity markets: Computational properties, in: 2018 Power
  Systems Computation Conference (PSCC), 2018, pp. 1--7.
\newblock \href {http://dx.doi.org/10.23919/PSCC.2018.8442914}
  {\path{doi:10.23919/PSCC.2018.8442914}}.

\bibitem{couraud_ISGT}
B.~Couraud, S.~Norbu, M.~Andoni, V.~Robu, H.~Gharavi, D.~Flynn, Optimal
  residential battery scheduling with asset lifespan consideration, in: 2020
  IEEE PES Innovative Smart Grid Technologies Europe (ISGT-Europe), 2020, pp.
  630--634.
\newblock \href {http://dx.doi.org/10.1109/ISGT-Europe47291.2020.9248889}
  {\path{doi:10.1109/ISGT-Europe47291.2020.9248889}}.

\bibitem{ofgem2020fit}
{Ofgem}, Fit scheme closure,
  \url{https://www.ofgem.gov.uk/publications/faq-fit-scheme-closure}, (Accessed
  on 26/10/2023) (2020).

\bibitem{ke2015control}
X.~Ke, N.~Lu, C.~Jin, Control and size energy storage systems for managing
  energy imbalance of variable generation resources, IEEE Transactions on
  Sustainable Energy 6~(1) (2015) 70--78.
\newblock \href {http://dx.doi.org/10.1109/TSTE.2014.2355829}
  {\path{doi:10.1109/TSTE.2014.2355829}}.

\bibitem{ukerc2017new}
{UKERC Energy Data Centre}, New thames valley vision - end point monitors,
  \url{https://ukerc.rl.ac.uk/DC/cgi-bin/edc_search.pl?WantComp=147}, (Accessed
  on 26/10/2023) (2017).

\bibitem{xu2018modeling}
B.~Xu, A.~Oudalov, A.~Ulbig, G.~Andersson, D.~S. Kirschen, Modeling of
  lithium-ion battery degradation for cell life assessment, IEEE Transactions
  on Smart Grid 9~(2) (2018) 1131--1140.
\newblock \href {http://dx.doi.org/10.1109/TSG.2016.2578950}
  {\path{doi:10.1109/TSG.2016.2578950}}.

\bibitem{IEEE_bus_system}
{European Low Voltage Test Feeder}, {Reseources | IEEE PES Test Feeders},
  \url{https://site.ieee.org/pes-testfeeders/resources}, (Accessed on
  26/10/2023) (2023).

\bibitem{ukpower2014smart}
{UK Power Networks}, Smartmeter energy consumption data in london households,
  \url{https://data.london.gov.uk/dataset/smartmeter-energy-use-data-in-london-households},
  (Accessed on 26/10/2023) (2014).

\bibitem{electronics10030290}
H.~C. Jeong, M.~Jang, T.~Kim, S.-K. Joo, Clustering of load profiles of
  residential customers using extreme points and demographic characteristics,
  Electronics 10~(3) (2021) 290.
\newblock \href {http://dx.doi.org/10.3390/electronics10030290}
  {\path{doi:10.3390/electronics10030290}}.

\bibitem{kwac2014household}
J.~Kwac, J.~Flora, R.~Rajagopal, Household energy consumption segmentation
  using hourly data, IEEE Transactions on Smart Grid 5~(1) (2014) 420--430.
\newblock \href {http://dx.doi.org/10.1109/TSG.2013.2278477}
  {\path{doi:10.1109/TSG.2013.2278477}}.

\bibitem{socialstudiesenergycommunities}
S.~Tuomela, T.~H{\"a}nninen, E.~Ruokamo, N.~Iivari, M.~Kopsakangas-Savolainen,
  R.~Svento, Energy Community Preferences of Solar Prosumers and Electricity
  Consumers in the Digital Energy Ecosystem, Springer International Publishing,
  2023, pp. 113--135.
\newblock \href {http://dx.doi.org/10.1007/978-3-031-21402-8_4}
  {\path{doi:10.1007/978-3-031-21402-8_4}}.

\bibitem{ReflexRSER}
B.~Couraud, M.~Andoni, V.~Robu, S.~Norbu, S.~Chen, D.~Flynn, Responsive
  {FLEXibility}: A smart local energy system, Renewable and Sustainable Energy
  Reviews 182 (2023) 113343.
\newblock \href {http://dx.doi.org/10.1016/j.rser.2023.113343}
  {\path{doi:10.1016/j.rser.2023.113343}}.

\bibitem{energycommunitieseurope}
D.~Frieden, A.~Tuerk, A.~R. Antunes, V.~Athanasios, A.-G. Chronis,
  S.~d’Herbemont, M.~Kirac, R.~Marouço, C.~Neumann, E.~Pastor~Catalayud,
  N.~Primo, A.~F. Gubina, Are we on the right track? collective
  self-consumption and energy communities in the european union, Sustainability
  13~(22) (2021) 12494.
\newblock \href {http://dx.doi.org/10.3390/su132212494}
  {\path{doi:10.3390/su132212494}}.

\bibitem{chapman_constraints1}
J.~Guerrero, A.~C. Chapman, G.~Verbič, Decentralized {P2P} energy trading
  under network constraints in a low-voltage network, IEEE Transactions on
  Smart Grid 10~(5) (2019) 5163--5173.
\newblock \href {http://dx.doi.org/10.1109/TSG.2018.2878445}
  {\path{doi:10.1109/TSG.2018.2878445}}.

\bibitem{chapman_constraints2}
D.~Gebbran, S.~Mhanna, Y.~Ma, A.~C. Chapman, G.~Verbič, Fair coordination of
  distributed energy resources with volt-var control and {PV} curtailment,
  Applied Energy 286 (2021) 116546.
\newblock \href {http://dx.doi.org/10.1016/j.apenergy.2021.116546}
  {\path{doi:10.1016/j.apenergy.2021.116546}}.

\bibitem{norbu_constraints}
S.~Norbu, B.~Couraud, V.~Robu, M.~Andoni, D.~Flynn, Modeling economic sharing
  of joint assets in community energy projects under {LV} network constraints,
  IEEE Access 9 (2021) 112019--112042.
\newblock \href {http://dx.doi.org/10.1109/ACCESS.2021.3103480}
  {\path{doi:10.1109/ACCESS.2021.3103480}}.

\bibitem{Kirli_etal_blockchain}
D.~Kirli, B.~Couraud, V.~Robu, M.~Salgado-Bravo, S.~Norbu, M.~Andoni,
  I.~Antonopoulos, M.~Negrete-Pincetic, D.~Flynn, A.~Kiprakis, Smart contracts
  in energy systems: A systematic review of fundamental approaches and
  implementations, Renewable and Sustainable Energy Reviews 158 (2022) 112013.
\newblock \href {http://dx.doi.org/10.1016/j.rser.2021.112013}
  {\path{doi:10.1016/j.rser.2021.112013}}.

\bibitem{Andoni_etal_blockchain}
M.~Andoni, V.~Robu, D.~Flynn, S.~Abram, D.~Geach, D.~Jenkins, P.~McCallum,
  A.~Peacock, Blockchain technology in the energy sector: A systematic review
  of challenges and opportunities, Renewable and Sustainable Energy Reviews 100
  (2019) 143--174.
\newblock \href {http://dx.doi.org/10.1016/j.rser.2018.10.014}
  {\path{doi:10.1016/j.rser.2018.10.014}}.

\bibitem{couraud_blockchain}
B.~Couraud, V.~Robu, D.~Flynn, M.~Andoni, S.~Norbu, H.~Quinard, Real-time
  control of distributed batteries with blockchain-enabled market export
  commitments, IEEE Transactions on Sustainable Energy 13~(1) (2022) 579--591.
\newblock \href {http://dx.doi.org/10.1109/TSTE.2021.3121444}
  {\path{doi:10.1109/TSTE.2021.3121444}}.

\bibitem{stroehle_all}
P.~Str\"{o}hle, E.~H. Gerding, M.~M. de~Weerdt, S.~Stein, V.~Robu,
  \href{https://dl.acm.org/doi/10.5555/2615731.2615803}{Online mechanism design
  for scheduling non-preemptive jobs under uncertain supply and demand}, in:
  Proceedings of the 2014 International Conference on Autonomous Agents and
  Multi-Agent Systems, AAMAS '14, International Foundation for Autonomous
  Agents and Multiagent Systems, 2014, pp. 437--444.
\newline\urlprefix\url{https://dl.acm.org/doi/10.5555/2615731.2615803}

\bibitem{roman_battery_SOH}
D.~Roman, S.~Saxena, V.~Robu, M.~Pecht, D.~Flynn, Machine learning pipeline for
  battery state-of-health estimation, Nature Machine Intelligence 3~(5) (2021)
  447--456.
\newblock \href {http://dx.doi.org/10.1038/s42256-021-00312-3}
  {\path{doi:10.1038/s42256-021-00312-3}}.

\bibitem{tang_al_ISCAS}
W.~Tang, M.~Andoni, V.~Robu, D.~Flynn, Accurately forecasting the health of
  energy system assets, in: 2018 IEEE International Symposium on Circuits and
  Systems (ISCAS), 2018, pp. 1--5.
\newblock \href {http://dx.doi.org/10.1109/ISCAS.2018.8351842}
  {\path{doi:10.1109/ISCAS.2018.8351842}}.

\bibitem{Bellman_2010}
R.~E. Bellman, Dynamic programming, Princeton University Press, Princeton, NJ,
  2010.
\newblock \href {http://dx.doi.org/10.1515/9781400835386}
  {\path{doi:10.1515/9781400835386}}.

\bibitem{Li_clustering}
K.~Li, Z.~Ma, D.~Robinson, J.~Ma, Identification of typical building daily
  electricity usage profiles using gaussian mixture model-based clustering and
  hierarchical clustering, Applied Energy 231 (2018) 331--342.
\newblock \href {http://dx.doi.org/10.1016/j.apenergy.2018.09.050}
  {\path{doi:10.1016/j.apenergy.2018.09.050}}.

\bibitem{GUO_datamining}
Z.~Guo, K.~Zhou, X.~Zhang, S.~Yang, Z.~Shao, Data mining based framework for
  exploring household electricity consumption patterns: A case study in china
  context, Journal of Cleaner Production 195 (2018) 773--785.
\newblock \href {http://dx.doi.org/10.1016/j.jclepro.2018.05.254}
  {\path{doi:10.1016/j.jclepro.2018.05.254}}.

\bibitem{Antonopoulos_EnergyAI}
I.~Antonopoulos, V.~Robu, B.~Couraud, D.~Flynn, Data-driven modelling of energy
  demand response behaviour based on a large-scale residential trial, Energy
  and AI 4 (2021) 100071.
\newblock \href {http://dx.doi.org/10.1016/j.egyai.2021.100071}
  {\path{doi:10.1016/j.egyai.2021.100071}}.

\bibitem{Yuan_KSelection}
C.~Yuan, H.~Yang, Research on k-value selection method of k-means clustering
  algorithm, J 2~(2) (2019) 226--235.
\newblock \href {http://dx.doi.org/10.3390/j2020016}
  {\path{doi:10.3390/j2020016}}.

\end{thebibliography}

\appendix
\section{Battery Usage} \label{app_sec:battery}

\section*{Additional symbols}
\noindent

{\(SoC_{\%}(t)\)}{\quad \quad  SoC level of the battery at time $t$ [\%]}
\\

{\(SoC_{\%}^{max}\)}{\quad \quad  Maximum SoC of the battery [\%]}
\\

{\(SoC_{\%}^{min}\)}{\quad \quad  Minimum SoC of the battery [\%]}
\\

{\(p^{\text{bat,max}}\)}{\quad \quad  Maximum (dis)charging power of the battery [\%kW]}
\\

{\(\Delta t\)}{\quad \quad  Duration of time period $t$ [hour]}
\\

{\(\eta^c\)} {\quad \quad  Charging efficiency of battery}
\\

{\(\eta^d\)}{\quad \quad  Discharging efficiency of battery}
\\

{\(\text{DoD}\)}{\quad \quad  Depth of discharge of battery [\%]}
\\

{\(\text{DF}^{\text{regular}}\)}{\quad \quad  Depreciation factor by regular cycles}
\\

{\(\text{DF}^{\text{irregular}}\)}{\quad \quad  Depreciation factor by irregular cycles}
\\

{\(L\)}{\quad \quad  Set of irregular cycles}
\\

{\(l\)}{\quad \quad  for irregular cycles}
\\

{\(SoC^{l,Start/End}_{\%}\)}{\quad \quad  Starting/ending SoC of cycle $l$ [\%]}
\\

{\(N^{\text{DoD,max}}_{cycles}\)}{\quad \quad  Maximum number of cycles allowed at specific DoD, provided in manufacturer specification}
\\

{\(n^{\text{DoD,regular}}_{cycles}\)}{\quad \quad  Number of regular cycles with the specific DoD}
\\

\section*{Battery Control Algorithm} \label{app_subsec:battery_control}

In this appendix, we provide the details of the heuristic-based battery control algorithm. Overall, our battery control algorithm follows the method first described in~\cite{norbu2021modelling} (and also used in~\cite{couraud_ISGT,Cremers_Shapley_APEN}, and interested readers can also consult that paper for the full details.

The charging of the battery is constrained by the maximum state of charge (SoC) level of the battery, $SoC_{\%}^{max}$, and the maximum (dis)charging power of the battery, $p^{\text{bat,max}}$. Note that the SoC level in this section indicates the battery's level of charge relative to its capacity in percentage, whereas in the main paper SoC level refers to how much of the battery is charged in kW. Therefore, the SoC level in percentage is denoted as $SoC_{\%}$ in the remaining of section for clarity. The constraints of the battery for any time $t$ while charging are defined as follows.

\begin{equation}
SoC_{\%}(t) \leq SoC_{\%}^{\text{max}}
\label{eq:SoCmax}
\end{equation}
\begin{equation}
\left | p^{\text{bat}}(t) \right | \leq p^{\text{bat, max}}
\label{eq:pbatconstraint1}
\end{equation}

Similar constraints apply while discharging the battery. The battery may only be discharged while the SoC level has not reached the minimum SoC level, $SoC_{\%}^{min}$. Furthermore, the discharging power may not exceed the maximum discharging power. These constraints are expressed as the following.

\begin{equation}
SoC_{\%}(t) \geq SoC_{\%}^{\text{min}}
\label{eq:SoCmin}
\end{equation}
\begin{equation}
\left | p^{\text{bat}}(t) \right | \leq p^{\text{bat, max}}
\label{eq:pbatconstraint2}
\end{equation}

Given the battery constraints above, the heuristic algorithm operates the battery according to certain rules. When the locally generated power exceeds the power demand, this excess power charges the battery with the charging efficiency $\eta^c$, while the SoC level has not reached its maximum capacity. When the excess power cannot be used to charge the battery due to the capacity limit or the charging power of the battery, the remaining power is sold to the grid (note that his algorithm does not consider uncertainty in future demand and supply, unlike e.g.~\cite{stroehle_all}). 
The updated power and SoC level of the battery and exported energy to the grid $e^s(t)$ at time $t$ with excess power are calculated as the following.
\begin{equation}
\begin{aligned}
p^{\text{bat}}(t) = -\min(\min((g(t) - d(t)), p^{\text{bat, max}}), \frac{SoC_{\%}^{\text{max}} - SoC_{\%}(t-1)}{\eta^c \Delta t})
\end{aligned}
\label{eq:pbatcharge}
\end{equation}
\begin{equation}
SoC_{\%}(t) = SoC_{\%}(t-1) - \eta^c p^{\text{bat}}(t) \Delta t
\label{eq:SoCcharge}
\end{equation}
\begin{equation}
e^s(t) = (g(t) - d(t) + p^{\text{bat}}(t)) \Delta t
\label{eq:exportEnergy}
\end{equation}
Where $\Delta t$ is the length of one time step in hours. In this study, $1/2$ hour time steps were used. The exported energy is sold to the grid with export tariff $\tau^{s}(t)$, generating a profit for the prosumer(s).

When the generated power falls short of the demand, on the hand, the discharged power from the battery with discharging efficiency $\eta^d$ is first used to cover the deficit. If the discharged battery is not enough to meet the demand, additional power $e^b(t)$ is imported from the grid with the import tariff $\tau^{b}(t)$. During the power deficit, the battery state and imported energy are determined as follows.

\begin{equation}
\begin{aligned}
p^{\text{bat}}(t) = \min(\min((d(t) - g(t)), p^{\text{bat, max}}), \frac{\eta^d}{\Delta t} ( SoC_{\%}(t-1) - SoC_{\%}^{\text{min}}))
\end{aligned}
\label{eq:pbatdischarge}
\end{equation}
\begin{equation}
SoC_{\%}(t) = SoC_{\%}(t-1) - \frac{p^{\text{bat}}(t)}{\eta^d} \Delta t
\label{eq:SoCdischarge}
\end{equation}
\begin{equation}
e^b(t) = (d(t) - g(t) - p^{\text{bat}}(t)) \Delta t
\label{eq:importEnergy}
\end{equation}

\section*{Battery Degradation} \label{app_subsec:battery_degradation}

In this section, we describe the computation of the battery's depreciation factor (DF) in \cref{eq:batteryCost}, as adapted from the work of Norbu et al.~\cite{norbu2021modelling}. Frequent charging and discharging of the battery, especially when the depth of discharge is deep, can accelerate degradation significantly. Manufacturers of batteries often specify a battery cycle life on a provided datasheet. In this study, we use the lithium-ion battery cycle life data specified by Xu et al.~\cite{xu2018modeling}, Roman et al.~\cite{roman_battery_SOH} and Tang~\cite{tang_al_ISCAS}. This cycle life specifies the expected number of charge/discharge cycles per depth of discharge (DoD) that can undergo before the performance drops below operable levels. A shortened lifetime of the battery can have an impact on the depreciation cost, hence the inclusion of the depreciation factor provides a more realistic estimate of the community cost.

The total depreciation factor of the battery can be found by computing the depreciation factors of regular and irregular cycles. A regular cycle starts its discharging phase from an SoC level of 100\%, and charged back to 100\% SoC level. The depreciation factor of regular cycles is then computed as follows:
\begin{equation}
    \text{DF}^{\text{regular}} = \sum_{DoD=0\%}^{100\%} \frac{n^{\text{DoD,regular}}_{cycles}}{N^{\text{DoD,max}}_{cycles}}
\end{equation}
where $n^{\text{DoD,regular}}_{cycles}$ is the number of regular cycles (starting from 100\% SoC) with the DoD value, and $N^{\text{DoD,max}}_{cycles}$ is the lifetime of the battery provided by the battery manufacturer in terms of the number of regular cycles the battery can go through with the DoD value.

However, not all cycles can start from 100\% SoC level. Cycles that have a starting point other than 100\% SoC are called irregular cycles. Regular and irregular cycles can have the same DoD, e.g., a regular cycle that starts discharging at 100\% SoC level until it reaches 60\% then charged back to 100\% and an irregular cycle with 80\% starting SoC discharging until 40\% SoC level and charged back to 80\% SoC, both cycles have the same DoD of 40\%. Yet, regular and irregular cycles have different impacts on the depreciation factor, and hence they are computed separately. Furthermore, an irregular cycle can either be a full or half cycle. A full cycle consists of both discharging and charging phases, whereas a half cycle is only one of the charging/discharging phases. The rain-flow cycle counting algorithm \cite{ke2015control, norbu2021modelling} is used to classify and count the cycles as regular or irregular, as well as full or half. 

For all the irregular cycles $L$ during the evaluation period, we compute the deprecation factor as follows:
\begin{equation}
    \text{DF}^{\text{irregular}} = \sum_{l \in L} \text{n}_l \times \left| \frac{1}{N^{\text{DoD}^{eq}(SoC^{l,Start}_{\%}),max}_{cycles}} - \frac{1}{N^{\text{DoD}^{eq}(SoC^{l,End}_{\%}),max}_{cycles}}\right|
\end{equation}
where $SoC^{l,Start}_{\%}$ and $SoC^{l,End}_{\%}$ are the starting and ending state of charges, respectively, for the irregular cycle $l$, and $\text{n}_l$ represents the type of the cycle ($1$ for full cycle, and $\frac{1}{2}$ for half cycle). Then, $N^{\text{DoD}^{eq}(SoC^{l,Start}_{\%}),max}_{cycles}$ corresponds to the maximum number of cycles the battery can perform for $\text{DoD}^{eq}(SoC^{l,Start}_{\%})$, i.e. a depth of discharge equivalent to a cycle starting at 100\% $SoC_{\%}$ and ending at $SoC^{l,Start}_{\%}$. This is computed as the following:
\begin{equation}
    \text{DoD}^{eq}(SoC^{l,Start}_{\%}) = 100 - \left(\frac{SoC^{l,Start}_{\%}}{SoC^{\text{max}}_{\%}} \times 100 \right)
\end{equation}
We compute $N^{\text{DoD}^{eq}(SoC^{l,End}_{\%}),max}_{cycles}$ using a similar notion.

Finally, Given the depreciation factors of regular and irregular cycles, the total deprecation factor of the battery is the sum of the two components: 
\begin{equation}
    \text{DF} = \text{DF}^{\text{regular}} + \text{DF}^{\text{irregular}}
\end{equation}

\section{Clustering \& Consumer Profiles} \label{app:clustering}
\label{appendix:clustering}

In this section, the method and steps used for clustering consumer profiles used in \cref{sec:experiments} are detailed. The data to be clustered originally consists of 48 half hours for 365 days, creating a demand profile with $365 \cdot 48 = 17520$ points and is generated from the Low Carbon London trial~\cite{ukpower2014smart}. However, due to the high dimensionality of the data, clustering directly on the profiles suffers from the curse of dimensionality \cite{Bellman_2010}. Instead, we reduce the number of points to consider by reducing the demand profile to a recognizable consumption pattern. Categorising consumption patterns is often done using daily average data from the winter months by utility companies and other studies \cite{Li_clustering,GUO_datamining,Antonopoulos_EnergyAI}. We apply a similar approach by considering the demands on the prosumers from December to March. Furthermore, demand data from the weekends (Friday to Sunday) and holidays are filtered out, as prosumers may display different consumption patterns. Then, the remaining data from December to March is averaged over the days. Each prosumer is then represented by a 48-dimensional vector representing the average 48 half hours of their typical daily energy demand. As the interest mainly lies in grouping the daily consumption pattern of the prosumers, the demand vectors are normalised using the L1-norm. These vectors are then clustered using K-means clustering. To determine the suitable number of clusters, the elbow method and the silhouette method \cite{Yuan_KSelection} were used.

\begin{figure}[H]
    \centering
     \includegraphics[width=0.6\textwidth]{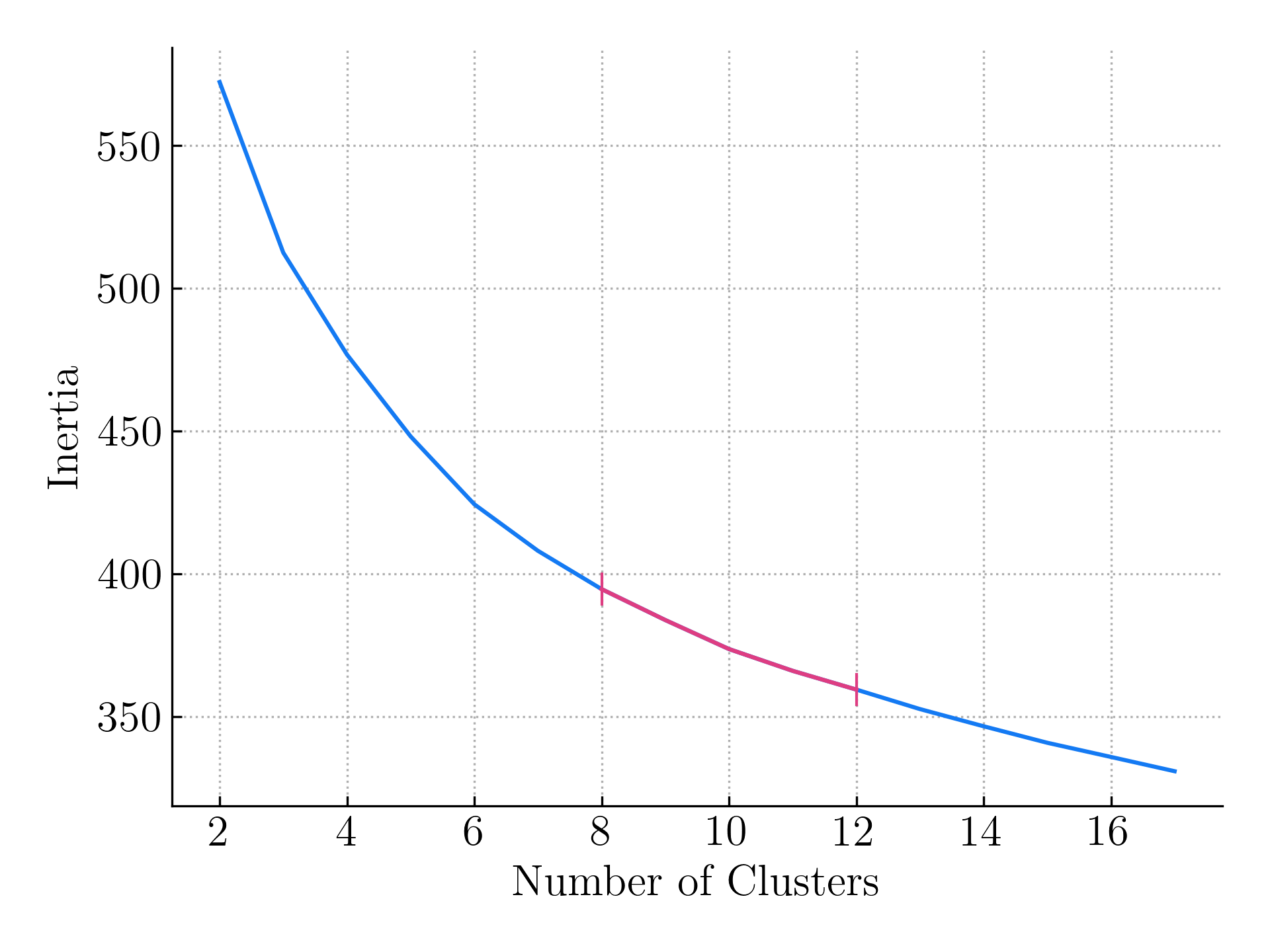}
    \caption{Elbow plot of K-means clustering on daily average demand profiles.}
    \label{fig:clustering:elbow}
\end{figure}

First, the elbow method identifies a range of reasonable cluster numbers. \Cref{fig:clustering:elbow} shows the result of the method. In this plot, the x-axis presents the number of clusters, and the y-axis is the average inertia of the clusters representing the sum of the squared distance to the centroid of the belonging cluster. 

\begin{figure}[t]
\centering
\begin{subfigure}{.33\textwidth}
  \centering
  \includegraphics[width=\linewidth]{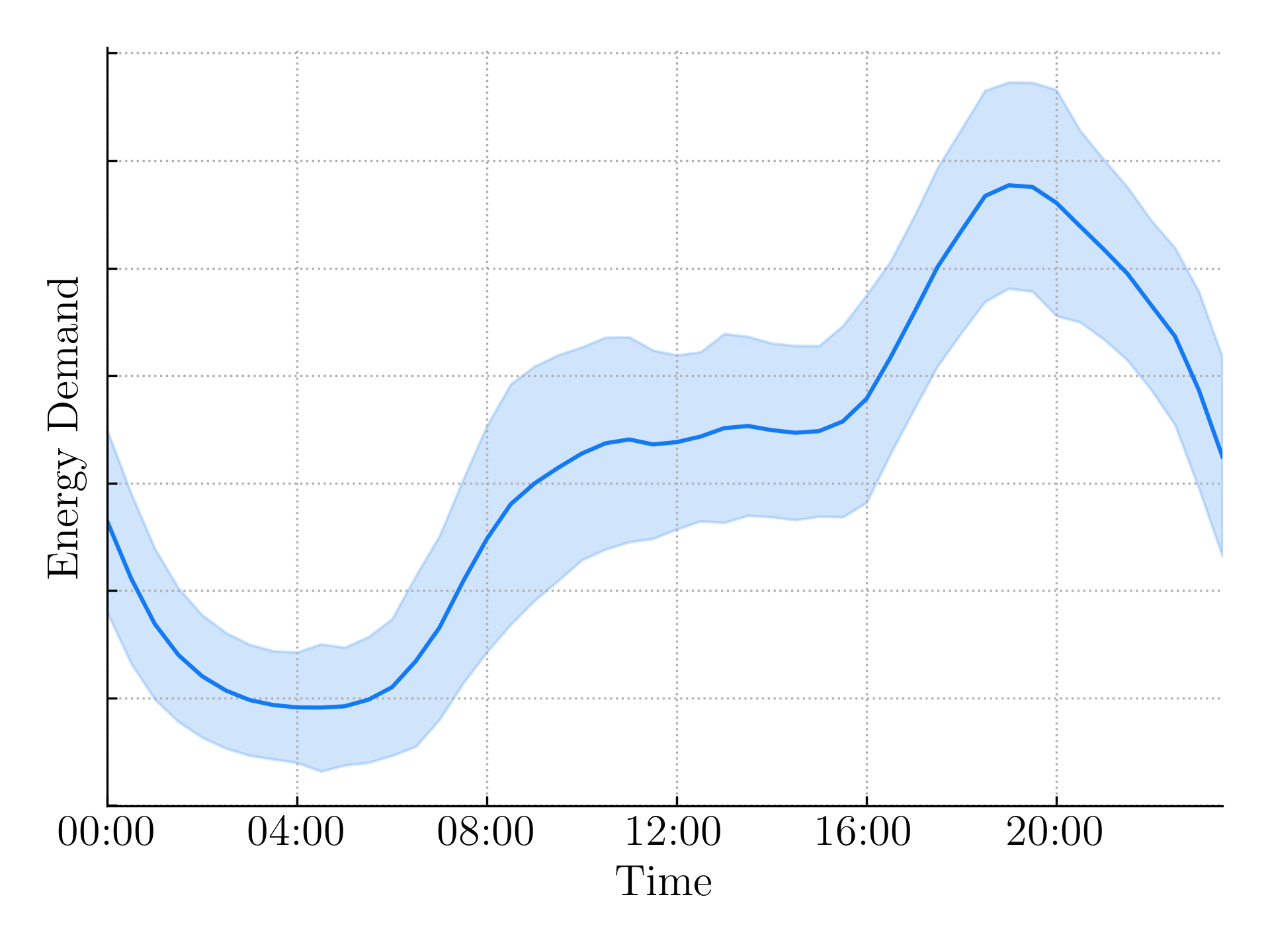}
  \caption{1299 profiles}
  \label{fig:O1}
\end{subfigure}%
\begin{subfigure}{.33\textwidth}
  \centering
  \includegraphics[width=\linewidth]{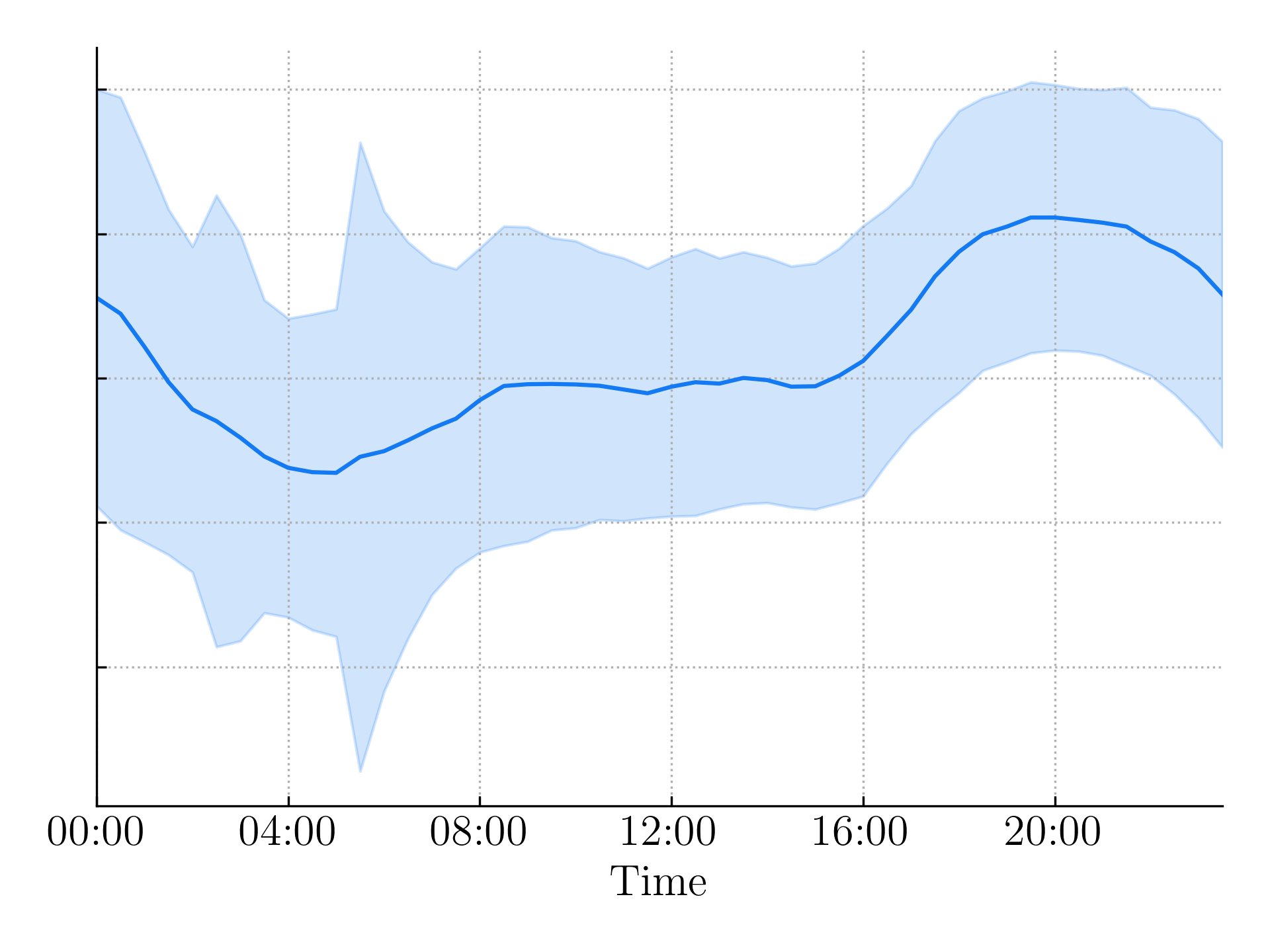}
  \caption{801 profiles}
  \label{fig:O2}
\end{subfigure}
\begin{subfigure}{.33\textwidth}
  \centering
  \includegraphics[width=\linewidth]{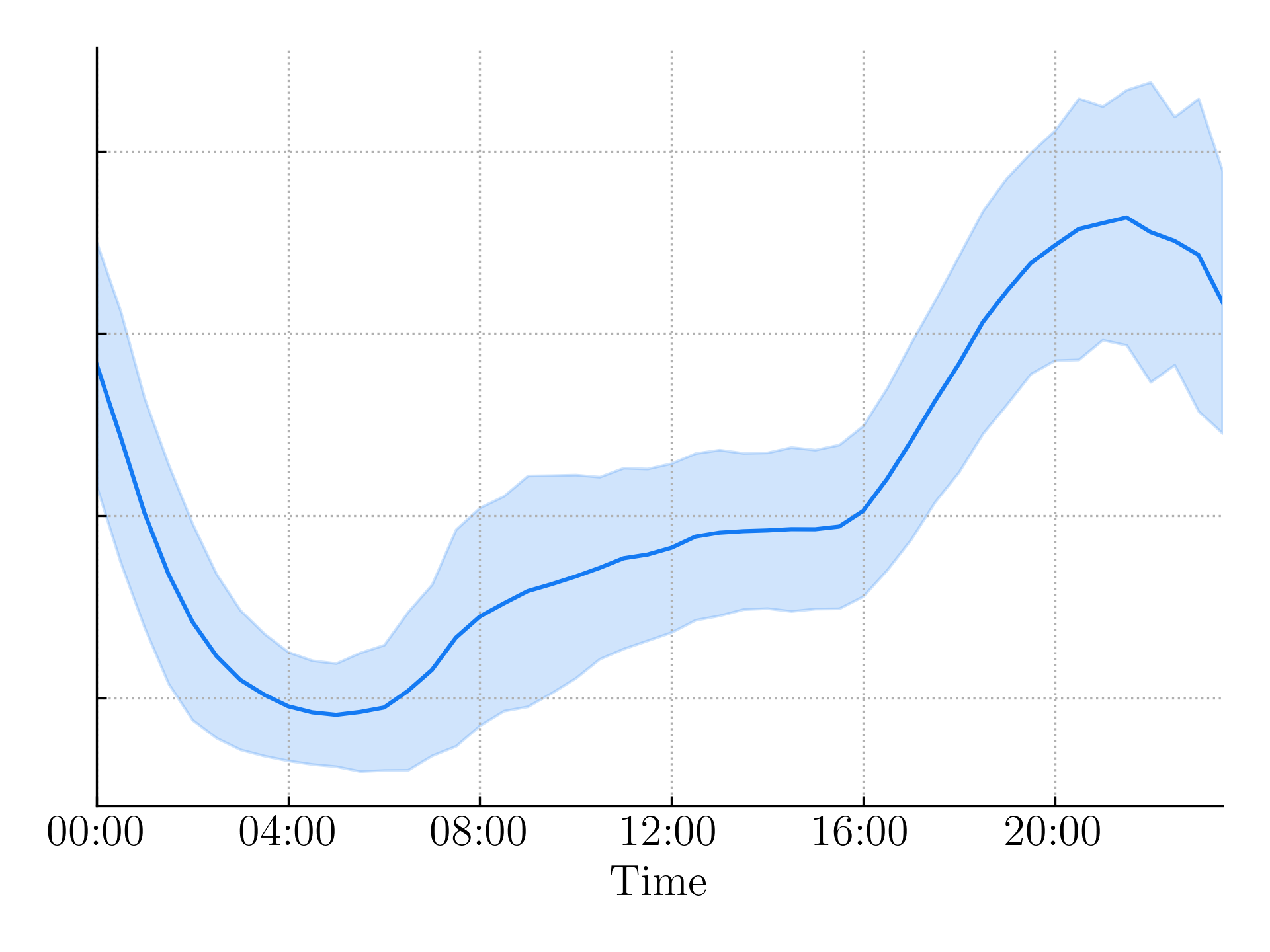}
  \caption{608 profiles}
  \label{fig:O3}
\end{subfigure}

\begin{subfigure}{.33\textwidth}
  \centering
  \includegraphics[width=\linewidth]{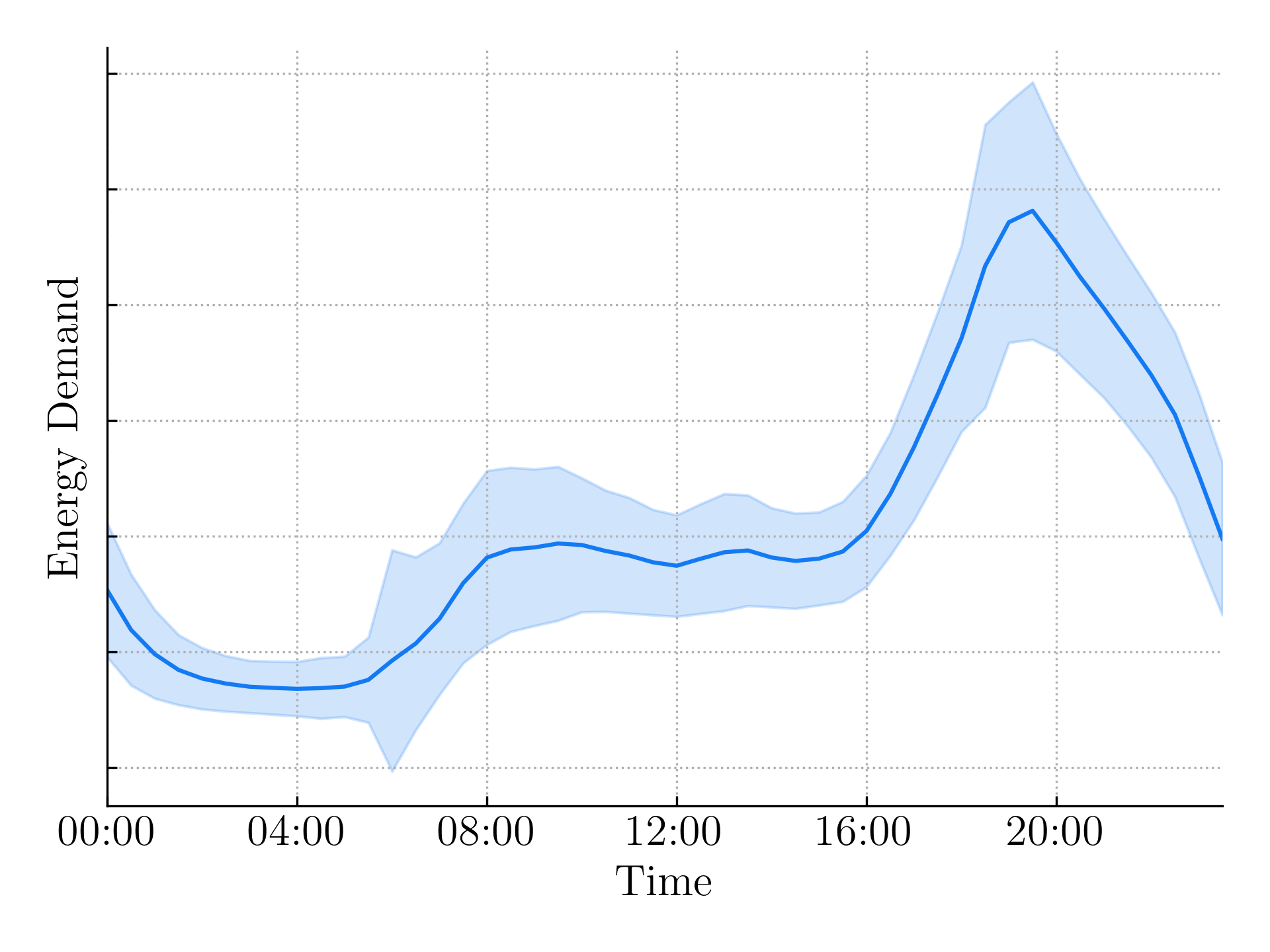}
  \caption{607 profiles}
  \label{fig:O4}
\end{subfigure}%
\begin{subfigure}{.33\textwidth}
  \centering
  \includegraphics[width=\linewidth]{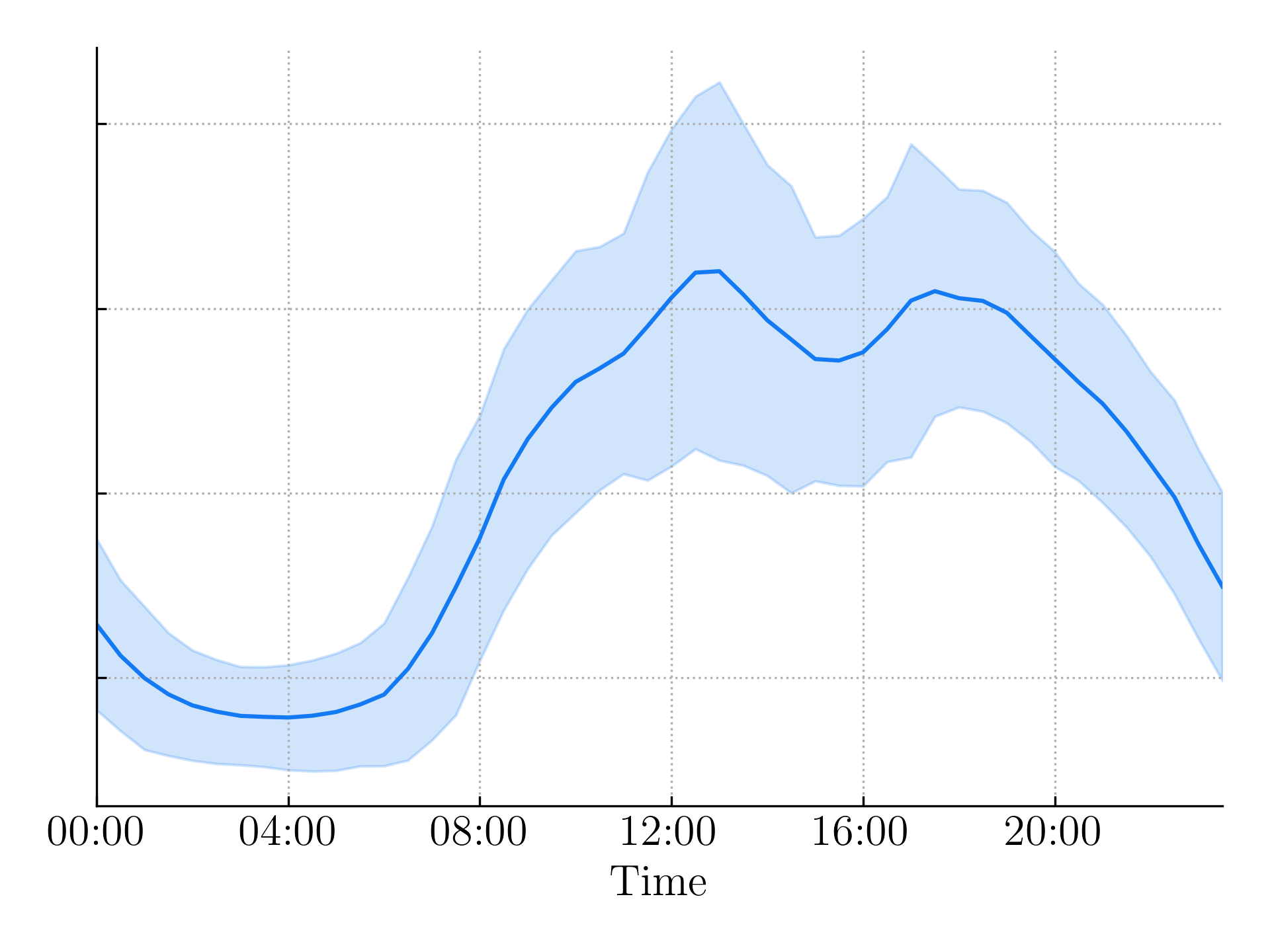}
  \caption{599 profiles}
  \label{fig:O5}
\end{subfigure}
\begin{subfigure}{.33\textwidth}
  \centering
  \includegraphics[width=\linewidth]{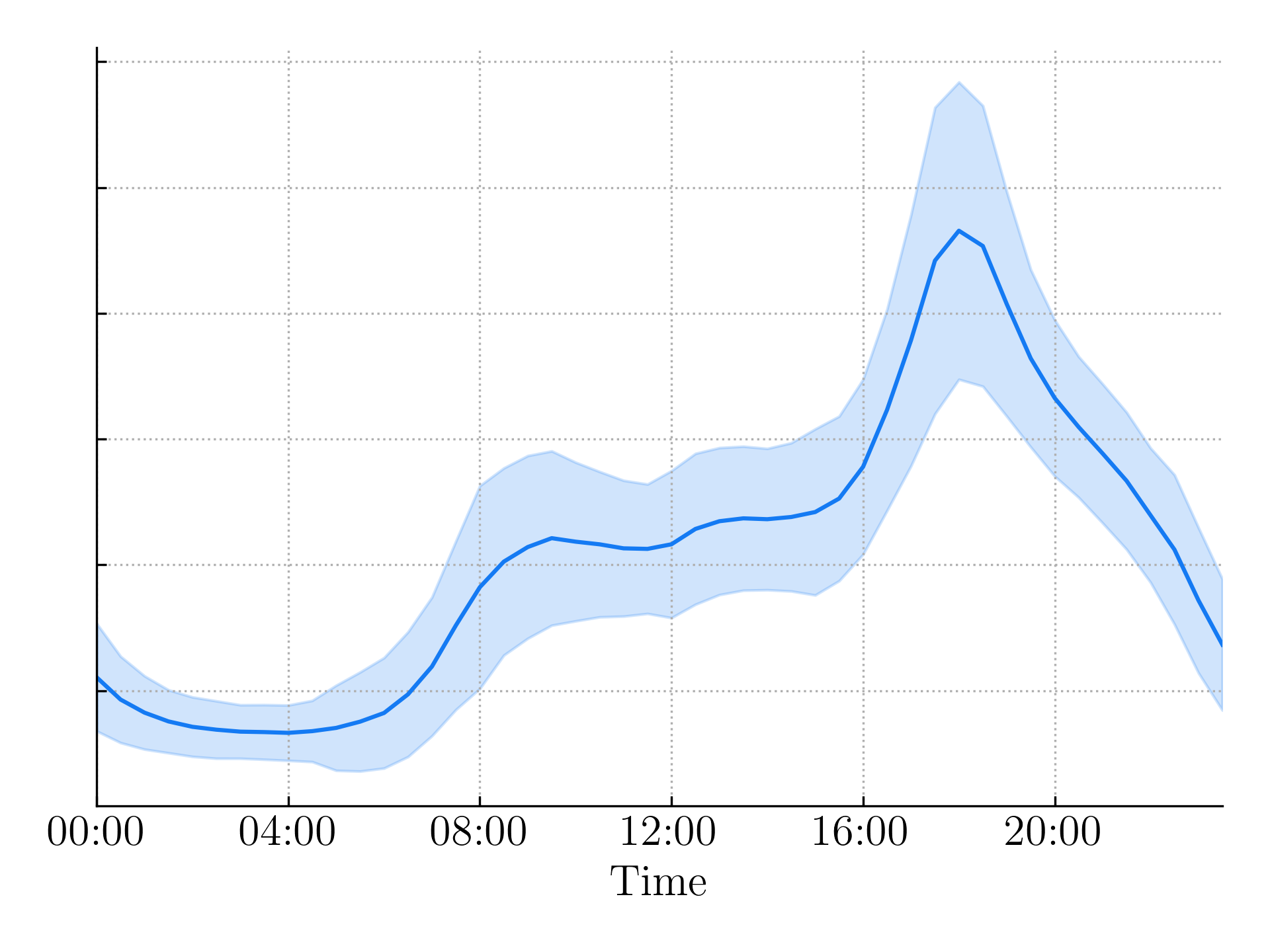}
  \caption{595 profiles}
  \label{fig:O6}
\end{subfigure}

\begin{subfigure}{.33\textwidth}
  \centering
  \includegraphics[width=\linewidth]{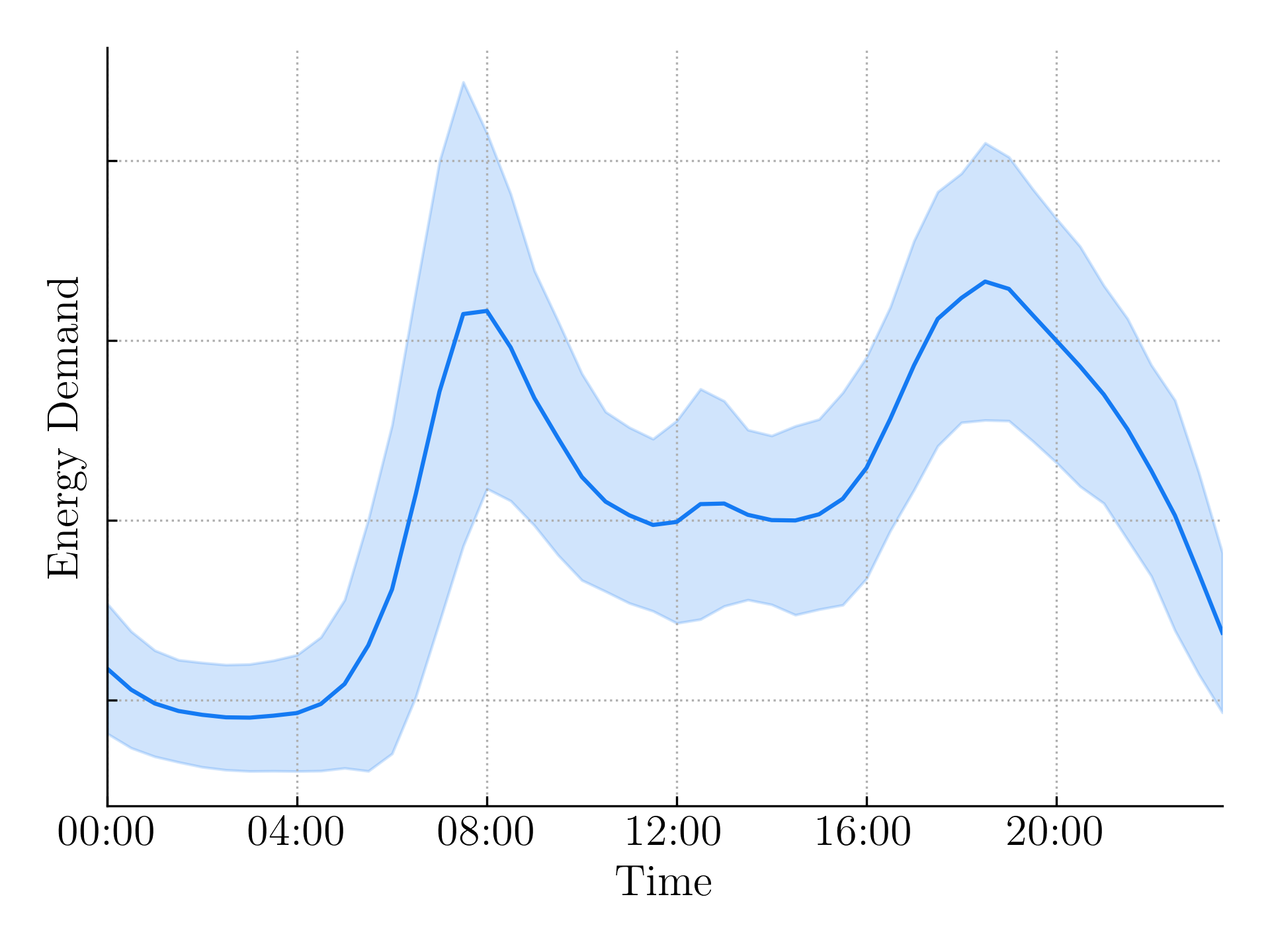}
  \caption{526 profiles}
  \label{fig:O7}
\end{subfigure}%
\begin{subfigure}{.33\textwidth}
  \centering
  \includegraphics[width=\linewidth]{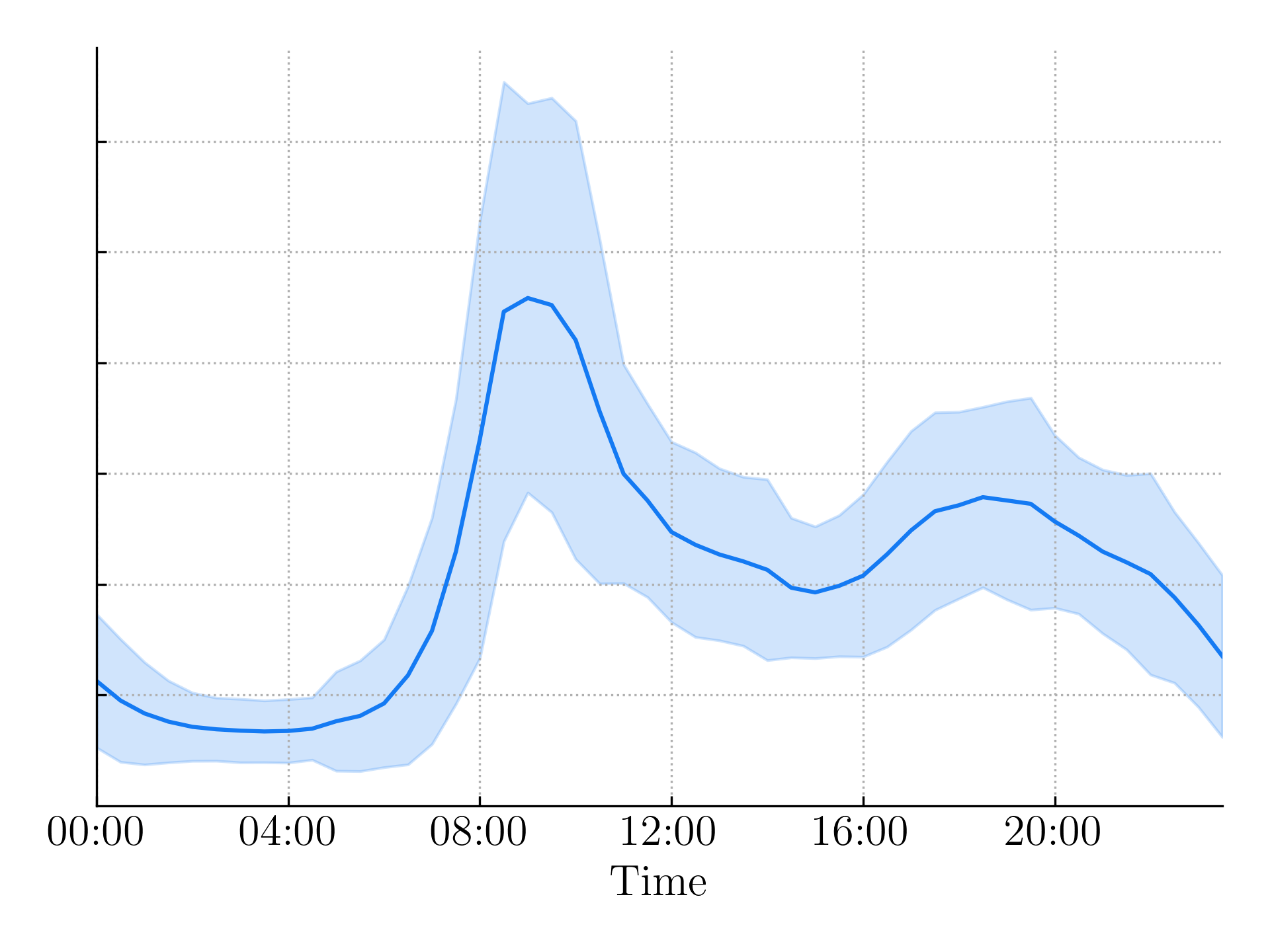}
  \caption{169 profiles}
  \label{fig:O8}
\end{subfigure}
\begin{subfigure}{.33\textwidth}
  \centering
  \includegraphics[width=\linewidth]{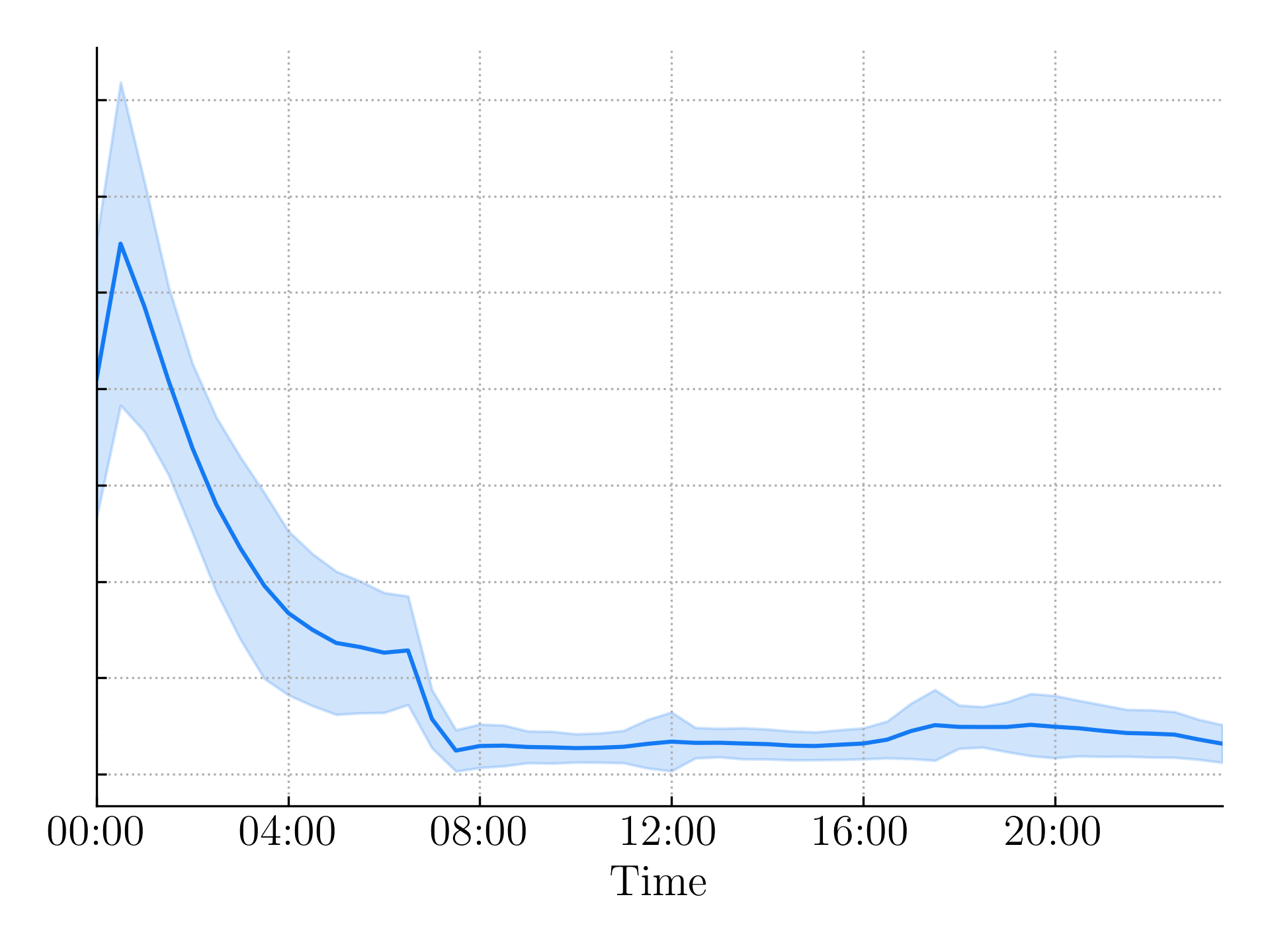}
  \caption{47 profiles}
  \label{fig:O9}
\end{subfigure}
\caption{Average daily demands for different consumer profile clusters. Error bars represent the standard deviation of the demand profiles within the cluster.}
\label{fig:o_clusters}
\end{figure}

Based on this analysis, we first identify a range of k values, i.e. 8-12, in which the inflexion point could be situated. This is further refined using the silhouette method, from which we find that the best fit value to be $k=9$ demand profile classes. These are shown in Figure~\ref{fig:o_clusters}.

For our P2P market simulation, we aimed to further restrict the number of clusters to a smaller number of more distinctive clusters, such as to be able to illustrate the effects of demand diversity. Looking in more depth at the mean demands of the 9 identified clusters in Figure~\ref{fig:o_clusters}, it can be seen that multiple clusters show the evening peak consumption pattern (i.e. profiles (a), (c) and (d) are rather similar), although there are some variations in which time the evening peak is located or the amplitude of the peak.

To reduce computation in the market simulation, all of these evening peak clusters are combined to create one large `evening peak' cluster. This finally results in the 5 clusters shown in \cref{fig:experiments:london:clusters} in \cref{sec:experiments}, which are used for the market simulation experiments.

Analogous to earlier works, we find that the majority of these profiles follow an evening peak pattern, where a noticeable peak in consumption can be observed during the early evening, often as a result of prosumers coming home from work. \Cref{fig:experiments:london:clusters:daily} shows another frequent pattern, where energy demands stay consistent throughout the day while still being low during the night. Since this is likely an effect of prosumers working from home and therefore using energy throughout the day, we call this group `work from home'. The final large group consists of prosumers that consume large quantities of energy during the morning and evening, likely pertaining to users that have a similar lifestyle to the `Evening peak' group, but also use a lot of energy in the mornings. The remaining two clusters do not have this peak consumption during the evening but at different moments during the day. The `morning peak' group consumes the majority of energy during the early mornings, while the `night owl' group consumes the majority of their energy overnight.

\end{document}